\documentclass[preprint,dvitex]{ptephy_v1}

\def\simge{\mathrel{%
       \rlap{\raise 0.511ex \hbox{$>$}}{\lower 0.511ex \hbox{$\sim$}}}}
\def\simle{\mathrel{
       \rlap{\raise 0.511ex \hbox{$<$}}{\lower 0.511ex \hbox{$\sim$}}}}

\preprintnumber{J-PARC-TH-0261, UTHEP-764} 

\begin{document}

\title{Scope and convergence of the hopping parameter expansion in finite temperature QCD with heavy quarks around the critical point}

\author[1]{Naoki Wakabayashi}
\affil[1]{Graduate School of Science and Technology, Niigata University, Niigata 950-2181, Japan}

\author[2]{Shinji Ejiri}
\affil[2]{Department of Physics, Niigata University, Niigata 950-2181, Japan}

\author[3]{Kazuyuki Kanaya}
\affil[3]{Tomonaga Center for the History of the Universe, University of Tsukuba, Tsukuba, Ibaraki 305-8571, Japan}

\author[4,5]{Masakiyo Kitazawa}
\affil[4]{Department of Physics, Osaka University, Toyonaka, Osaka 560-0043, Japan}
\affil[5]{J-PARC Branch, KEK Theory Center, Institute of Particle and Nuclear Studies, KEK, 203-1, Shirakata, Tokai, Ibaraki, 319-1106, Japan}

\date{January 19, 2022}

\begin{abstract}
Hopping parameter expansion is a useful tool to investigate heavy dynamical quarks in lattice QCD, while the range of its applicability has been sometimes questioned. 
We study the convergence and the valid range of the hopping parameter expansion in the determination of the critical point (critical quark mass) of QCD with heavy quarks at finite temperature and density.
On lattices with sufficiently large spatial extent, the terms in the hopping parameter expansion are classified into Wilson loop terms and Polyakov-type loop terms.
We first study the case of the worst convergence in which all the gauge link variables are unit matrices and thus the Wilson loops and the Polyakov-type loops get their maximum values.
We perform explicit calculation up to more than 100$^{\rm th}$ order of the hopping parameter expansion.
We show that the hopping parameter expansion is convergent up to the chiral limit of free Wilson quarks.
We then perform a Monte-Carlo simulation to measure correlation among Polyakov-type loop terms up to the 20$^{\rm th}$ order of the hopping parameter expansion. 
In previous studies, strong correlation between the leading order Polyakov loop term and the next-to-leading order bent Polyakov loop terms was reported and used to construct an effective theory to incorporate the next-to-leading order effect by a shift of the leading order coupling parameter.
We establish that the strong correlation among Polyakov-type loop terms holds also at higher orders of the hopping parameter expansion, and extend the effective theory to incorporate higher-order effects up to high orders.
Using the effective theory, we study the truncation error of the hopping parameter expansion. 
We find that the previous next-to-leading order result of the critical point for $N_t=4$ are well reliable. 
For $N_t \ge 6$, we need to incorporate higher-order effects in the effective theory. 
\end{abstract}

\maketitle

\section{Introduction}
\label{intro}

Quantum chromodynamics (QCD) is the fundamental theory of quarks and gluons. 
When all quarks are infinitely heavy, QCD tends to be pure gauge SU(3) Yang-Mills theory (quenched QCD), which is confining at low temperatures but turns into a deconfining phase by a first order deconfinement phase transition.
This transition becomes weaker as the quark mass decreases, and eventually changes to a crossover at a critical point (critical quark mass).
Identification of the critical point is important in understanding the phase structure of QCD.

In the determination of the critical point in lattice QCD with heavy quarks, the hopping parameter expansion has played important roles: 
It enables us to carry out several analytic studies and also helps us in reducing much the computational demands. 
On the other hand, the range of its applicability has been questioned sometimes.
In particular, we need to estimate the systematic error caused by the truncation of higher-order terms in the hopping parameter expansion.

To determine the phase transition point and to investigate the properties of the phase transition, we need to calculate the order parameter as a continuous function of temperature and quark mass. 
The reweighting method is useful for calculating physical quantities as continuous functions by Monte Carlo simulations \cite{rew}. 
However, in order to apply the reweighting method to investigate the quark mass dependence, it is necessary to calculate the quark determinant, which requires large computational costs. 
Therefore, in Refs.~\cite{Allton:2002zi,Ejiri:2004yw,Iwami:2015eba}, the reweighting method was performed by Taylor-expanding the quark determinant with the hopping parameters and calculating the expansion coefficients.
To investigate the heavy quark region, the Taylor expansion around zero hopping parameter (hopping parameter expansion) is useful.
The reweighting factor can be given by calculating Wilson loops and Polyakov loops, and
the reweighting factor adds the effect of dynamic quarks to the quenched QCD simulations.

In Refs.~\cite{Saito:2011fs,Saito:2013vja,Ejiri:2019csa,Kiyohara:2021smr}, 
the method by the hopping parameter expansion was adopted to investigate the critical point of QCD with heavy quarks on $N_t=4$ and 6 lattices, where $N_t$ is the temporal extent of the lattice.
This method was extended to finite densities in Refs.~\cite{Ejiri:2012rr,Saito:2013vja}.
In Refs.~\cite{Saito:2013vja,Ejiri:2019csa}, an effective method was proposed to incorporate the next-to-leading order effect of bent Polyakov loops by a shift of the coefficient for the leading-order Polyakov loop term, and the truncation error was estimated from the difference between the leading-order and the next-to-leading order calculations.
In Ref.~\cite{Kiyohara:2021smr}, a finite size scaling analysis was adopted to obtain the critical point in the thermodynamic limit, where, to overcome the overlapping problem on spatially large lattices, the configurations are generated by an action incorporating the leading-order effect by the Polyakov loop term, and the next-to-leading order effect is taken into account by the reweighting. 
The hopping parameter expansion was applied also to QCD with mixed heavy and light quarks~\cite{Ejiri:2012rr,Ejiri:2015vip}.

From these studies, it was found that the hopping parameter for the critical point $\kappa_c$ increases as $N_t$ increases. 
Similar trend is observed also in recent full QCD studies~\cite{Cuteri:2020yke,Kara:2021btt}.
At the critical point of $N_t = 4$, the effect of next-to-leading order was found to be small, 
while at that of $N_t = 6$, the effect of next-to-leading order turned out to be significant. 
As $N_t$ increases further, the truncation error of higher-order terms will increase in the determination of the critical point.
This means that, when $N_t$ exceeds a certain value, the critical point would move into a region where the hopping parameter expansion is not applicable. 
Therefore, it is important to confirm the reliability of the resulting $\kappa_c$.

In this paper, we study the convergence and the valid range of the hopping parameter expansion in the determination of the critical point in QCD with heavy quarks. 
In Sec.~\ref{hopping}, we introduce our model and the hopping parameter expansion, 
and discuss our method to calculate the expansion terms.
To obtain the lower bound of the convergence radius of the hopping parameter expansion, in Sec.~\ref{coefficients}, 
we first consider the case of the worst convergence by setting all link variables to unit matrices, 
and study the convergence radius by explicitly calculating the expansion terms up to high orders in this case. 
In~Sec.~\ref{higheroder}, we then study the effect of high order expansion terms and the systematic error due to truncation of the expansion series, and confirm if the previous calculation of the critical point for $N_t=4$ and 6 lattices by the hopping parameter expansion is within the valid range. 
We also discuss that calculation of $\kappa_c$ for $N_t=8$ will be also possible when we combine the calculation with an effective theory which incorporates higher-order effects of the hopping parameter expansion.
The effective theory is an extension of the effective theory developed in~Refs.~\cite{Saito:2013vja,Ejiri:2019csa}
to high orders, and is based on the strong correlation among Polyakov-type loops. 
By performing Monte Carlo simulations in~Sec.~\ref{correlation}, we explicitly show that the Polyakov-type loops are strongly correlated up to high orders, and calculate the coefficients for the effective theory.
Using the coefficients thus obtained, we then calculate the critical point in $2+1$-flavor QCD incorporating the higher-order effect.
Section~\ref{sec:nonzeromu} is devoted to a discussion for the case of non-zero densities.
Finally, we summarize and conclude in~Sec.~\ref{conclusion}.

\section{Hopping parameter expansion}
\label{hopping}

We study lattice QCD with $N_{\rm f}$ flavors of quarks on an $N_s^3 \times N_t$ lattice.
For gluons, we adopt the standard plaquette gauge action given by  
\begin{eqnarray}
S_g = - 6 N_{\rm site} \,\beta \, \hat{P},
\label{eq:Sg}
\end{eqnarray} 
with $\beta = 2N_c/g^2$ the gauge coupling parameter,
$N_{\rm site} = N_s^3 N_t$ the space-time lattice volume,
$N_c=3$ the number of colors, 
and $\hat{P}$ the plaquette operator defined by 
\begin{eqnarray}
\hat{P}= \frac{1}{6 N_{\rm site} N_c} \displaystyle \sum_{x,\,\mu < \nu} 
 {\rm Re \ tr_c} \left[ U_{x,\mu} U_{x+\hat{\mu},\nu}
U^{\dagger}_{x+\hat{\nu},\mu} U^{\dagger}_{x,\nu} \right] ,
\label{eq:SgP}
\end{eqnarray} 
where $U_{x,\mu}$ is the gauge link variable in the $\mu$ direction at site $x$,  
$x+ \hat\mu$ is the next site in the $\mu$ direction from $x$, 
and ${\rm tr_c}$ means the trace over the color index. 
For quarks, we adopt the standard Wilson quark action given by 
\begin{eqnarray}
S_q = \sum_{f=1}^{N_{\rm f}} \sum_{x,\,y} \bar{\psi}_x^{(f)} \, M_{xy} (\kappa_f) \, \psi_y^{(f)} ,
\label{eq:Sq}
\end{eqnarray} 
where $M_{xy}$ is the Wilson quark kernel 
\begin{eqnarray}
M_{xy} (\kappa_f) &=& \delta_{xy} 
-\kappa_f \sum_{\mu=1}^4 
\left[ (1-\gamma_{\mu})\,U_{x,\mu}\,\delta_{y,x+\hat{\mu}} 
+ (1+\gamma_{\mu})\,U_{y,\mu}^{\dagger}\,\delta_{y,x-\hat{\mu}} \right]
\nonumber \\
&\equiv&  \delta_{xy} - \kappa_f B_{xy}
\label{eq:Mxy}
\end{eqnarray} 
and $\kappa_f$ is the hopping parameter for the $f$th flavor which is related to the bare quark mass $m_f$ by
$
\kappa_f = 1/(2am_f+8)
$
with $a$ the lattice spacing.

For simplicity, we mainly consider the case of degenerate $N_{\rm f}$ flavors in this paper unless otherwise stated.
Then the expectation value of an operator ${\cal O}$ is given by 
\begin{equation}
\langle {\cal O} \rangle_{(\beta, \kappa)}
= \frac{ \int {\cal D} U \, {\cal O} \,
[\det M(\kappa)]^{N_{\rm f}} \, e^{6 \beta N_{\rm site} \hat{P}} }{
 \int {\cal D} U \, [\det M(\kappa)]^{N_{\rm f}} \, e^{6 \beta N_{\rm site} \hat{P}} } .
\label{eq:<O>}
\end{equation}
%

To study QCD in the vicinity of the heavy quark limit $\kappa=0$,
we perform the hopping parameter expansion of $\det M(\kappa)$ around $\kappa=0$. 
For the effective quark action $\ln \det M(\kappa)$ we find 
\begin{eqnarray}
\ln \det M(\kappa) \; = \;
\ln \det M(0) + N_{\rm site} \sum_{n=1}^{\infty} D_{n} \kappa^{n} , 
\label{eq:tayexp}
\end{eqnarray}
where
\begin{eqnarray}
D_n &=& \frac{1}{N_{\rm site} \ n!} \left[ \frac{\partial^n \ln \det M(\kappa)}{\partial \kappa^n} \right]_{\kappa=0}
\;=\; \frac{(-1)^{n+1} (n-1)!}{N_{\rm site} \ n!} \; {\rm Tr} 
\left[ \left( M^{-1} \, \frac{\partial M}{\partial \kappa} \right)^n \right]_{\kappa=0} .
\label{eq:derkappa0}
\end{eqnarray}
Since $M(0)=1$ for the Wilson fermion, the first term $\ln \det M(0)$ vanishes in~Eq.~(\ref{eq:tayexp}), and $M^{-1}$ in~Eq.~(\ref{eq:derkappa0}) can be neglected.
We thus find
\begin{eqnarray}
D_n &=& \frac{-1}{N_{\rm site} \ n} \; {\rm Tr} \left[ B^n \right] ,
\label{eq:derkappa}
\end{eqnarray}
with $B_{xy}=-(\partial M/\partial \kappa)_{xy}$ the hopping term defined in~Eq.~(\ref{eq:Mxy}).

Non-vanishing contributions to the trace of Eq.~(\ref{eq:derkappa}) appear only when the product of the hopping terms forms a connected closed loop in the space-time. 
We classify the closed loops in $D_n$ by the winding number $m$ which counts the number of windings in the temporal direction without distinguishing the positive and negative directions, and decompose $D_n$ as
\begin{eqnarray}
D_n &=& W(n) + \sum_{m=1}^{\infty} L_m (N_t, n) \equiv W(n) + L(N_t, n),
\label{eq:loopex}
\end{eqnarray}
where the first term $W(n)$ for $m=0$ is the summation of various $n$-step Wilson loops, and $L_m (N_t, n)$ is the summation of $n$-step Polyakov-type loops with the winding number $m$.
Here, $L_m (N_t, n)$ can be further decomposed as $L_m = L_m^+ + L_m^-$ with $L_m^+$ going in the positive direction and $L_m^-$ going in the negative direction.
These are complex numbers with $L_m^- = (L_m^+)^*$, and have the properties of
$L_m = L_m^+ + L_m^- = 2{\rm Re} L_m^+$ and 
$L_m^+ - L_m^- = 2i\ {\rm Im} L_m^+$. 

In Eq.~(\ref{eq:loopex}), the range of $m$ is actually finite for each $n$:  $L_m (N_t, n)$ vanishes when $n< m N_t$. 
For simplicity, we assume that $N_t$ is even in the followings, though extension to odd $N_t$'s is straightforward.
When $N_t$ is even, $L_m (N_t, n)=0$ also at odd $n$'s, and thus we have non-zero $D_n$ only at even $n$'s.

In the calculation of closed loops, when the loop winds around the lattice, we need to take into account the effect of boundary conditions:
When the hopping term $B$ is multiplied by the $4N_c N_{\rm site}$-component pseudo-quark field $\psi$, $\psi$ 
satisfies the following anti-periodic boundary condition in the temporal direction
\begin{eqnarray}
\psi (x_1, x_2, x_3, x_4-N_t) &=& - \psi (x_1, x_2, x_3, x_4) .
\label{eq:apbc}
\end{eqnarray}
As a result of Eq.~(\ref{eq:apbc}), one must apply a factor of $(-1)^m$ in the calculation of Polyakov-type loops with the winding number $m$.
For the influence of spatial boundary conditions, 
we assume that $N_s$ is sufficiently large such that we do not need to consider that through spatial windings.
When $n$ is smaller than $N_s$, the expression of $D_n$ in Eq.~(\ref{eq:loopex}) written in the Wilson and Polyakov-type loop operators does not explicitly depend on $N_s$.
On the other hand, the expectation values of loop operators and their probability distributions in actual Monte Carlo simulations have $N_s$ dependence through the finite volume effect in gauge configurations.
These non-trivial spatial volume dependencies affect the finite size scaling analysis.

\begin{figure}[tb]
\begin{center}
\begin{minipage}{0.48\hsize}
\begin{center}
\includegraphics[width=7.5cm]{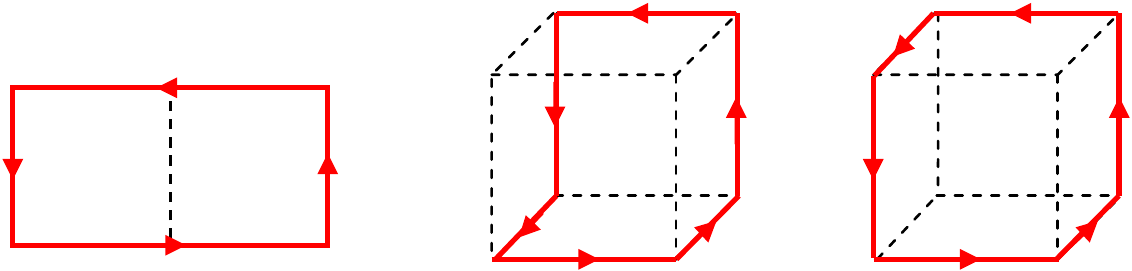}
\end{center}
\caption{Six-step Wilson loops; rectangle (left), chair-type (middle) and crown-type (right).}
\label{fig:6stepwilson}
\end{minipage}
\hspace{3mm}
\begin{minipage}{0.48\hsize}
\begin{center}
\includegraphics[width=6.2cm]{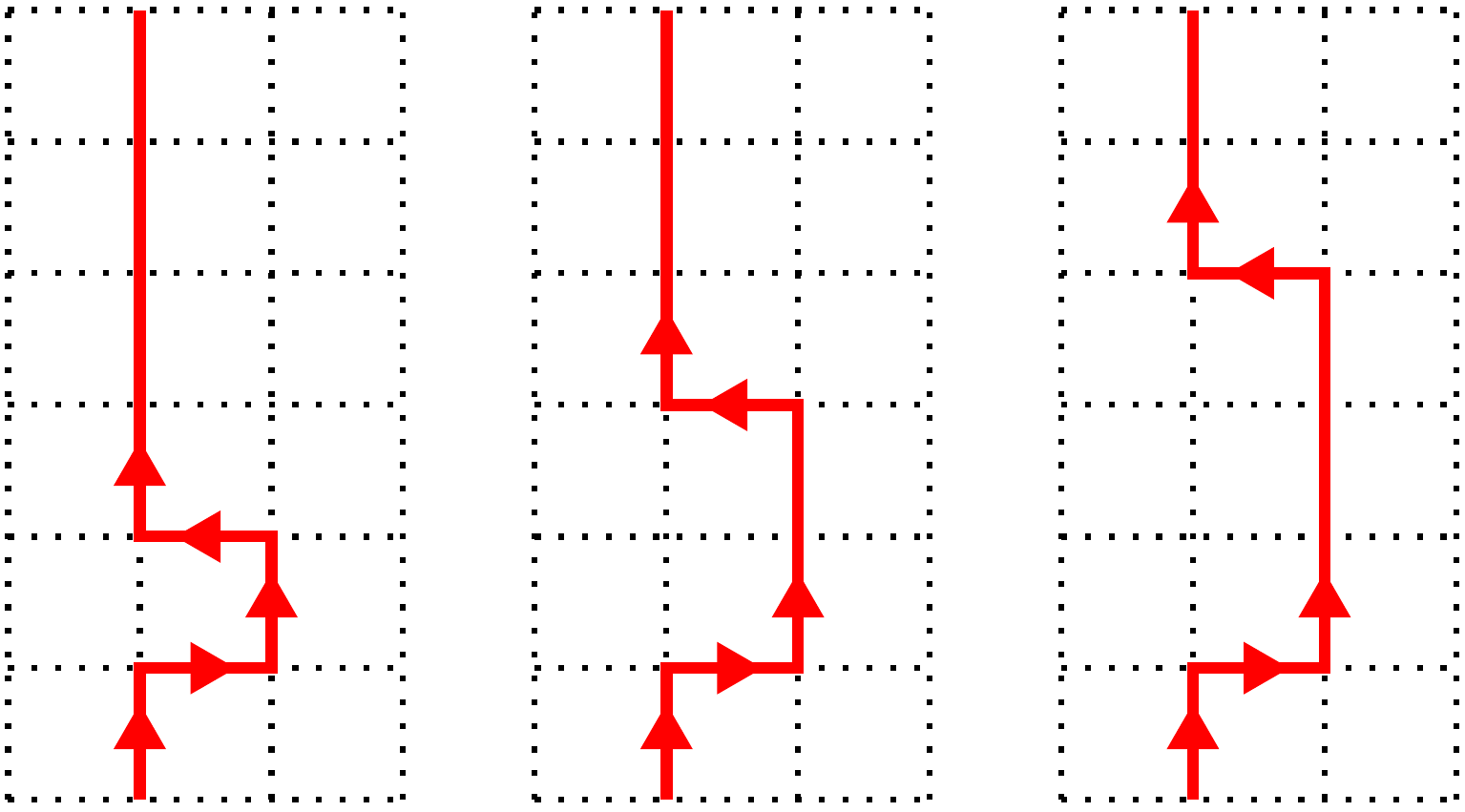}
\end{center}
\caption{$(N_t +2)$-step bent Polyakov loops,  $\hat{\Omega}_1$ (left), $\hat{\Omega}_2$ (middle) and $\hat{\Omega}_3$ (right) for $N_t=6$. The vertical direction is the temporal direction.}
\label{fig:bendedpoly}
\end{minipage}
\end{center}
\end{figure}

We define each of the Wilson and Polyakov-type loops as the average over the space-time position and normalize them such that they become unity when all $U_{x, \mu}$'s are set to the unit matrix. 
The first several terms of Eq.~(\ref{eq:loopex}) are then given by
\begin{eqnarray}
W(4) &=& 96 N_c \hat{P}, 
\label{eq:hpenlo1}
\\
W(6) &=& 256 N_c \left( 3 \hat{W}_{\rm rec}+6 \hat{W}_{\rm chair}+2 \hat{W}_{\rm crown} \right),
 \\
L_1 (N_t, N_t) &=& \frac{4 N_c \times 2^{N_t}}{N_t} {\rm Re} \hat\Omega, 
\\
L_1 (N_t, N_t +2) &=& 12 N_c \times 2^{N_t} \left( 2 \! \sum_{k=1}^{N_t/2-1} {\rm Re} \hat\Omega_k + {\rm Re} \hat\Omega_{N_t/2} \right) .
\label{eq:hpenlo}
\end{eqnarray}
Here, $\hat{W}_{\rm rec}$, $\hat{W}_{\rm chair}$, and $\hat{W}_{\rm crown}$ are 6-step Wilson loops with rectangular, chair-shaped, and crown-shaped loops as illustrated in Fig.~\ref{fig:6stepwilson}, respectively, which are averaged over the space-time position on each configuration.
$\hat\Omega$ is the Polyakov loop defined as
\begin{equation}
\hat\Omega = \frac{1}{N_c N_s^3}
\displaystyle \sum_{\vec{x}} {\rm tr_c } \left[ 
U_{\vec{x},4} U_{\vec{x}+\hat{4},4} U_{\vec{x}+2 \cdot \hat{4},4} 
\cdots U_{\vec{x}+(N_t -1) \cdot \hat{4},4} \right] 
\label{eq:ploop}
\end{equation}
with $\sum_{\vec{x}}$ for a summation over the sites on a time slice, 
and $\hat\Omega_k$'s are bent Polyakov loops with $N_t +2$ steps, shown in Fig.~\ref{fig:bendedpoly} 
\cite{Ejiri:2019csa,Kiyohara:2021smr}.

\subsection{Calculation of $W(n)$ and $L_m(N_t, n)$}
\label{sec:twist}

Effects of $W(n)$ and $L_m(N_t, n)$ can be in part separated by adopting different temporal boundary conditions. 
For the case of periodic boundary condition at the temporal boundary, the $n$th coefficient $D_n$ of the hopping parameter expansion of $\ln \det M(\kappa)$ is changed to  
\begin{eqnarray}
D_n^+ = W(n) + \sum_{m=1}^{\infty} (-1)^m L_m (N_t, n).
\label{eq:loopexp}
\end{eqnarray}
Here, $(-1)^m$ is multiplied because $L_m(N_t,n)$ is defined with the anti-periodic boundary condition.
Combining $D_n$ and $D_n^+$, we can separate the contributions of even and odd winding numbers as 
\begin{eqnarray}
\frac{D_n + D_n^+}{2} &=& W(n) + \sum_{m=1}^{\infty} L_{2m} (N_t, n), 
\label{eq:l2p}
\\
\frac{D_n - D_n^+}{2} &=& \sum_{m=1}^{\infty} L_{2m-1} (N_t, n).
\label{eq:l2m}
\end{eqnarray}
These relations enables us to determine $W(n)$ and $L_1(N_t, n)$ for small $n$:
\begin{eqnarray}
W(n) &=& \frac{D_n + D_n^+}{2}  \ \ \ \ {\rm for} \ \  n < 2N_t ,
\label{eq:wddp}
\\
L_1(N_t, n) &=& \frac{D_n - D_n^+}{2}   \ \ \ \ {\rm for} \ \  n < 3N_t .
\label{eq:l1ddp}
\end{eqnarray}

We can further decompose the contributions of Wilson and Polyakov-type loop terms by introducing twisted boundary conditions at the temporal boundary. 
To show the case to decompose $m$ in modulus 4, 
we impose a boundary condition 
$\psi (x_1, x_2, x_3, x_4-N_t) = i \psi (x_1, x_2, x_3, x_4)$ or equivalently $\psi (x_1, x_2, x_3, x_4+N_t) = -i \psi (x_1, x_2, x_3, x_4)$.
Then the $n$th coefficient of $\ln \det M(\kappa)$ with this boundary condition reads
\begin{eqnarray}
D_n^i &=& W(n) + \sum_{m=1}^{\infty} (-i)^m L_m^+ (N_t, n) + \sum_{m=1}^{\infty} i^m L_m^- (N_t, n)
\nonumber \\
&=& W(n) + \sum_{m=1}^{\infty} (-1)^m L_{2m} (N_t, n) - 2 \sum_{m=1}^{\infty} (-1)^m {\rm Im} L_{2m-1}^+ (N_t, n).
\label{eq:loopexi}
\end{eqnarray} 
Similarly, when we impose the boundary condition 
$\psi (x_1, x_2, x_3, x_4-N_t) = -i \,\psi(x_1, x_2, x_3, x_4)$, 
we find
\begin{eqnarray}
D_n^{-i} = W(n) + \sum_{m=1}^{\infty} (-1)^m L_{2m} (N_t, n) + 2 \sum_{m=1}^{\infty} (-1)^m {\rm Im} L_{2m-1}^+ (N_t, n)
\label{eq:loopexmi}
\end{eqnarray}
Combining them, we obtain relations including
\begin{eqnarray}
\frac{D_n + D_n^+ + D_n^i + D_n^{-i}}{4} &=& W(n) + \sum_{m=1}^{\infty} L_{4m} (N_t, n), 
\\ 
\frac{D_n + D_n^+ - D_n^i - D_n^{-i}}{4} &=& \sum_{m=1}^{\infty} L_{4m-2} (N_t, n) .
\label{eq:l4m}
\end{eqnarray}
From the last equation, we find that $L_2(N_t,n)=(D_n + D_n^+ - D_n^i - D_n^{-i})/4$ when $n<6N_t$.

The separation of $D_n$ into $W(n)$ and $L_m(N_t,n)$
can be established more generally by combining the values of
${\rm Tr}[B^n]$ calculated with various twisted boundary conditions.
If we impose the boundary condition: 
\begin{eqnarray}
\psi (x_1, x_2, x_3, x_4-N_t) &=& e^{i \theta} \,\psi(x_1, x_2, x_3, x_4),  
\label{eq:twisttheta}
\end{eqnarray}
the expansion term, denoted by $D_n^{\theta}$, becomes
\begin{eqnarray}
D_n^{\theta} &=& W(n) + \sum_{m=1}^{\infty} (-1)^m e^{i m \theta} L_m^+ (N_t, n) + \sum_{m=1}^{\infty}(-1)^m e^{-i m \theta} L_m^- (N_t, n) 
\nonumber \\
&=& W(n) + \sum_{m=1}^{\infty} (-1)^m \cos(m \theta)\, L_m (N_t, n) + 2\sum_{m=1}^{\infty}(-1)^m \sin(m \theta)\, {\rm Im}\, L_m^+ (N_t, n).
\label{eq:looptheta}
\end{eqnarray}
Then,
\begin{eqnarray}
\frac{D_n^{\theta} + D_n^{2 \pi -\theta}}{2}
&=& W(n) + \sum_{m=1}^{\infty} (-1)^m \cos(m \theta) \, L_m (N_t, n).
\label{eq:looptheta2}
\end{eqnarray}

Let $D_{n,w}$ be the value of $(-1/N_{\rm site}n){\rm Tr}[B^n]$ calculated with the twisted boundary condition with $\theta = \pi w/Y$, 
i.e., $\psi (x_1, x_2, x_3, x_4-N_t) = e^{i \pi w/Y} \,\psi(x_1, x_2, x_3, x_4)$, 
where $Y$ is the maximum winding number, i.e., the largest integer satisfying $Y \le n/N_t$, and
$w=0$, 1, $\cdots$, $2Y-1$.
Then $W(n)$ and $\tilde{L}_m(N_t,n)=(-1)^m L_m(N_t,n)$ are
related to $D_{n,w}$ as
\begin{eqnarray}
  \begin{bmatrix} D_{n,0} \\
   (D_{n, 1} + D_{n, 2Y-1})/2 \\
   \vdots \\ 
   (D_{n, Y-1} + D_{n, Y+1})/2 \\
   D_{n,Y} \end{bmatrix}
  = G
  \begin{bmatrix} W(n) \\ \tilde{L}_1 (N_t,n) \\ \vdots \\ 
  \tilde{L}_{Y-1} (N_t,n) \\ \tilde{L}_Y (N_t,n) \end{bmatrix},
  \label{eq:B=ML}
\end{eqnarray}
where $G$ is a $(Y+1)\times(Y+1)$ matrix whose components are given by 
$G_{jk} = \cos(\pi jk/Y)$ with $j,k=0$, 1, $\cdots$, $Y$.
By inversely solving Eq.~(\ref{eq:B=ML}) one obtains $W(n)$ and $L_m(N_t,n)$
for $m=1$, $\cdots$, $Y$. 
In particular, $W(n)$ is given by 
\begin{eqnarray}
W(n) = \frac{1}{2Y} \sum_{w=0}^{2Y-1} D_{n, w} .
\label{eq:lmtheta}
\end{eqnarray}

\section{Expansion coefficients and convergence radius for the case $U_{x, \mu}=\mathbf{1}$}
\label{coefficients}

In this section, we calculate the Wilson loop terms $W (n)$ and the Polyakov-type loop terms $L_m (N_t, n)$ in the weak coupling limit, i.e., the case that all link variables are set to the unit matrix, $U_{x, \mu}=\mathbf{1}$.
The $n^{\rm th}$-order term of hopping parameter expansion is given as a summation of Wilson and Polyakov loops of various shapes with length $n$.
The Wilson and Polyakov loops in $W (n)$ and $L_m (N_t, n)$ get their maximum value one in this case.
We denote $W (n)$ and $L_m (N_t, n)$ for the case $U_{x, \mu}=\mathbf{1}$ 
as $W^0 (n)$ and $L_m^0 (N_t, n)$, respectively.

In order for the hopping parameter expansion to converge, it is required that 
\begin{eqnarray}
|W(n-2) \kappa^{n-2}| > |W(n) \kappa^n| \ \  {\rm and} \ \  
|L(N_t,n-2) \kappa^{n-2}| > |L(N_t,n) \kappa^n|
\end{eqnarray}
in the limit where $n$ is large.
Therefore, the applicable range is estimated by 
$\kappa < \sqrt{|W(n-2)/W(n)|}$ and $\kappa < \sqrt{|L(N_t,n-2)/L(N_t,n)|}$.
Because the variety of $n$-step loops increase rapidly with $n$, 
the absolute values of $W (n)$ and $L_m (N_t, n)$ increase rapidly as $n$ increases.
On the other hand, because the values of the Wilson and Polyakov loops on actual configurations at $\beta < \infty$ decrease exponentially as the loop length $n$ increases, 
the ratios $|W/W^0|$ and $|L_m/L_m^0|$ are at most one and are decreasing functions of $n$.
Then, the expansion converges in the region of $\kappa$ that satisfies the condition given by 
\begin{eqnarray}
\kappa < \sqrt{\left| \frac{W^0 (n-2)}{W^0 (n)} \right| } < \sqrt{\left| \frac{W(n-2)}{W(n)} \right| }  \ \ {\rm and} \ \ 
\kappa < \sqrt{\left| \frac{L^0 (N_t,n-2)}{L^0 (N_t,n)} \right| } < \sqrt{\left| \frac{L (N_t,n-2)}{L (N_t,n)} \right|}.
\end{eqnarray}
Thus, information of $W^0 (n)$ and $L_m^0 (N_t, n)$ gives lower bounds for the convergence radius of the hopping parameter expansion.
As shown below, $W^0 (n)$ and $L_m^0 (N_t, n)$ can be easily calculated numerically up to very higher-order terms, and the convergence condition can be estimated.
We then study the ratios $|W/W^0|$ and $|L_m/L_m^0|$ on actual configurations by a Monte-Carlo simulation in~Sec.~\ref{correlation}.

\subsection{Numerical values for $W^0(n)$ and $L_m^0(N_t, n)$}

Setting $N_c=3$, the first several terms are obtained from Eqs.~(\ref{eq:hpenlo1})--(\ref{eq:hpenlo}) as
\begin{eqnarray}
&&
W^0(4) = 288, \hspace{7mm}
W^0(6) = 768 \times (3+6+2) = 8448,
\label{eq:hpeu1w}
\\
&&
L_1^0 (N_t, N_t) = \frac{12\times 2^{N_t}}{N_t}, \hspace{7mm}
L_1^0 (N_t, N_t +2) = 36 \times 2^{N_t} (N_t -1).
\label{eq:hpeu1}
\end{eqnarray}
For the present case of $U_{x, \mu}=\mathbf{1}$, we can also show 
\begin{eqnarray}
L_m^0 (N_t, n) = (-1)^{m-1} L_1^0 (m N_t, n).
\label{eq:lmu1}
\end{eqnarray}
Using this property, we can calculate some $L_m^0$ for $m> 1$ by substituting $L_1^0$ of Eq.~(\ref{eq:hpeu1}).

We calculate $D_n$ for $U_{x, \mu}= \mathbf{1}$ on a lattice with an $N_t$ and sufficiently large $N_s$.
Since the link variables are uniform in this case, calculation of one diagonal element for the position index is sufficient.
For the color and spinor indexes, we take the trace over them.
More concretely, we prepare a pseudo-fermion field $\vec{e}_i$ having non-vanishing element only at a position and at the $i$th combination of the color and spinor indexes, we calculate the diagonal element $[B^n]_{ii}= {\vec{e}_i}^{\;\dag} B^n \vec{e}_i$ for all combinations of $i$. We then calculate $D_n$ from $\sum_{i=1}^{4N_c} [B^n]_{ii}$.

To compute $D_n$ up to $n=n_{max}$ on lattices with $N_t=4$, 6, $\cdots$, $n_{max}/2+1$, the spatial lattice size $N_s=n_{max}+2$ is sufficient. 
We also compute $D_n^+$ with the periodic boundary condition.
For $n < 2 N_t$ and $n < N_s$, $W^0(n)$ is given by Eq.~(\ref{eq:wddp}). 
We compute $D_n$ and $D_n^+$ on a $56^3 \times 28$ lattice.
The results of $W^0(n)$ are listed in Table~\ref{tab:wn} up to $n=50$. 
The sum of the Polyakov-type loop terms, 
\begin{eqnarray}
L^0(N_t, n) &=& \sum_{m=1}^{\infty} L_m^0 (N_t, n),
\end{eqnarray}
is then obtained by 
$L^0(N_t, n) = D_n (N_t) - W^0(n)$ .
The terms corresponding to each loop in $W(n)$ and $L_m(N_t,n)$ can take both positive and negative signs depending on the product of gamma matrices of the hopping term Eq.~(\ref{eq:Mxy}), and, in the case of Polyakov-type loops, also on the temporal boundary condition.
The total sign of $W(n)$ and $L_m(N_t,n)$ is determined by which sign is dominant.

\begin{table}[t]
\begin{center}
\caption{Wilson loop terms $W^0(n)$ for the case $U_{x, \mu}= \mathbf{1}$.}
\label{tab:wn}
\scalebox{1.0}{
\begin{tabular}{cr||cr||cr}
\hline
$W^0(4)$  & $288$            \ & $W^0(20)$ &  $1.54422361 \times 10^{14}$ \ & $W^0(36)$ & $-5.58410362 \times 10^{27}$ \\
$W^0(6)$  & $8\ 448$           \ & $W^0(22)$ &  $2.83682900 \times 10^{15}$ \ & $W^0(38)$ & $-2.91018925 \times 10^{29}$ \\
$W^0(8)$  & $245\ 952$         \ & $W^0(24)$ & $-2.40028584 \times 10^{16}$ \ & $W^0(40)$ & $-1.50223497 \times 10^{31}$ \\
$W^0(10)$ & $7\ 372\ 800$        \ & $W^0(26)$ & $-6.88836562 \times 10^{18}$ \ & $W^0(42)$ & $-7.71380102 \times 10^{32}$ \\
$W^0(12)$ & $225\ 232\ 896$      \ & $W^0(28)$ & $-5.41133954 \times 10^{20}$ \ & $W^0(44)$ & $-3.95168998 \times 10^{34}$ \\
$W^0(14)$ & $6\ 906\ 175\ 488$     \ & $W^0(30)$ & $-3.39122203 \times 10^{22}$ \ & $W^0(46)$ & $-2.02386871 \times 10^{36}$ \\
$W^0(16)$ & $208\ 431\ 502\ 848$   \ & $W^0(32)$ & $-1.93668514 \times 10^{24}$ \ & $W^0(48)$ & $-1.03783044 \times 10^{38}$ \\
$W^0(18)$ & $6.00259179 \times 10^{12}$ \ & $W^0(34)$ & $-1.05424635 \times 10^{26}$ \ & $W^0(50)$ & $-5.33468075 \times 10^{39}$ \\
\hline
\end{tabular}
}
\end{center}
\end{table}

For each of the Polyakov-type loop terms $L_m^0(N_t, n)$, we measure $D_n$ and $D_n^+$ up to $n=30$ for the case $U_{x, \mu}= \mathbf{1}$ on lattices  with $N_s=32$ and $N_t=4$--24.
We first calculate $L_1^0 (N_t, n)$ for $n < 3N_t$ by Eq.~(\ref{eq:l1ddp}) for each $N_t$, 
and then $L_m^0 (N_t, n)$ for $m \geq 2$ as much as possible by Eq.~(\ref{eq:lmu1}). 
When $n < 5N_t$, we also have the equation $L_1^0 (N_t, n) = (D_n - D_n^+)/2 - L_1 (2N_t, n)$, and when $n < 7N_t$, $L_1^0 (N_t, n) = (D_n - D_n^+)/2 - L_1^0 (2N_t, n) - L_1^0 (3N_t, n)$.
By repeating this procedure, $L_m^0 (N_t, n)$ is calculated for all values of $m$.
The results for $n \leq 30$ are summarized in Table~\ref{tab:lmn}.
In this table, the theoretical value of Eq.~(\ref{eq:hpeu1}) is given when available.
We find that, for these values of $(N_t,n)$, the term with $m=1$ is dominant and the contributions of the terms with $m=2$--4 are small.

\begin{table}[t]
\begin{center}
\caption{Polyakov-type loop expansion terms $L^0_m (N_t, n)$ for the case $U_{x, \mu}= \mathbf{1}$.}
\label{tab:lmn}
\scalebox{0.96}{
\begin{tabular}{cr||cr||crc}
\hline
$L^0_1(4,4)$   & \hspace{26mm} 48 & $L^0_1(10,10)$ & \hspace{18mm}  1 228.8 & $L^0_1(18,18)$ &  \hspace{12mm} 174 762.67 & \\
$L^0_1(4,6)$   &         1 728 & $L^0_1(10,12)$ &      331 776 & $L^0_1(18,20)$ &   160 432 128 & \\
$L^0_1(4,8)$   &        45 792 & $L^0_1(10,14)$ &    52 862 976 & $L^0_1(18,22)$ & 75 497 472 000 & \\
$L^0_1(4,10)$  &       645 120 & $L^0_1(10,16)$ &  6 258 180 096 & $L^0_1(18,24)$ & $2.36626 \times 10^{13}$ & \\
$L^0_1(4,12)$  &    $-$26 224 128 & $L^0_1(10,18)$ & $5.99330 \times 10^{11}$ & $L^0_1(18,26)$ & $5.50232 \times 10^{15}$ & \\
$L^0_1(4,14)$  &  $-$3 201 067 008 & $L^0_1(10,20)$ & $4.87727 \times 10^{13}$ & $L^0_1(18,28)$ & $1.01809 \times 10^{18}$ & \\
$L^0_1(4,16)$  & $-2.14087 \times 10^{11}$ & $L^0_1(10,22)$ & $3.47446 \times 10^{15}$ & $L^0_1(18,30)$ & $1.57315 \times 10^{20}$ & \\
$L^0_1(4,18)$  & $-1.19007 \times 10^{13}$ & $L^0_1(10,24)$ & $2.20156 \times 10^{17}$ & $L^0_1(20,20)$ &    629 145.6 & \\
$L^0_1(4,20)$  & $-6.00757 \times 10^{14}$ & $L^0_1(10,26)$ & $1.24531 \times 10^{19}$ & $L^0_1(20,22)$ &   717 225 984 & \\
$L^0_1(4,22)$  & $-2.84486 \times 10^{16}$ & $L^0_1(10,28)$ & $6.20798 \times 10^{20}$ & $L^0_1(20,24)$ & $4.11140 \times 10^{11}$ & \\
$L^0_1(4,24)$  & $-1.28105 \times 10^{18}$ & $L^0_1(10,30)$ & $2.59861 \times 10^{22}$ & $L^0_1(20,26)$ & $1.54445 \times 10^{14}$ & \\
$L^0_1(4,26)$  & $-5.50874 \times 10^{19}$ & $L^0_1(12,12)$ &        4 096 & $L^0_1(20,28)$ & $4.24543 \times 10^{16}$ & \\
$L^0_1(4,28)$  & $-2.25576 \times 10^{21}$ & $L^0_1(12,14)$ &     1 622 016 & $L^0_1(20,30)$ & $9.17892 \times 10^{18}$ & \\
$L^0_1(4,30)$  & $-8.69402 \times 10^{22}$ & $L^0_1(12,16)$ &   360 603 648 & $L^0_1(22,22)$ &  2 287 802.18 & \\
$L^0_1(6,6)$   &          128 & $L^0_1(12,18)$ & 57 416 810 496 & $L^0_1(22,24)$ &  3 170 893 824 & \\
$L^0_1(6,8)$   &        11 520 & $L^0_1(12,20)$ & $7.19497 \times 10^{12}$ & $L^0_1(22,26)$ & $2.17478 \times 10^{12}$ & \\
$L^0_1(6,10)$  &       716 544 & $L^0_1(12,22)$ & $7.51820 \times 10^{14}$ & $L^0_1(22,28)$ & $9.64167 \times 10^{14}$ & \\
$L^0_1(6,12)$  &     35 891 712 & $L^0_1(12,24)$ & $6.80443 \times 10^{16}$ & $L^0_1(22,30)$ & $3.09123 \times 10^{17}$ & \\
$L^0_1(6,14)$  &   1 464 910 848 & $L^0_1(12,26)$ & $5.46987 \times 10^{18}$ & $L^0_1(24,24)$ &     8 388 608 & \\
$L^0_1(6,16)$  &  43 817 011 200 & $L^0_1(12,28)$ & $3.96931 \times 10^{20}$ & $L^0_1(24,26)$ & 13 891 534 848 & \\
$L^0_1(6,18)$  & $ 3.17933 \times 10^{11}$ & $L^0_1(12,30)$ & $2.62442 \times 10^{22}$ & $L^0_1(24,28)$ & $1.12307 \times 10^{13}$ & \\
$L^0_1(6,20)$  & $-8.54676 \times 10^{13}$ & $L^0_1(14,14)$ &    14 043.43 & $L^0_1(24,30)$ & $5.80075 \times 10^{15}$ & \\
$L^0_1(6,22)$  & $-9.18906 \times 10^{15}$ & $L^0_1(14,16)$ &     7 667 712 & $L^0_1(26,26)$ & 30 973 321.85 & \\
$L^0_1(6,24)$  & $-6.76634 \times 10^{17}$ & $L^0_1(14,18)$ &  2 263 154 688 & $L^0_1(26,28)$ & 60 397 977 600 & \\
$L^0_1(6,26)$  & $-4.25366 \times 10^{19}$ & $L^0_1(14,20)$ & $4.64539 \times 10^{11}$ & $L^0_1(28,28)$ & 115 043 766.9 & \\
$L^0_1(6,28)$  & $-2.43350 \times 10^{21}$ & $L^0_1(14,22)$ & $7.33145 \times 10^{13}$ & $L^0_1(28,30)$ & $2.60919 \times 10^{11}$ & \\
$L^0_1(6,30)$  & $-1.30192 \times 10^{23}$ & $L^0_1(14,24)$ & $9.47783 \times 10^{15}$ & $L^0_1(30,30)$ & 429 496 729.6 & \\
\cline{5-7}
$L^0_1(8,8)$   &          384 & $L^0_1(14,26)$ & $1.04744 \times 10^{18}$ & & & \\
$L^0_1(8,10)$  &        64 512 & $L^0_1(14,28)$ & $1.01916 \times 10^{20}$ & \multicolumn{3}{l}{ $L^0_1(8,n)=-L^0_2(4,n)$ } \\
$L^0_1(8,12)$  &      6 842 880 & $L^0_1(14,30)$ & $8.91247 \times 10^{21}$ & \multicolumn{3}{l}{ $L^0_1(12,n)=-L^0_2(6,n)=L^0_3(4,n)$ }  \\
$L^0_1(8,14)$  &    563 816 448 & $L^0_1(16,16)$ &       49 152 & \multicolumn{3}{l}{ $L^0_1(16,n)=-L^0_2(8,n)=-L^0_4(4,n)$ } \\
$L^0_1(8,16)$  &  38 644 455 168 & $L^0_1(16,18)$ &    35 389 440 & \multicolumn{3}{l}{ $L^0_1(18,n)=L^0_3(6,n)$ } \\
$L^0_1(8,18)$  & $2.27266 \times 10^{12}$ & $L^0_1(16,20)$ & 13 373 669 376 & \multicolumn{3}{l}{ $L^0_1(20,n)=-L^0_2(10,n)=L^0_5(4,n)$ } \\
$L^0_1(8,20)$  & $1.15216 \times 10^{14}$ & $L^0_1(16,22)$ & $3.43220 \times 10^{12}$ & \multicolumn{3}{l}{ $L^0_1(24,n)=-L^0_2(12,n)=L^0_3(8,n)$ } \\
$L^0_1(8,22)$  & $4.87450 \times 10^{15}$ & $L^0_1(16,24)$ & $6.64230 \times 10^{14}$ & \multicolumn{3}{l}{ \hspace{13mm} $=-L^0_4(6,n)=-L^0_6(4,n)$ } \\
$L^0_1(8,24)$  & $1.48769 \times 10^{17}$ & $L^0_1(16,26)$ & $1.03670 \times 10^{17}$ & \multicolumn{3}{l}{ $L^0_1(28,n)=-L^0_2(14,n)=L^0_7(4,n)$ } \\
$L^0_1(8,26)$  & $3.87212 \times 10^{17}$ & $L^0_1(16,28)$ & $1.36624 \times 10^{19}$ & \multicolumn{3}{l}{ $L^0_1(30,n)=L^0_3(10,n)=L^0_5(6,n)$ } \\
$L^0_1(8,28)$  & $-4.23050 \times 10^{20}$ & $L^0_1(16,30)$ & $1.57009 \times 10^{21}$ & & & \\
$L^0_1(8,30)$  & $-4.60409 \times 10^{22}$ & & & & & \\
\hline
\end{tabular}
}
\end{center}
\end{table}

\subsection{Alternative method to calculate $W^0(n)$ and $L_m^0(N_t, n)$}
\label{sec:alt}

In Sec.~\ref{sec:twist}, we have shown that
the values of $L_m(N_t,n)$ for individual winding numbers $m$ 
can be calculated by combining $D_n^\theta$ obtained with 
various twisted boundary conditions Eq.~(\ref{eq:twisttheta}).
For the case $U_{x,\mu}=1$, because of the uniformity of the system
one finds by extending this idea that $W^0(n)$ and $L_m^0(N_t, n)$ are
calculable on the lattice with temporal extent of any divisor of $N_t$.
Pursuing this idea leads to the conclusion that 
the calculation can be carried out on the $N_s^3\times1$ lattice
by combining $D_n^\theta$ with $\theta=\pi w/N_t Y$
with $w=0,1,2,\cdots,2N_t Y$.

There is no reason not to apply the same idea to all spatial directions.
Then, after folding all the spatial directions 
one finally finds that the calculation is feasible
just on $1^4$ lattice.
Since the spatial coordinates no longer exist in this limit,
in this case the ``hopping'' term is given by the following $4\times4$ matrix:
\begin{eqnarray}
  b(\varphi) = 
  b( \varphi_1, \varphi_2, \varphi_3, \varphi_4 )
  = \sum_{\mu=1}^4 \Big[ (1-\gamma_\mu) e^{i \varphi_\mu} + (1+\gamma_\mu) e^{-i \varphi_\mu} \Big],
\label{eq:b}
\end{eqnarray}
where $\varphi_\mu$ is the phase of the ``twisted boundary conditions''
for the $\mu$th direction.
Using Eq.~(\ref{eq:b}), the value of $D_n$ on the $N_s^4$ lattice
with the periodic boundary conditions for all directions is calculated to be
\begin{eqnarray}
  D_{n;N_s^4} = - \frac{N_c}{n} \frac1{N_s^4}
  \sum_{k_1,k_2,k_3,k_4=1}^{N_s}  {\rm tr_D}
  \left[ b\Big( \frac{2\pi k_1}{N_s}, \frac{2\pi k_2}{N_s}, \frac{2\pi k_3}{N_s} ,\frac{2\pi k_4}{N_s}\Big)^n \right],
\label{eq:Dn1}
\end{eqnarray}
where $ {\rm tr_D}$ is the trace over the Dirac index.
This result leads to 
\begin{eqnarray}
  W^0(n) = D_{n;N_s^4} \qquad ( n<N_s ).
  \label{eq:w0onesite}
\end{eqnarray}
Next, the value of $D_n^\theta$ on the $N_s^3\times N_t$ lattice 
is similarly obtained as
\begin{eqnarray}
  D_{n;N_s^3\times N_t}^\theta = - \frac{N_c}{n} \frac1{N_s^4}
  \sum_{k_1,k_2,k_3=1}^{N_s} 
  \sum_{k_4=1}^{N_t} {\rm tr_D}
  \left[ b\Big( \frac{2\pi k_1}{N_s}, \frac{2\pi k_2}{N_s}, \frac{2\pi k_3}{N_s} ,\frac{2\pi k_4+ \theta}{N_t}\Big)^n \right].
  \label{eq:Dntheta1}
\end{eqnarray}
From Eq.~(\ref{eq:Dntheta1}) one can construct $L_m(N_t,n)$
with a similar manipulation as Eq.~(\ref{eq:B=ML}).
Since $L_m^0(N_t,n)$ is always real, one has $D_n^\theta = D_n^{-\theta}$
and thereby only $Y+1$ calculations of $D_n^\theta$ are enough in this analysis.
In the large $N_s$ limit, the sum over $k_\mu$ in Eqs.~(\ref{eq:Dn1}) and
(\ref{eq:Dntheta1}) is replaced with an integral:
For example, Eq.~(\ref{eq:Dn1}) in this limit reads
\begin{eqnarray}
  D_{n;N_s^4} \underset{N_s\to\infty}{\longrightarrow} - \frac{N_c}{n} 
  \int_0^{2\pi} \frac{d^4\varphi}{(2\pi)^4}  {\rm tr_D}
  \left[ b( \varphi )^n \right].
\label{eq:Dn1int}
\end{eqnarray}

The same result is obtained by directly expanding $\ln\det M(\kappa)$.
By Fourier transforming Eq.~(\ref{eq:Mxy}), one has
\begin{eqnarray}
  \ln\det M(\kappa)
  = N_c \int_{-\pi}^{\pi} \frac{d^4k}{(2\pi)^4} {\rm tr}_D \ln \tilde{M}_{k}(\kappa)
  \label{eq:lndetMk}
\end{eqnarray}
with the $4\times4$ matrix in the Dirac-spinor space
\begin{eqnarray}
  \tilde{M}_{k}(\kappa)
  = 1 - \kappa b(k)
  = 1 - 2 \kappa
  \Big( \sum_\mu \cos k_\mu - i \sum_\mu \gamma_\mu \sin k_\mu \Big),
\end{eqnarray}
for $U_{x,\mu}=1$.
Then, one easily finds that the Taylor expansion of Eq.~(\ref{eq:lndetMk})
with respect to $\kappa$ at $\kappa=0$ gives Eq.~(\ref{eq:Dn1int}).

The calculation of ${\rm tr_D} [ (b(\varphi))^n ]$ in Eqs.~(\ref{eq:Dn1}), 
(\ref{eq:Dntheta1}) and (\ref{eq:Dn1int}) is simplified
using the fact that the two degenerate eigenvalues of the $4\times4$
matrix $b$ are given by
\begin{eqnarray}
  \lambda_\pm(\varphi) = 2\sum_\mu \cos\varphi_\mu \pm 2i \sqrt{ \sum_\mu \sin^2 \varphi_\mu }.
  \label{eq:lambda_pm}
\end{eqnarray}
Using Eq.~(\ref{eq:lambda_pm}) one finds
\begin{align}
 {\rm tr_D} \left[ b(\varphi)^n \right]
 = 2 \Big[ \big(\lambda_+(\varphi)\big)^n + \big(\lambda_-(\varphi)\big)^n \Big].
\end{align}
The eigenvalues Eq.~(\ref{eq:lambda_pm}) have
the maximum absolute value $|\lambda_\pm|=8$ at
$\varphi_1=\varphi_2=\varphi_3=\varphi_4=0$.
Therefore, Eq.~(\ref{eq:Dn1int}) grows as $8^n$ for $n\to\infty$.
This shows that the radius of convergence of the hopping parameter
expansion for $U_{x,\mu}=1$ is $1/8$, i.e., up to the chiral limit of free Wilson fermions.

The above procedure to calculate $W^0(n)$ and $L_m^0(N_t,n)$ 
enables us to increase $n$ and $N_t$ up to extremely large values
without worrying about the memory limitations.
We have numerically checked that $W^0(n)$ and $L_m^0(N_t,n)$ obtained
from Eqs.~(\ref{eq:Dn1}) and (\ref{eq:Dntheta1})
give exactly the same results with those given
in Tables~\ref{tab:wn} and \ref{tab:lmn}.

\begin{figure}[tb]
\begin{center}
\vspace{0mm}
\includegraphics[width=8.0cm]{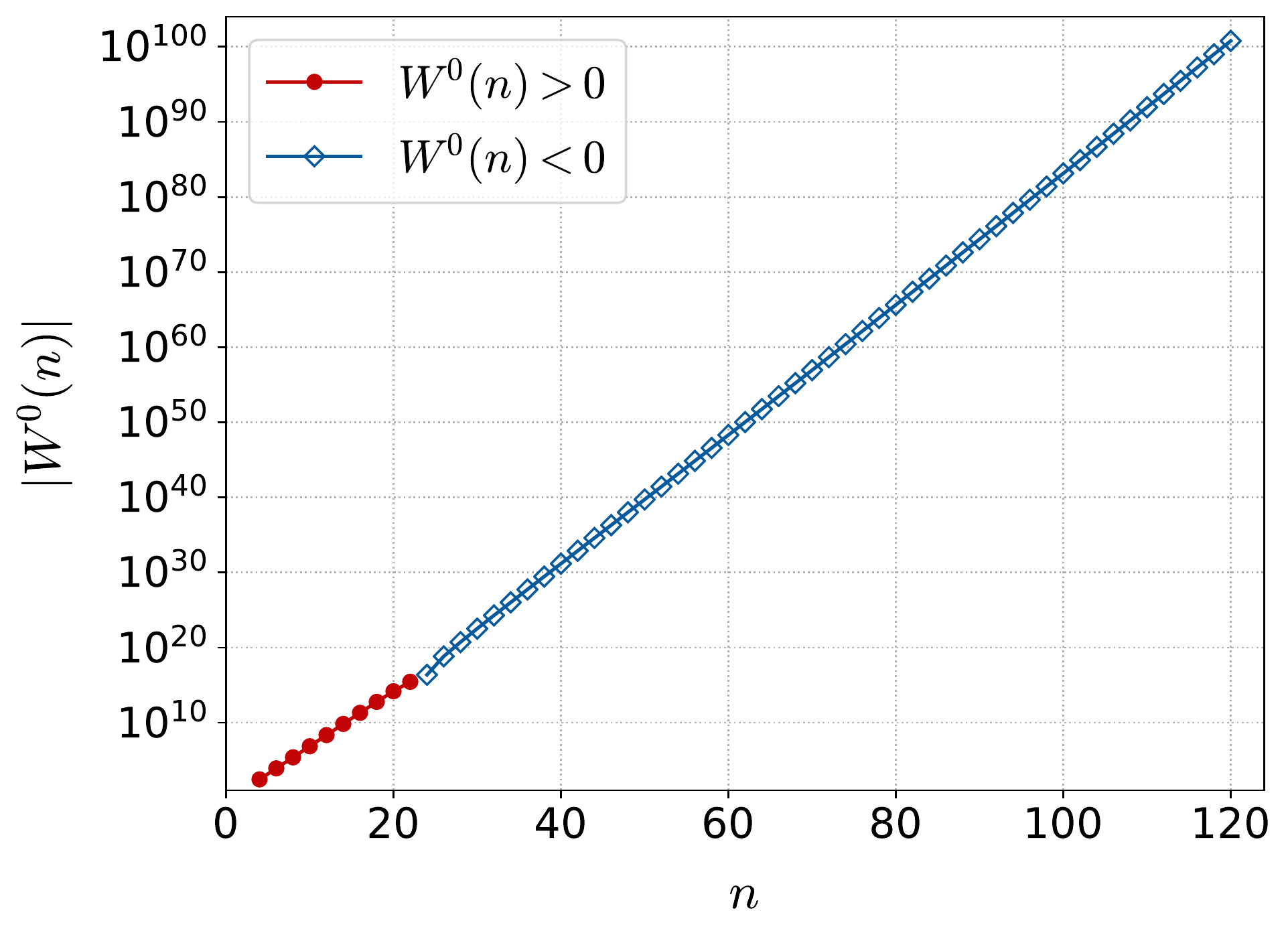}
\hspace*{2mm}
\includegraphics[width=7.8cm]{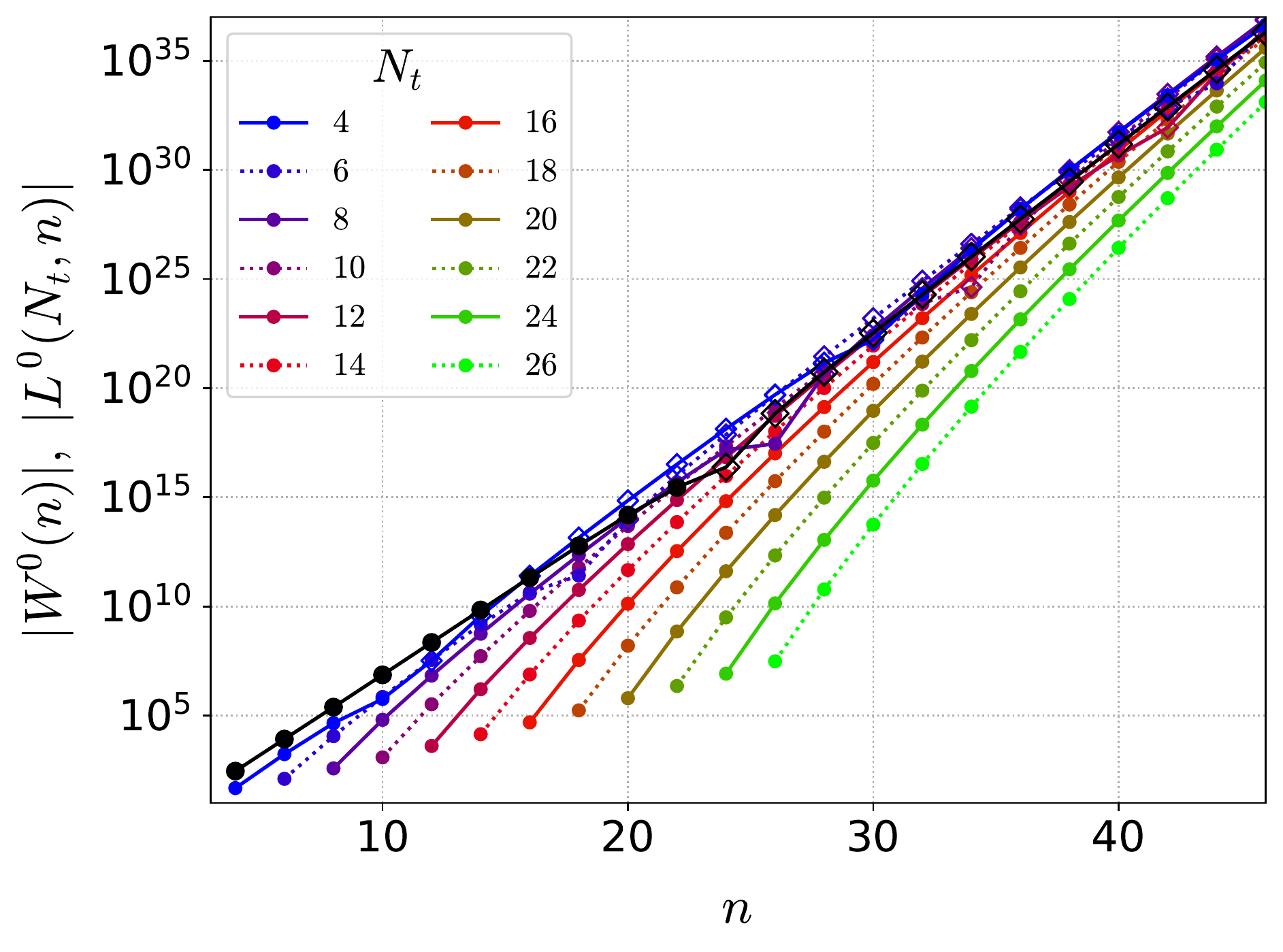}
\vspace{0mm}
\caption{Absolute values of the Wilson loop term $W^0(n)$ (left) and the sum of Polyakov-type loop terms $L^0(N_t, n)$ (right), computed with setting all $U_{x, \mu.}= \mathbf{1}$. In the right panel, results of $|W^0(n)|$ is also shown by black symbols. The closed circle and open diamond symbols mean positive and negative values, respectively.}
\label{fig:wnln}
\end{center}
\end{figure}

\begin{figure}[tb]
\begin{center}
\vspace{0mm}
\includegraphics[width=7.6cm]{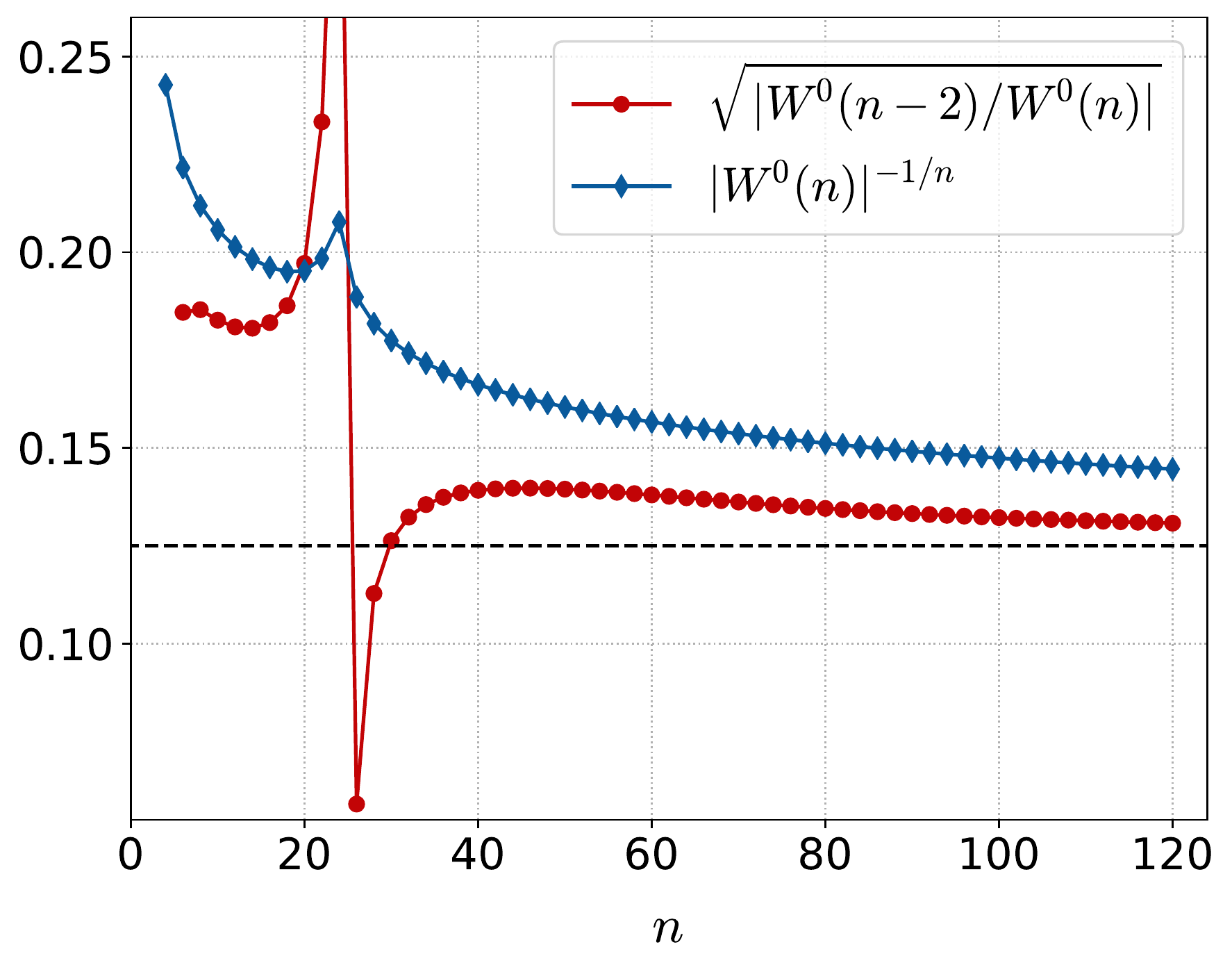}
\hspace*{2mm}
\includegraphics[width=8.2cm]{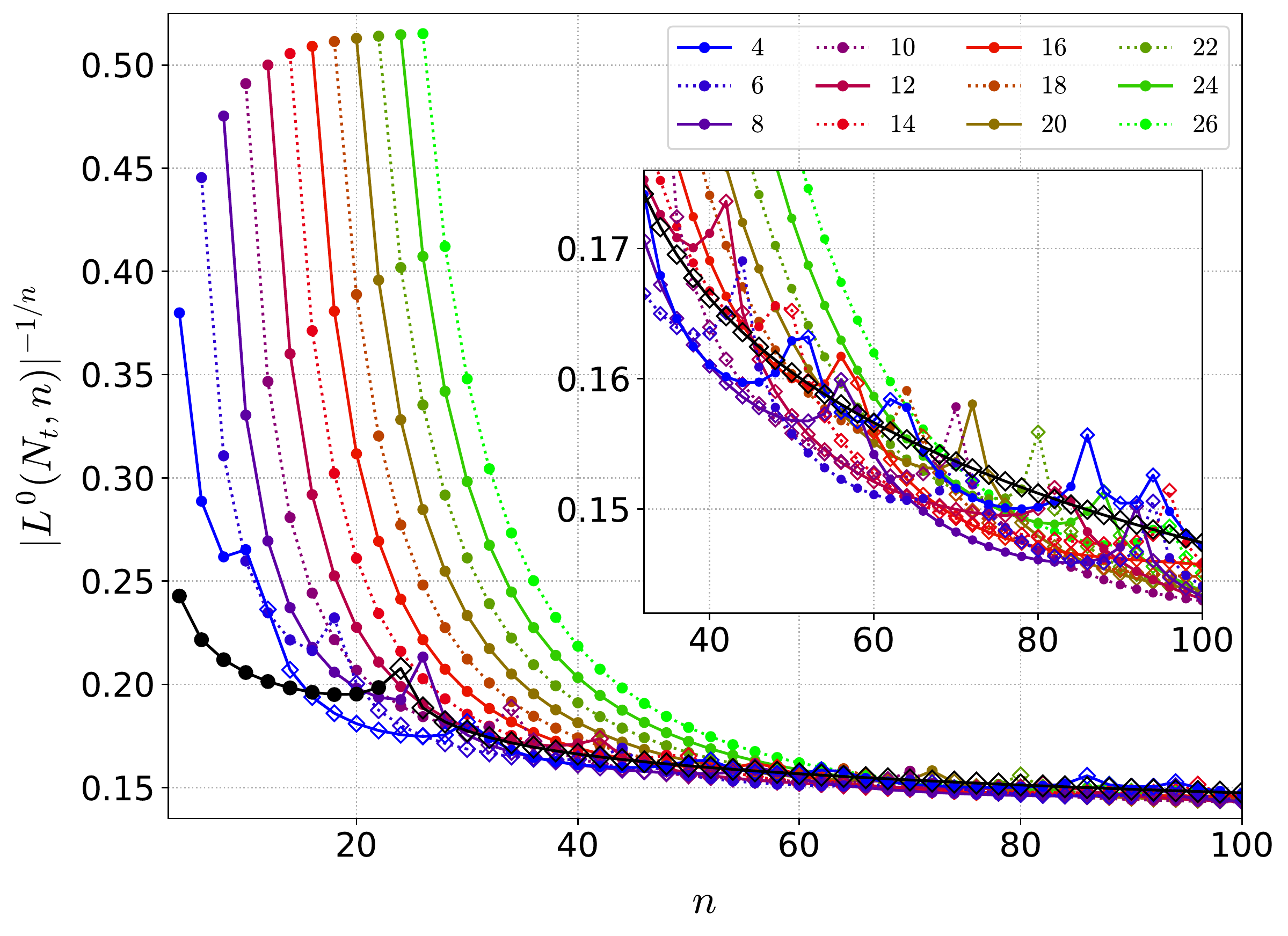}
\vspace{0mm}
\caption{Test of convergence in the case $U_{x, \mu}=\mathbf{1}$.
\textbf{Left}: $\kappa_{\rm dA}(W^0;n) = \sqrt{\left| W^0(n-2)/W^0(n) \right|}$ (red symbols) based on the d'Alembert's test and $\kappa_{\rm CH}(W^0;n) = \left| W^0(n) \right|^{-1/n} $ (blue symbols) based on the Cauchy-Hadamard's test, for Wilson loop terms. 
The horizontal dashed line in black represents the chiral limit $\kappa=1/8$ for free Wilson fermions.
\textbf{Right}: $\kappa_{\rm CH}(L^0;n) =  \left| L^0(N_t, n) \right|^{-1/n}$ for Polyakov-type loop terms (colored symbols), together with $\kappa_{\rm CH}(W^0;n)$ (black symbols).
The inset in the right panel is a close-up of the range $n=32$--100.}
\label{fig:conv}
\end{center}
\end{figure}

\subsection{Convergence radius} 
\label{sec:convrad}

We show the results of $| W^0(n) |$ in the left panel of Fig.~\ref{fig:wnln}.
The vertical axis is logarithmic. 
$W^0(n)$ changes its sign at $n = 24$.
The closed circle symbol means a positive $W^0(n)$, and the open diamond symbol a negative $W^0(n)$.
The results of the absolute value of $| L^0(N_t, n) |$ are given in the right panel of Fig.~\ref{fig:wnln}. 
The sign changes for $N_t=4$ at $n=12$, 32 and 52; for $N_t=6$ at $n=20$ and 44; for $N_t=8$ at $n=28$; for $N_t=10$ at $n=34$;  for $N_t=12$ at $n=42$; and for $N_t=14$ at $n=50$. 
The open diamond symbol means a negative value, again.
As seen in the left and right panels of Fig.~\ref{fig:wnln}, $W^0(n)$ and $L^0(N_t, n)$ start from the values of Eqs.~(\ref{eq:hpeu1w}) and (\ref{eq:hpeu1}) at $n=4$ and $N_t$, respectively, and their absolute values increase exponentially with $n$.

Since $W^0(n)$ and $L^0(N_t, n)$ increase rapidly with $n$, the hopping parameter expansion does not converge unless $| W^0(n-2) \kappa^{n-2} | \, > \, | W^0(n) \kappa^n |$ and $| L^0(N_t, n-2) \kappa^{n-2} | \, > \, | L^0(N_t, n) \kappa^n |$ for large $n$. 
This leads to d'Alembert's ratio test of convergence, 
\begin{eqnarray}
\kappa < \kappa_{\rm dA}(W^0;n) = \sqrt{\left| \frac{W^0(n-2)}{W^0(n)} \right|}, \;\;
\kappa < \kappa_{\rm dA}(L^0;n) = \sqrt{\left| \frac{L^0(N_t, n-2)}{L^0(N_t, n)} \right|} \quad \textrm{for large} \; n.
\label{eq:convergence}
\end{eqnarray}
Another conventional test of convergence is based on the Cauchy-Hadamard's convergence radius given by
\begin{eqnarray}
\kappa < \kappa_{\rm CH}(W^0;n) = \left| W^0(n) \right|^{-1/n}, \;\;
\kappa < \kappa_{\rm CH}(L^0;n) = \left| L^0(N_t, n) \right|^{-1/n}  \quad \textrm{for large} \; n.
\label{eq:cauchy}
\end{eqnarray}
Because the values of the Wilson loops and the Polyakov-type loops on actual configurations decrease exponentially as $n$ increases, 
the convergence radius in reality should be larger than the right hand sides of Eqs.~(\ref{eq:convergence}) and (\ref{eq:cauchy}), 
i.e., $\kappa_{\rm dA}(X;n)$ and $\kappa_{\rm CH}(X;n)$ at large $n$ provide us with lower bounds for the convergence radius.

In the left panel of~Fig.~\ref{fig:conv}, we show $\kappa_{\rm dA}(W^0;n)$ (red symbols ) and $\kappa_{\rm CH}(W^0;n)$ (blue symbols) as functions of $n$.
The oscillating behavior of the red symbols and the peak of the blue symbols at $n = 24$ is due to the sign change of $W(n)$ there, as shown in the left panel of Fig.~\ref{fig:wnln}.
This figure shows that, in the large $n$ limit,  $\kappa_{\rm dA}(W^0;n)$ and $\kappa_{\rm CH}(W^0;n)$ approach $\kappa=1/8$ shown by the dashed line in the figure. 
This is in accordance with the discussion in Sec.~\ref{sec:alt}.

In the right panel of Fig.~\ref{fig:conv}, we show $\kappa_{\rm CH}(L^0;n)$ for Polyakov-type loop terms.
Since $L^0(N_t,n)$ changes its sign frequently, we plot the results of $\kappa_{\rm CH}(L^0;n)$ only, because their $n$-dependence is milder than $\kappa_{\rm dA}(L^0;n)$.
This figure is also consistent with the expectation that $\kappa_{\rm CH}(L^0;n)$ approaches $1/8$ in the large $n$ limit.


\section{Effect of high order terms}
\label{higheroder}

Though the convergence radius of the hopping parameter expansion turned out to be not small, 
in practice, we need to truncate the expansion at some finite order and have to take into account the systematic error due to the truncation.
In this section, we study the effect of higher-order terms more closely and estimate the magnitude of the truncation error considering the case of the worst convergence.
We also introduce an effective theory to incorporate the effect of high order term to reduce the truncation error of the hopping parameter expansion.

From Eq.~(\ref{eq:<O>}), the expectation value at the simulation point $(\beta, \kappa)$ can be expressed in terms of those at a different simulation point $(\beta_0, \kappa_0)$ by the reweighting method~\cite{rew}:
\begin{equation}
\langle {\cal O} \rangle_{(\beta, \kappa)}
= \frac{ \langle {\cal O} \,
[\det M(\kappa)/ \det M(\kappa_0)]^{N_{\rm f}} \, e^{6 (\beta - \beta_0) N_{\rm site} \hat{P}} \rangle_{(\beta_0, \kappa_0)} }{
 \langle [\det M(\kappa)/ \det M(\kappa_0)]^{N_{\rm f}} \, e^{6 (\beta - \beta_0) N_{\rm site} \hat{P}} \rangle_{(\beta_0, \kappa_0)} }.
\label{eq:<O>rew}
\end{equation} 
Setting $\kappa_0=0$, we find from Eqs.~(\ref{eq:tayexp}) and (\ref{eq:loopex}) that the hopping parameter expansion of the reweighting factor is given by
\begin{equation}
\det M(\kappa)= \exp \left[ N_{\rm site} \sum_{n=4}^{\infty} W(n) \kappa^{n} +
 N_{\rm site} \sum_{n=N_t}^{\infty} \sum_{m=1}^{\infty} L_m (N_t, n) \kappa^{n} \right].
 \label{eq:hpeIV}
\end{equation}

\clearpage

\begin{figure}[tb]
\begin{center}
\vspace{0mm}
\includegraphics[width=7.8cm]{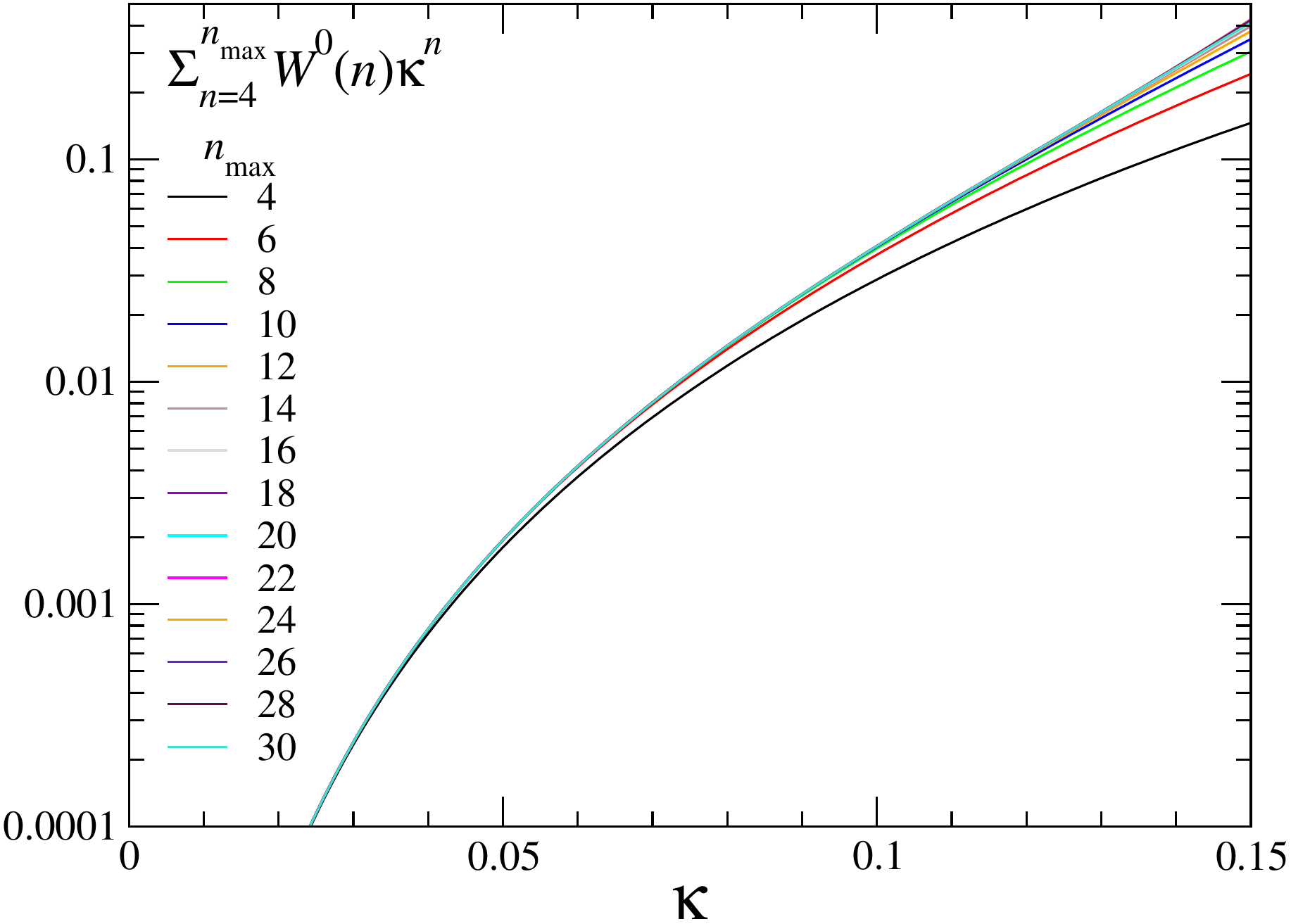}
\vspace{0mm}
\caption{$n_{\rm max}$ dependence of $\sum_{n=4}^{n_{\rm max}} W^0(n) \kappa^n$ for the case $U_{x, \mu.}= \mathbf{1}$.
The vertical axis is in a logarithmic scale.
}
\label{fig:wnnmax}
\end{center}
\end{figure}

\subsection{Effect of Wilson loop terms}
\label{wilsonloop}

We first discuss the effect of the Wilson loop terms, $\sum_{n=4}^{\infty} W(n)\,\kappa^n$, in Eq.~(\ref{eq:hpeIV}).
In practice, we need to truncate the the hopping parameter expansion by introducing a highest power of $\kappa$, say $n_{\rm max}$. 
To see an upper bound for the truncation error of the the hopping parameter expansion, we study the case of worst convergence $U_{x, \mu}=\mathbf{1}$ discussed in~Sec.~\ref{coefficients}.
Using the results of~Table~\ref{tab:wn}, we compute $\sum_{n=4}^{n_{\rm max}} W^0(n)\,\kappa^n$ as functions of $\kappa$. 
The results are shown in Fig.~\ref{fig:wnnmax} for $n_{\rm max}=4$--30.
We see that the truncation error in the sum of Wilson loop terms is small up to $\kappa \simeq 0.125$ when we choose $n_{\rm max} \simge 10$.

Next, we discuss the effect of each Wilson loop term $W(n)$.
Since the first term $W(4)$ is proportional to $\hat{P}$, its effects can be reproduced by a shift $\beta \rightarrow \beta^* = \beta + 48 N_{\rm f} \kappa^4$ in the gauge action. 
Though Wilson loops of $n\ge6$ do not appear in the standard plaquette gauge action, they can also be regarded as lattice expression of the gauge action. 
In fact, improved gauge actions contain terms with such longer Wilson loops. 
Therefore, the effect of the Wilson loop terms can be reproduced by the following modification of the lattice gauge action,
\begin{eqnarray}
6 \beta \hat{P} &\longrightarrow& 6 \beta \hat{P} + N_{\rm f} [ W(4)\ \kappa^4 +W(6)\ \kappa^6 + W(8)\ \kappa^8 + W(10)\ \kappa^{10} + \cdots ]
\nonumber \\
 &=& 6 ( \beta +48 N_{\rm f}  \kappa^4) \hat{P} + N_{\rm f} [ 8448 \kappa^6 \hat{P}_6 + 245952 \kappa^8 \hat{P}_8 + 7372800 \kappa^{10} \hat{P}_{10} + \cdots ] ,
\label{eq:wltermseffects}
\end{eqnarray}
where, $\hat{P}_n$ is the linear combination of $n$-step Wilson loops in $W(n)$ and is normalized to one when $U_{x, \mu}= \mathbf{1}$.
When we view this shift of the gauge action as a shift in improvement parameters in the parameter space of improved gauge actions, we find that, at least in these low-order terms, the magnitude of the shift is much smaller than those for typical improved actions:
For example, in the Iwasaki improved gauge action, $S_g = -6N_{\rm site} \beta (c_0 \hat{P} + 2c_1 \hat{R})$ with $\hat{R}$ the $1 \times 2$ Wilson loop, the improvement parameters are $c_0 = 3.648$ and $c_1 = -0.331$, i.e., the  ratio of the absolute values of the 4-step and 6-step terms is about $5.5 : 1$~\cite{iwasaki}.
On the other hand, the ratio of the 4-step term to the 6-step term in Eq.~(\ref{eq:wltermseffects}) is about $4000 : N_{\rm f}$ even at $\kappa \sim 0.1$.
The magnitude of coupling parameters is similar also in other improved gauge actions~\cite{symanzik,dbw2,perfect}.
Therefore, the shift in gauge coupling parameters due to the dynamical quark effect is quite small at $\kappa \simle 0.1$.\footnote{We also note that, since the sign of the 6-step loop term of typical improved actions is negative while that of the $W (6)$ term is positive, a slight unimprovement is required to reproduce the dynamic quark effect.}
Because a slight shift in improvement parameters mainly affects the lattice discretization errors, the Wilson loop terms will not affect characteristic features of the system in the continuum limit, 
though the convergence of the hopping parameter expansion may rapidly worsen close to the chiral limit.
In contrast, the Polyakov-type loop terms affect like external magnetic fields in spin models, and thus can change the nature of the phase transition.

\begin{figure}[tb]
\begin{center}
\vspace{0mm}
\includegraphics[width=7.8cm]{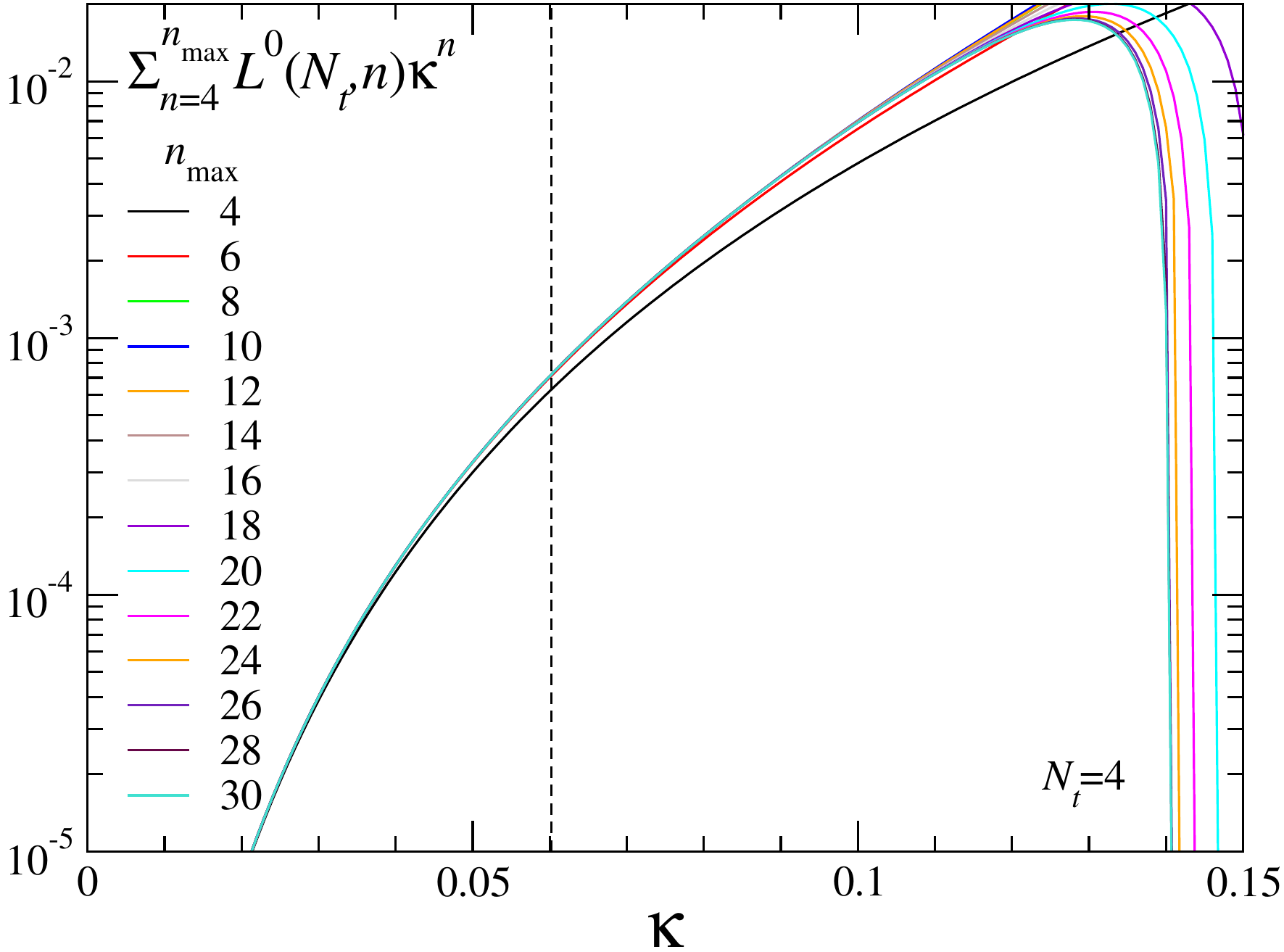}
\hspace*{2mm}
\includegraphics[width=7.8cm]{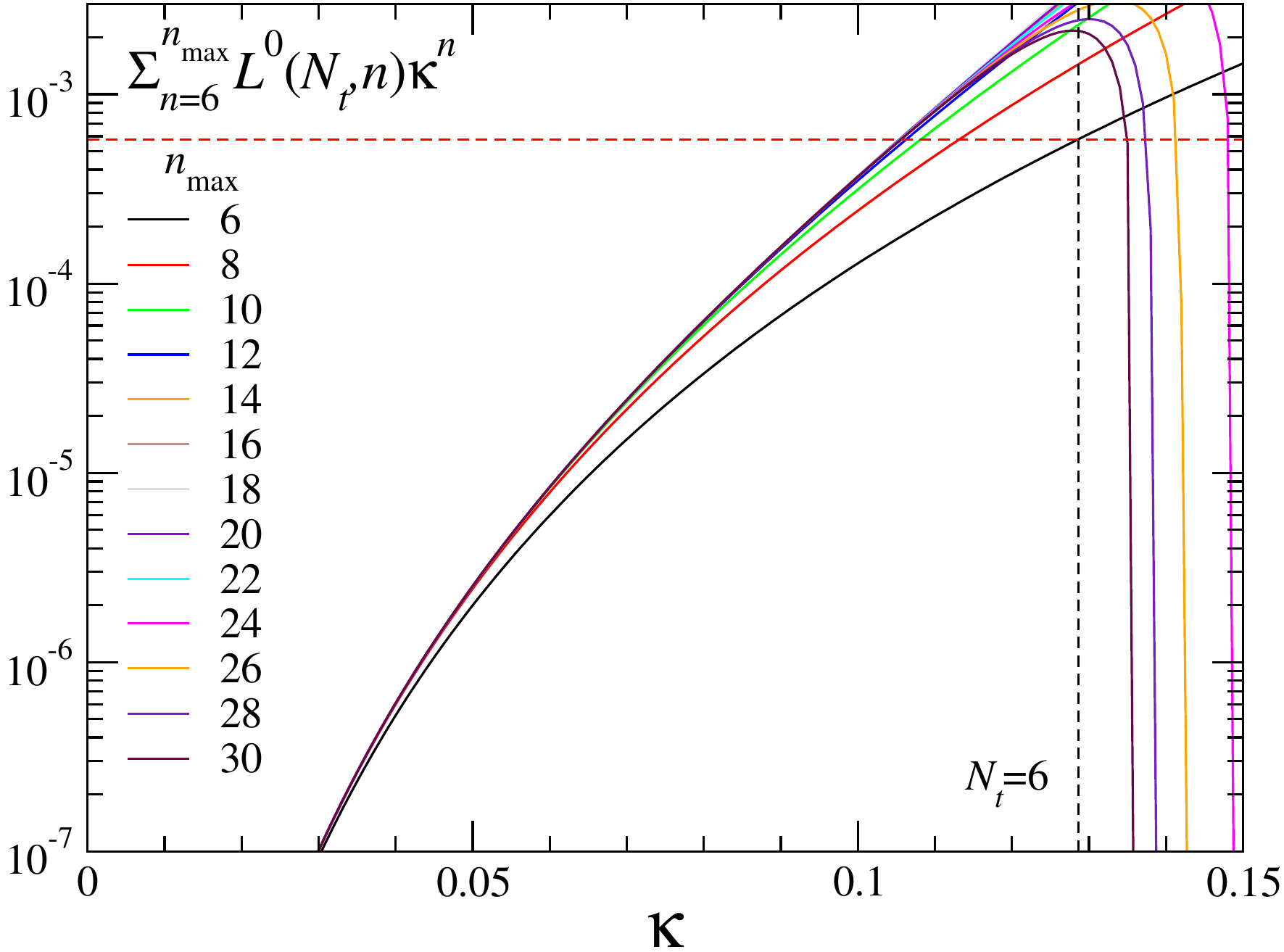}
\vspace{0mm}
\caption{$n_{\rm max}$ dependence of $\sum_{n=N_t}^{n_{\rm max}} L^0(N_t,n) \kappa^n$ at $N_t=4$ (left) and $N_t=6$ (right) for the case $U_{x, \mu.}= \mathbf{1}$.
}
\label{fig:lnnt46}
\end{center}
\end{figure}

\begin{figure}[tb]
\begin{center}
\vspace{0mm}
\includegraphics[width=7.8cm]{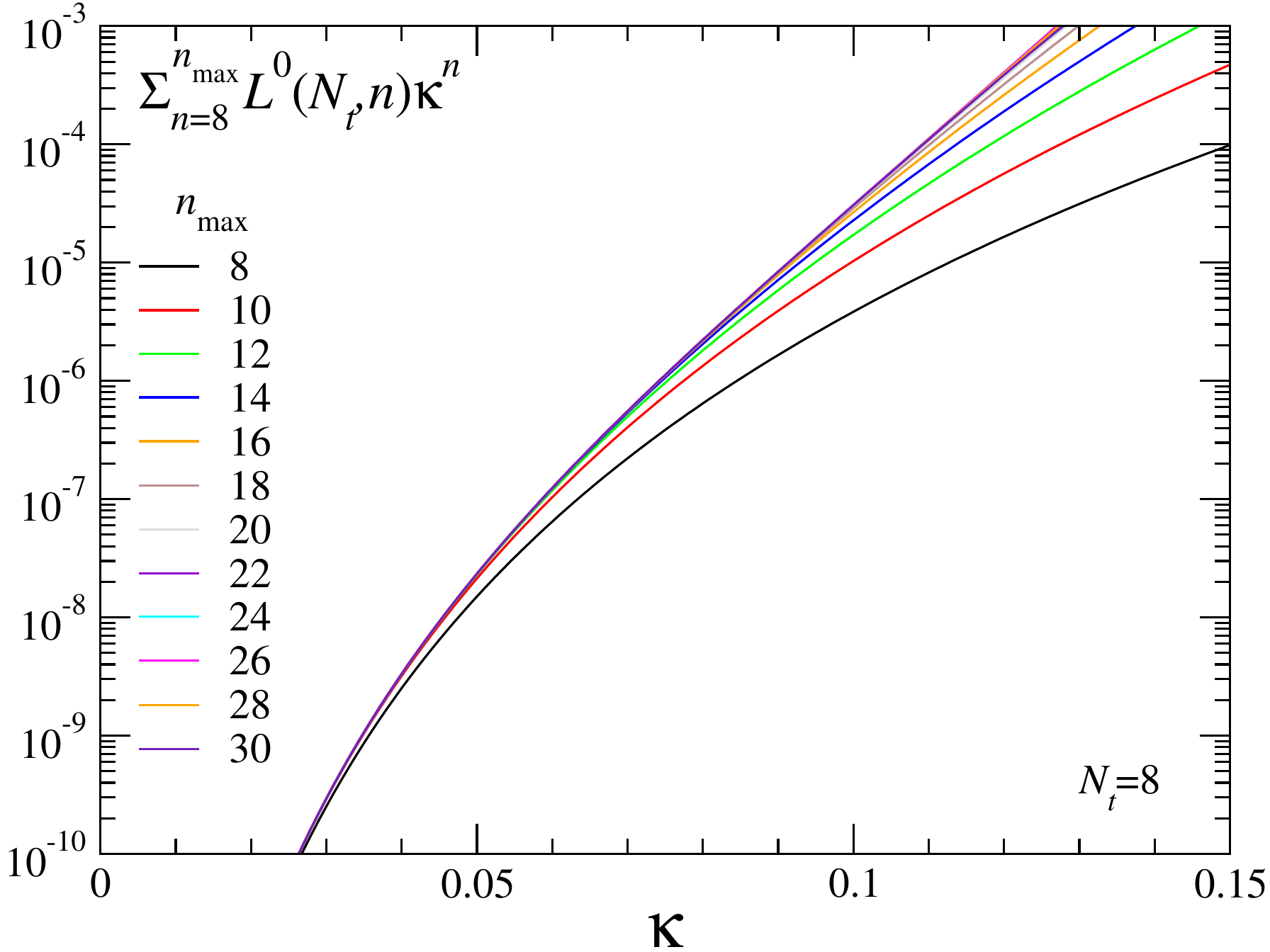}
\hspace*{2mm}
\includegraphics[width=7.8cm]{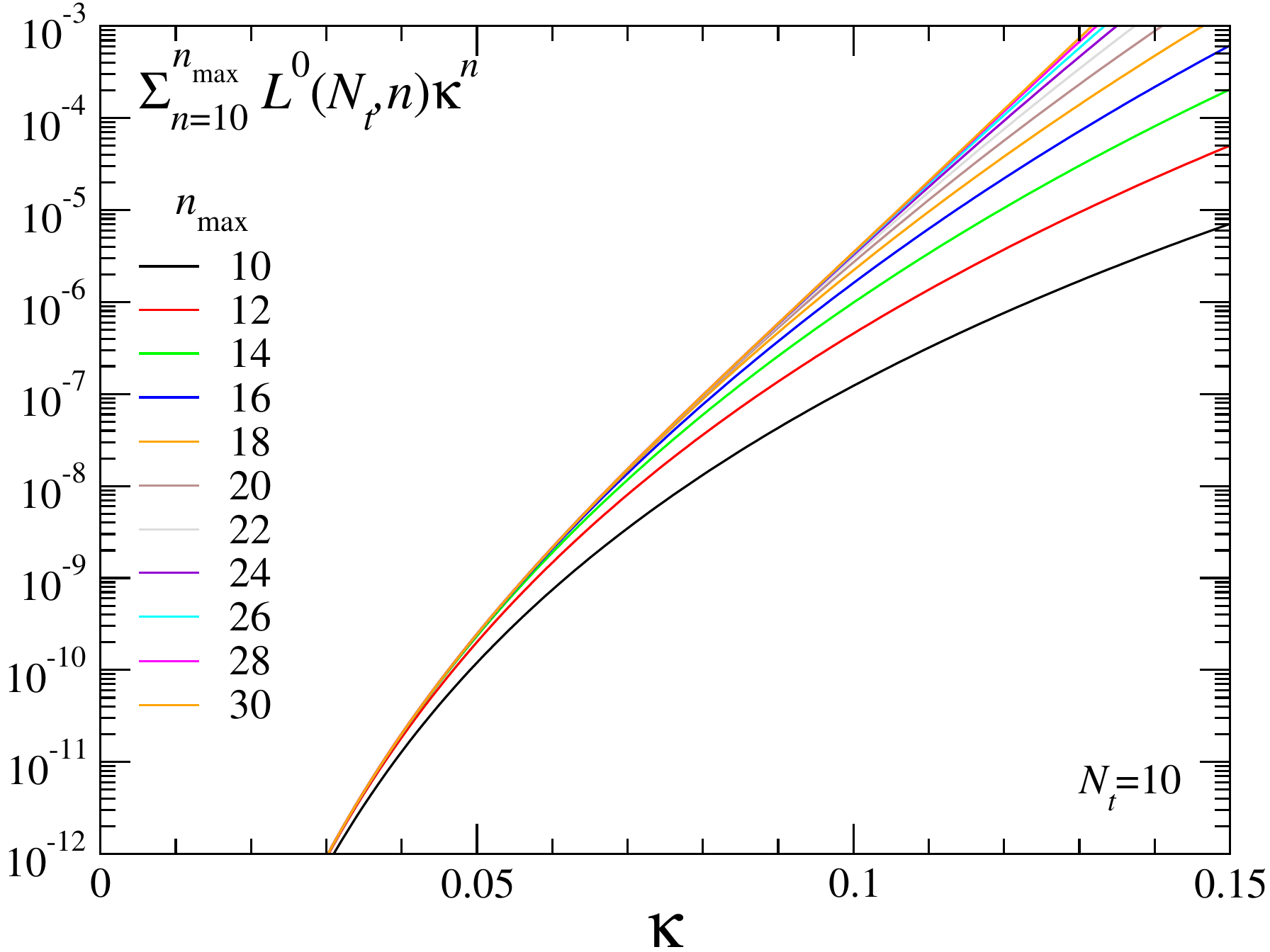} \\
\vspace*{2mm}
\includegraphics[width=7.8cm]{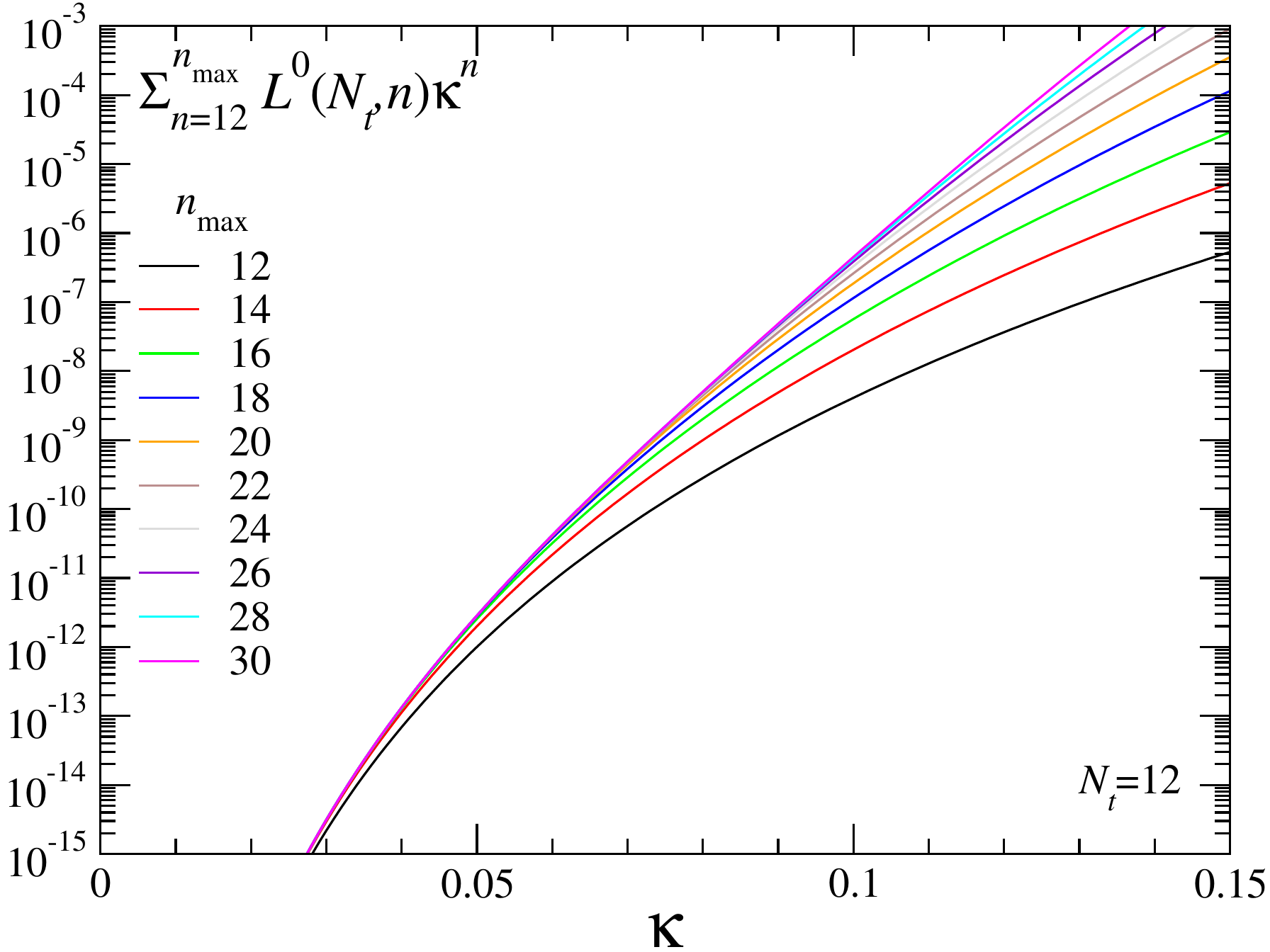}
\hspace*{2mm}
\includegraphics[width=7.8cm]{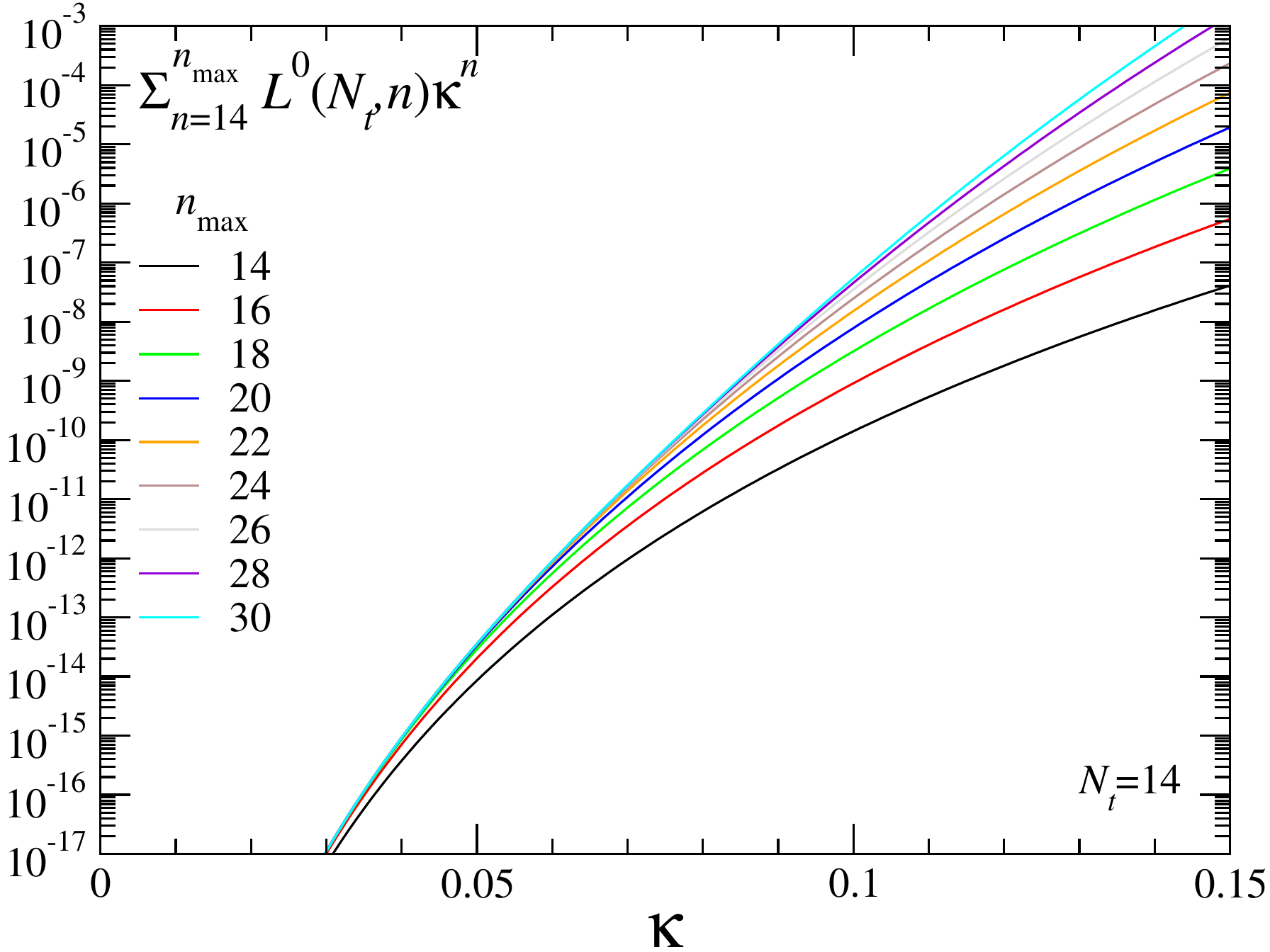}
\vspace{0mm}
\caption{The same as Fig.~\ref{fig:lnnt46}, but for $N_t=8$ (top left), 10 (top right), 12 (bottom left), and 14 (bottom right).}
\label{fig:lnnt814}
\end{center}
\end{figure}

\subsection{Truncation error of higher-order Polyakov-type loop terms}
\label{sec:truncationerror}

We now study the effect of Polyakov-type loop terms. 
To see an upper bound for the truncation error of the hopping parameter expansion, we first study the case of worst convergence, $U_{x, \mu}=\mathbf{1}$, discussed in~Sec.~\ref{coefficients}:
\begin{eqnarray}
\left[ \, \ln \det M(\kappa) - {\rm (Wilson\ loop\ terms)} \, \right]_{U_{x, \mu}=\mathbf{1}}
\simeq N_{\rm site} \sum_{n=N_t}^{n_{\rm max}} L^{0}(N_t, n) \kappa^{n} ,
\end{eqnarray}
where $n_{\rm max}$ is the highest power of $\kappa$ to truncate the hopping parameter expansion.
In Figs.~\ref{fig:lnnt46} and \ref{fig:lnnt814}, 
we plot $\sum_{n=N_t}^{n_{\rm max}} L^0(N_t, n) \kappa^{n}$ changing $n_{\rm max} = N_t$, $N_t+2$, $\cdots$ up to 30
on $N_t=4$--14 lattices.

The left panel of Fig.~\ref{fig:lnnt46} shows the results for $N_t=4$. 
The vertical dashed line represents $\kappa_c = 0.0602(4)$ for two-flavor QCD~\cite{Kiyohara:2021smr}, which was obtained for infinitely large spatial volume by a finite size scaling analysis including the next-to-leading order effect, i.e., $n_{\rm max}=N_t+2$.
The location of the critical point for three-flavor QCD is similar~\cite{Saito:2011fs,Kiyohara:2021smr}.
From this plot, we see that, around $\kappa_c$ of two- and three-flavor QCD, the $\kappa^4$ term is dominant and the effect of the higher-order terms is negligible. 
Hence, the determination of $\kappa_c$ using the hopping parameter expansion is reliable for $N_t=4$.
We also see that the convergence of the hopping parameter expansion suddenly worsens at $\kappa \simge 0.13$, suggesting a lower bound of the convergence radius around there. 

In the right panel of Fig.~\ref{fig:lnnt46}, we show the corresponding results for $N_t=6$. 
The vertical dashed line shows the critical point $\kappa_c = 0.1286 (40)$ for two-flavor QCD obtained by looking at the shape of the histogram of the Polyakov loop on a $32^3 \times 6$ lattice \cite{Ejiri:2019csa}, which was calculated with the leading-order term, i.e., $n_{\rm max}=N_t$.
Here, we notice that the critical point determined by the histogram has a large systematic error due to the finite volume effect.
The critical point determined by the histogram is about 10\% larger than that determined by a finite size scaling analysis on $N_t = 4$ lattices~\cite{Kiyohara:2021smr}.
From a full QCD simulation on $N_t = 6$ lattices, $\kappa_c = 0.0877(9)$ is obtained by a finite size scaling analysis~\cite{Cuteri:2020yke}.

Unlike the case of $N_t=4$ shown in the left panel of Fig.~\ref{fig:lnnt46}, we see significant effect from high order terms for $N_t=6$ around $\kappa_c$ determined on a $32^3 \times 6$ lattice by a leading-order calculation.
The results for $N_t=8$, 10, 12, and 14 are shown in Fig.~\ref{fig:lnnt814}. 
We see that the convergence worsens further as $N_t$ increases. 
Note, however, that these results are obtained for the case $U_{x, \mu}=\mathbf{1}$.
As  discussed in Sec.~\ref{coefficients}, because the values of Polyakov loops on actual configurations become exponentially small as $n$ increases, the convergence in actual simulations should be better than the case $U_{x, \mu}=\mathbf{1}$.
In the next subsection, we also discuss that the convergence is much improved by incorporating higher-order effects --- using an effective theory incorporating high order terms, we can reliably determine the critical point $\kappa_c$ for the case of $N_t=6$.

\subsection{Effective theory incorporating high order terms}
\label{keffect}

As noted in Refs.~\cite{Saito:2013vja,Ejiri:2019csa}, the Polyakov loop $\hat\Omega$ and the bent Polyakov loops $\hat{\Omega}_k$ are strongly correlated on each configuration. 
In Ref.~\cite{Ejiri:2019csa}, this correlation was used to construct an effective theory in which the next-to-leading order effect of $\hat{\Omega}_k$ is effectively absorbed into a shift of the coupling for the leading order term $\hat\Omega$. 
In Sec.~\ref{correlation}, we show more generally that $L (N_t, n)$ and $\hat\Omega$ are strongly correlated with each other on each configuration.
We may thus approximate $L (N_t, n)$ by
\begin{eqnarray}
L (N_t, n) \approx L^0 (N_t, n) \, c_n \, {\rm Re} \hat{\Omega}, 
\label{eq:lco}
\end{eqnarray}
where $c_n$ is a constant to be determined by measuring the correlation between $L (N_t, n)$ and $\hat\Omega$ by a Monte Carlo simulation at each simulation point, while $c_n=1$ for $n = N_t$, and $c_n=0$ if $n$ is odd or $n <N_t$.
We have $c_{N_t + 2}=0.7996 (7)$ around the phase transition point for $N_t = 4$, and $0.8130 (3)$ for $N_t = 6$~\cite{Ejiri:2019csa}.
In Sec.~\ref{correlation}, we determine $c_n$ up to $n=20$ around the phase transition point for $N_t=6$ and 8. 

From Eq.~(\ref{eq:lco}), the Polyakov-type loop terms in the effective quark action $\ln\det M(\kappa)$ are approximated as
\begin{eqnarray}
\sum_{n=N_t}^{n_{\rm max}} L (N_t, n) \, \kappa^n \approx \left[ \sum_{n=N_t}^{n_{\rm max}} L^0 (N_t, n) \,c_n \kappa^n \right] {\rm Re} \hat{\Omega}.
\label{eq:slco}
\end{eqnarray}
Equation~(\ref{eq:slco}) means that the effect of high order Polyakov-type loop terms can be effectively incorporated to the leading order calculation by a shift of $\kappa$ appearing in front of ${\rm Re}\hat\Omega$ of the leading-order correction term:
\begin{eqnarray}
L^0 (N_t, N_t)\, \kappa^{N_t} \; \longrightarrow \;
L^0 (N_t, N_t)\, (\kappa^*)^{N_t} 
= \sum_{n=N_t}^{n_{\rm max}} L^0 (N_t, n) \, c_n \kappa^n
\label{eq:replaceL}
\end{eqnarray}
in the leading order calculation.
For example, when the critical point is determined as $\kappa_{\rm c, LO}$ by a leading-order calculation, then we obtain the critical point $\kappa_{\rm c, eff}$ effectively incorporating higher-order effects up to $n_{\rm max}$th order by solving
\begin{eqnarray}
\sum_{n=N_t}^{n_{\rm max}} L^0 (N_t, n) \, c_n \kappa_{\rm c, eff}^n = L^0 (N_t, N_t)\, \kappa_{\rm c, LO}^{N_t} .
\label{eq:keff}
\end{eqnarray}

We now revisit the case of worst convergence discussed in Sec.~\ref{sec:truncationerror}. 
In the right panel of Fig.~\ref{fig:lnnt46}, $L^0 (N_t, N_t)\, \kappa_{\rm c, LO}^{N_t}$ for $N_t = 6$ is shown by the red dashed line.
The colored solid lines represent the left-hand-side of Eq.~(\ref{eq:keff}) for various $n_{\rm max}$ with substituting $c_n =1$.
The red dashed line intersects each colored solid line at $\kappa=\kappa_{\rm c, eff}$ for the corresponding $n_{\rm max}$. 
From the right panel of Fig.~\ref{fig:lnnt46}, we find that, even in the case of worst convergence with $c_n=1$, a stable and reliable $\kappa_{\rm c, eff}$ is obtained for $N_t = 6$ when $n_{\rm max} \simge 10$. 
Since $c_n <1$ in practice, the correction of $\kappa_{\rm c, eff}$ from higher-order terms is smaller than the case of $c_n = 1$ shown in~Fig.~\ref{fig:lnnt46}.
Thus the next-to-leading order calculation of $\kappa_{\rm c, eff}$ for $N_t = 6$ with $n_{\rm max}=8$~\cite{Ejiri:2019csa} may not be so wrong.

For $N_t = 8$, though $\kappa_{\rm c, LO}$ is not available yet\footnote{
$\kappa_c=0.1135(8)$ was reported for $N_t=8$ by a full QCD simulation of~Ref.~\cite{Cuteri:2020yke}.}, 
the top left panel of~Fig.~\ref{fig:lnnt814} shows that higher-order effects are well suppressed up to $\kappa \simle 0.125$ when we choose $n_{\rm max } \simge 20$. 
As $N_t$ increases, the convergence gradually worsens and thus a larger value of $n_{\rm max}$ will be required to obtain a reliable $\kappa_c$ by the hopping parameter expansion.

\section{Correlation among expansion terms}
\label{correlation}

\subsection{Calculation of expansion terms by numerical simulation}
\label{sec:MC}

We now perform Monte Carlo simulations of $SU(3)$ lattice gauge theory (quenched QCD) around the phase transition point to calculate the expansion terms $W(n)$ and $L_m(N_t,n)$ and study correlations among them.
As discussed in Sec.~\ref{hopping}, $W(n)$ and $L_m(N_t,n)$ can be extracted by calculating $D_n = (-1/N_{\rm site} n) \, {\rm Tr} [B^n]$ defined by~Eq.~(\ref{eq:derkappa}) with various boundary conditions.

To calculate $D_n$, we adopt the noise method for the trace over the position index:
For each of the $i$th set of the $3 \times 4$ color and spinor indexes, 
we generate a series of random numbers $(\vec{\eta}_j)_x$ with random complex phase at each position $x$, 
where the random vectors $\vec{\eta}_j$ satisfy
\begin{eqnarray}
\lim_{N_{\rm noise} \to \infty} \frac{1}{N_{\rm noise}} \sum_{j=1}^{N_{\rm noise}} (\vec{\eta}_j^{\; *})_y (\vec{\eta}_j)_x = \delta_{y,x} .
\end{eqnarray}
Multiplying $B^n$ to $\vec{\eta}_j$ and calculating the inner product with the complex conjugate $\vec{\eta}_j^{\; *}$ of the original vector, 
we obtain ${\rm Tr}[B^n]$ as 
\begin{eqnarray}
\lim_{N_{\rm noise} \to \infty} \frac{1}{N_{\rm noise}} \sum_{j=1}^{N_{\rm noise}} \sum_i  \sum_{x,y}
(\vec{\eta}_j^{\; *})_y [B^n]_{ii, yx} (\vec{\eta}_j)_x 
= \sum_i \sum_x [B^n]_{ii, xx}  
= {\rm Tr} [B^n]
\end{eqnarray}

We calculate $D_n$ with the following four types of boundary conditions in the temporal direction, while we impose periodic boundary condition in the spatial directions.
$D_n$ for periodic and anti-periodic boundary conditions are given by Eqs.~(\ref{eq:loopexp}) and (\ref{eq:loopex}), respectively.
Next, we prepare a lattice with a size of $N_s^3 \times (2N_t)$ by copying configurations generated on the $N_s^ 3 \times N_t$ lattice twice along the temporal direction.
Since Polyakov-type loops do not close for $n<2N_t$ on this lattice, $D_n$ in this case with periodic boundary condition $D_n^{2+}$ and anti-periodic boundary condition $D_n^{2-}$ are given by 
\begin{eqnarray}
D_n^{2+} &=& W(n) + \sum_{m=1}^{\infty} L_{2m} (N_t, n) = \frac{D_n + D_n^+}{2},
\\
D_n^{2-} &=& W(n) + \sum_{m=1}^{\infty} (-1)^m L_{2m} (N_t, n) = \frac{D_n^i + D_n^{-i}}{2},
\label{eq:loopex2}
\end{eqnarray}
respectively.
These relations lead
\begin{eqnarray}
(D_n + D_n^{+} + 2D_n^{2-})/4 &=& W(n) + L_4 (N_t, n) + \cdots ,
\\
(D_n + D_n^{+} - 2D_n^{2-})/4 &=& L_2 (N_t, n) + L_6 (N_t, n) + \cdots ,
\\
(D_n - D_n^{+})/2 &=& L_{1} (N_t, n) + L_3 (N_t, n) + L_5 (N_t, n) +\cdots .
\end{eqnarray}
Using these relations, we calculate $W(n)$ etc.\ for $n < 4N_t$.
As discussed in~Sec.~\ref{keffect}, to determine $\kappa_c$ for $N_t=8$, a calculation up to about $O(\kappa^{20})$ may be required.
We thus calculate $L_m(N_t,n)$ for $n$ from $N_t$ up to $20$.
For $n\leq20$, because $L_m(6,n)=0$ for $m \geq 4$ and $L_m(8,n)=0$ for $m \geq 3$, we have 
$W(n)=(D_n + D_n^{+} + 2D_n^{2-})/4$,
$L_1(N_t,n) + L_3 (N_t,n)=(D_n - D_n^{+})/2$,
$L_2(N_t,n) = (D_n + D_n^{+} - 2D_n^{2-})/4$, and
$L(N_t, n) = L_1(N_t,n)+L_2(N_t,n)+L_3(N_t,n)$.
We thus compute $D_n$, $D_n^{+}$, and $D_n^{2-}$.

We perform simulations on a $32^3 \times 6$ lattice at $\beta = 5.8810$ and $5.9000$, 
and on a $32^3 \times 8$ lattice at $\beta = 6.0320$ and $6.0660$.
These $\beta$ values are chosen to be slightly below and above the phase transition point.\footnote{
The transition point determined by the peak of the Polyakov loop susceptibility locates at $\beta_{\rm trans}=5.89383(24)$ on a $48^3\times6$ lattice and at $6.06160(18)$ on a $48^3\times8$ lattice~\cite{Shirogane:2016zbf}.
}
Details of the simulations are the same as in our previous studies~\cite{Saito:2011fs,Ejiri:2012rr,Saito:2013vja,Ejiri:2015vip,Ejiri:2019csa}.
Independent 50 configurations are generated at each $\beta$.
The number of noises $N_{\rm noise}$ is $1000$ for each configuration.

\begin{figure}[tb]
\begin{center}
\vspace{0mm}
\includegraphics[width=5.8cm]{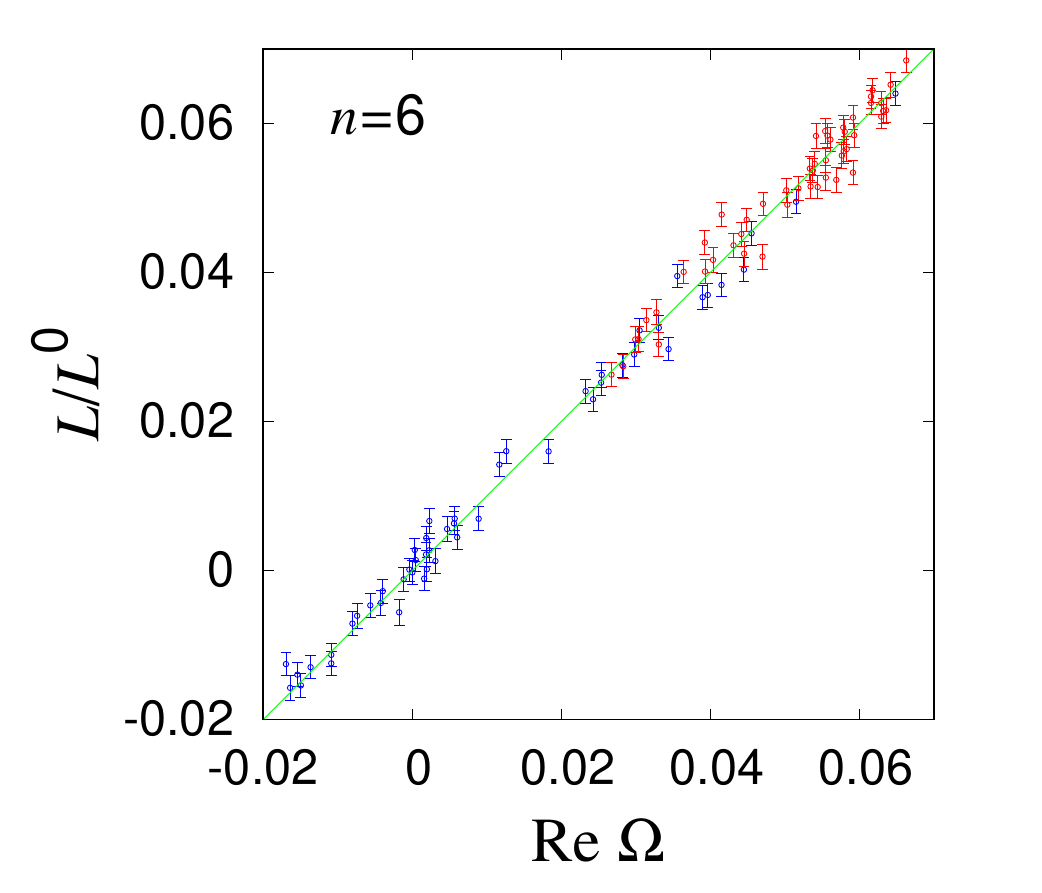}
\hspace*{-7mm}
\includegraphics[width=5.8cm]{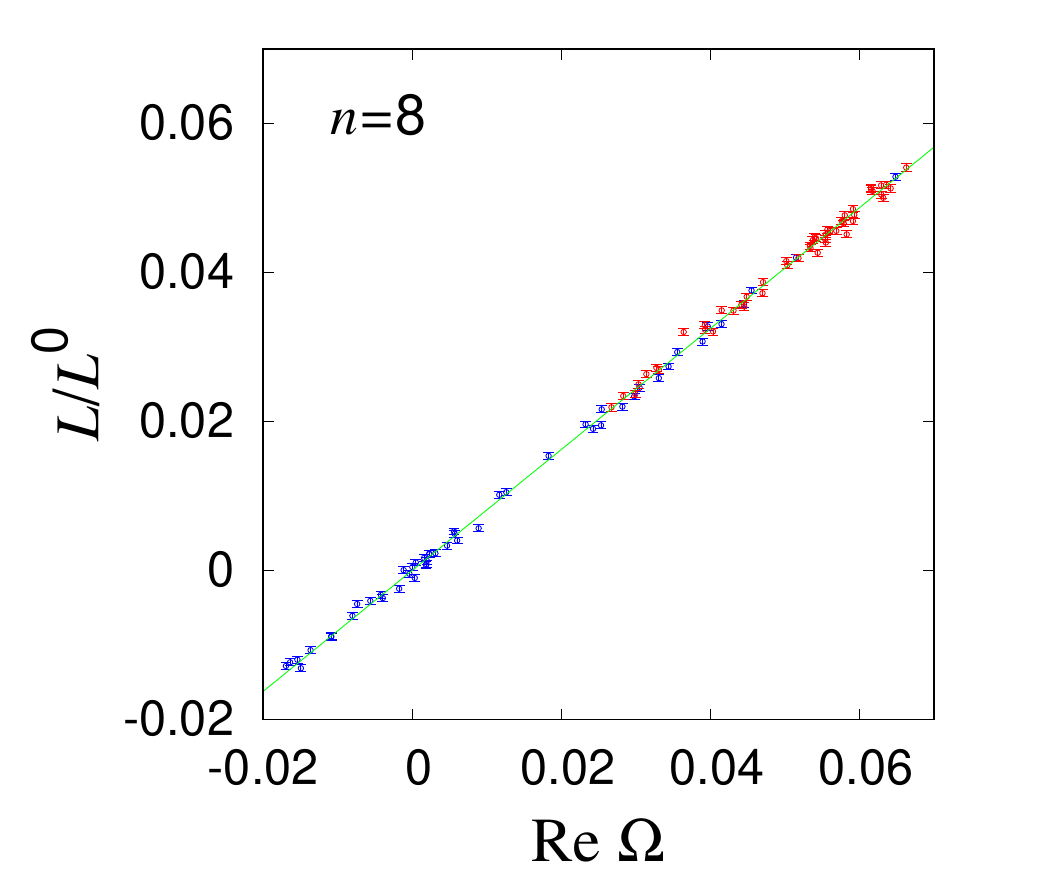}
\hspace*{-7mm}
\includegraphics[width=5.8cm]{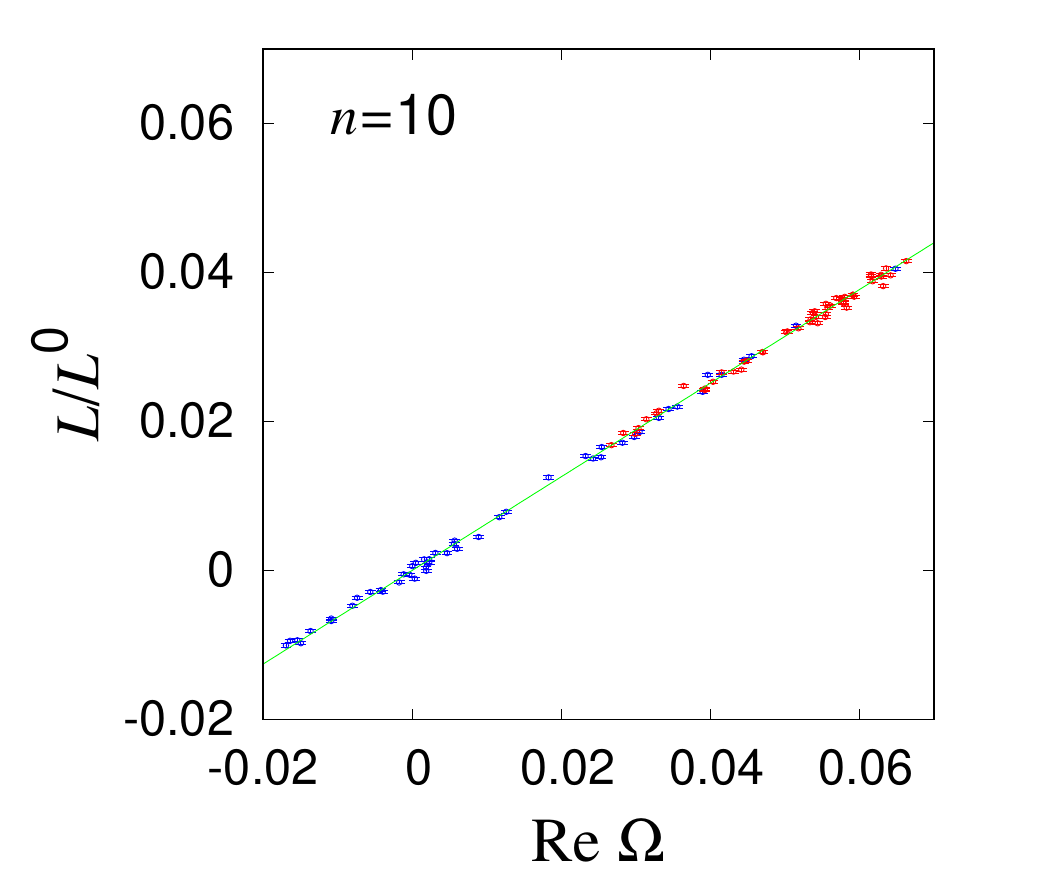} \\
\includegraphics[width=5.8cm]{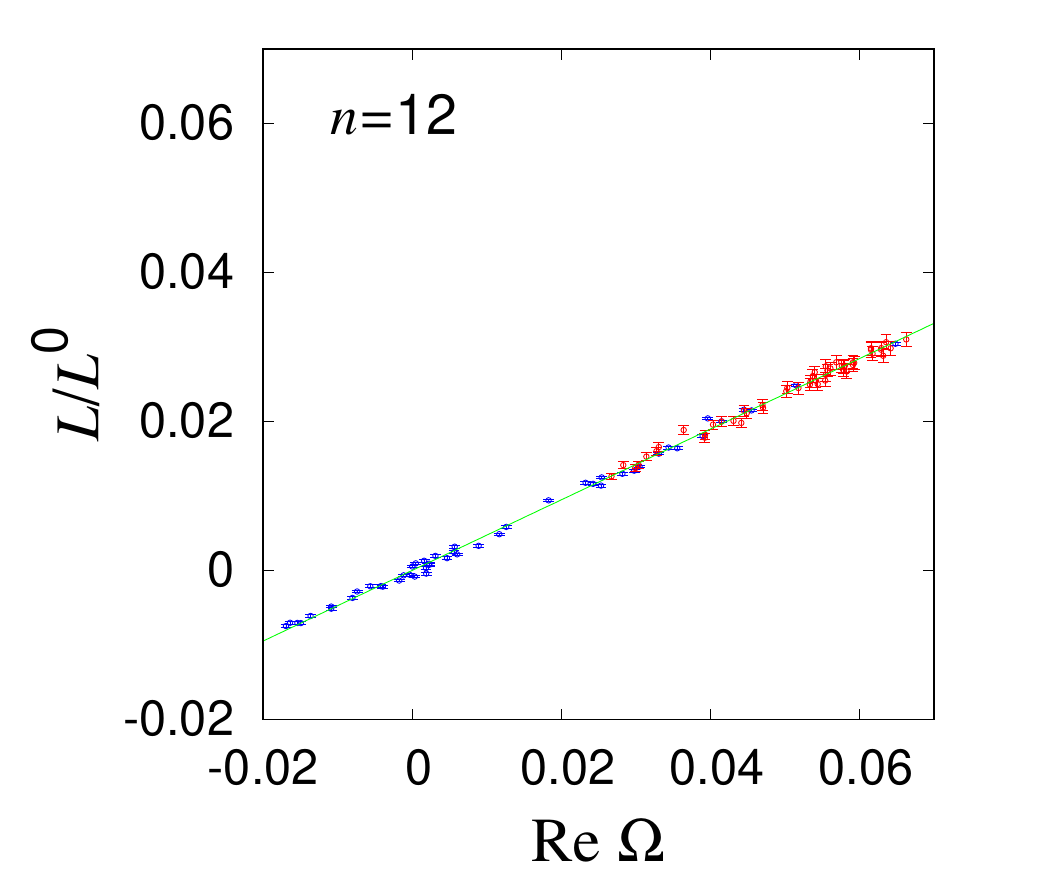}
\hspace*{-7mm}
\includegraphics[width=5.8cm]{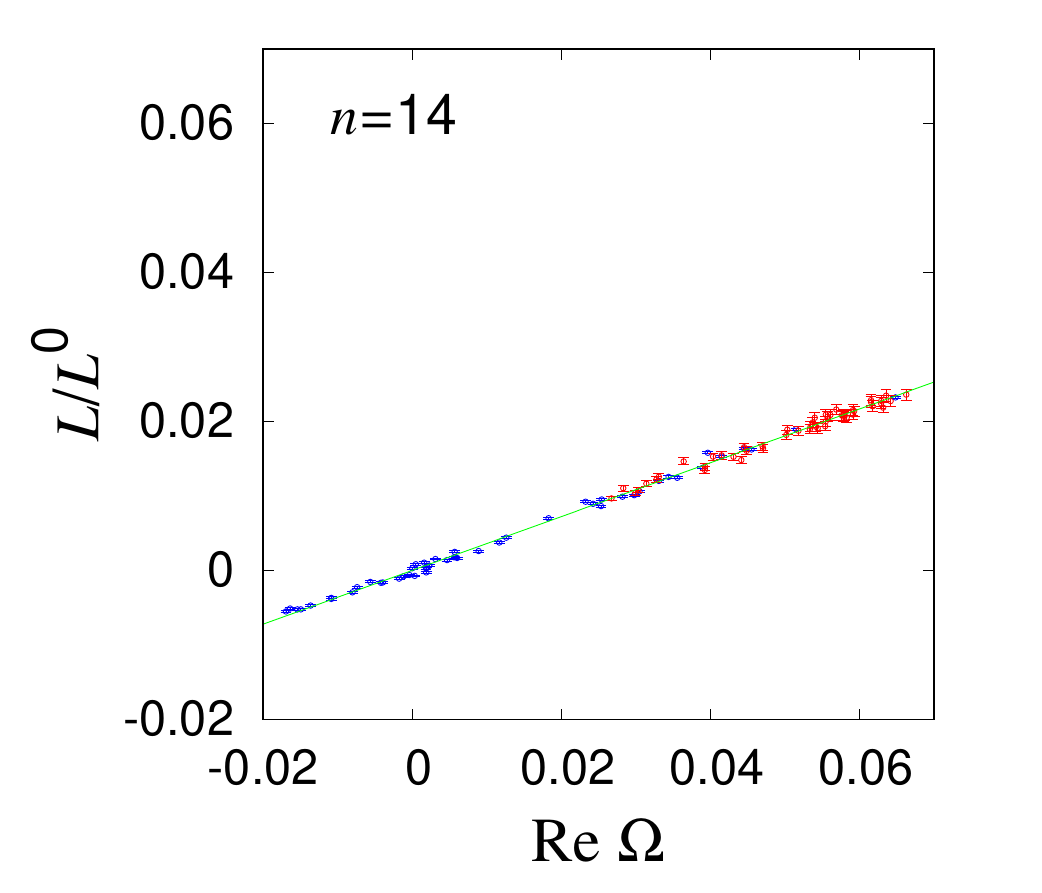}
\hspace*{-7mm}
\includegraphics[width=5.8cm]{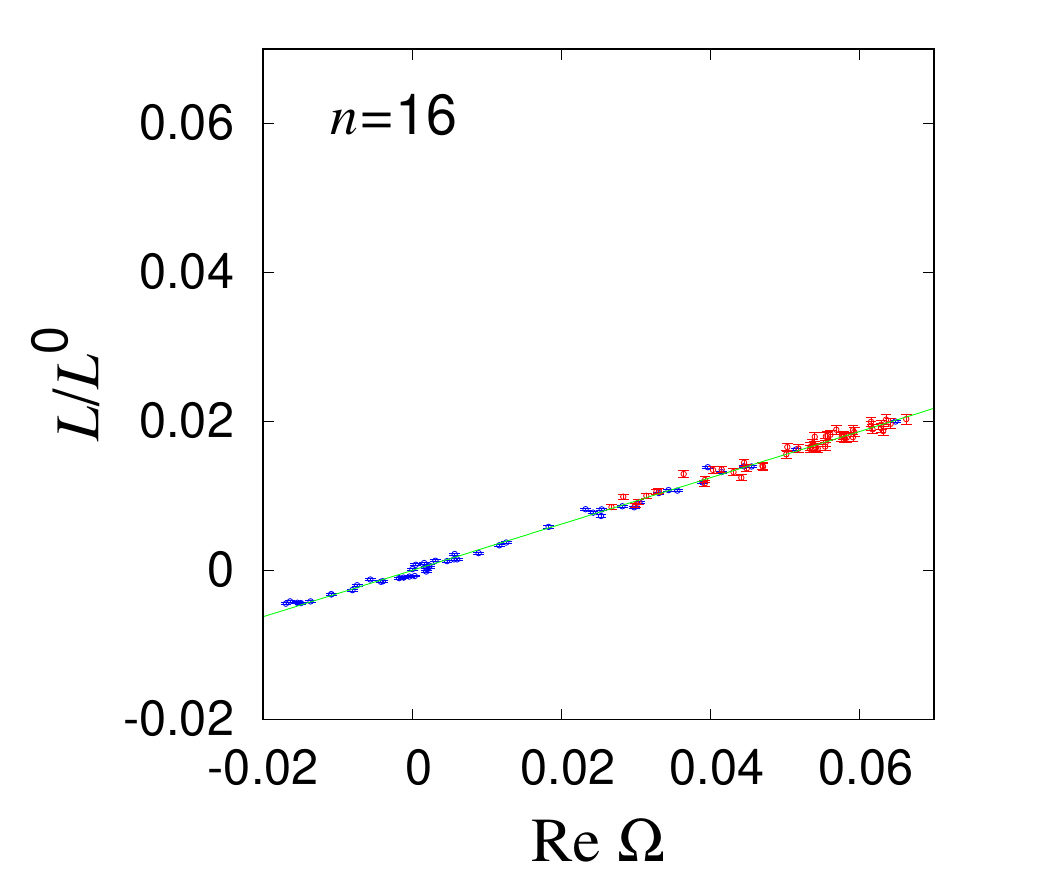} \\
\includegraphics[width=5.8cm]{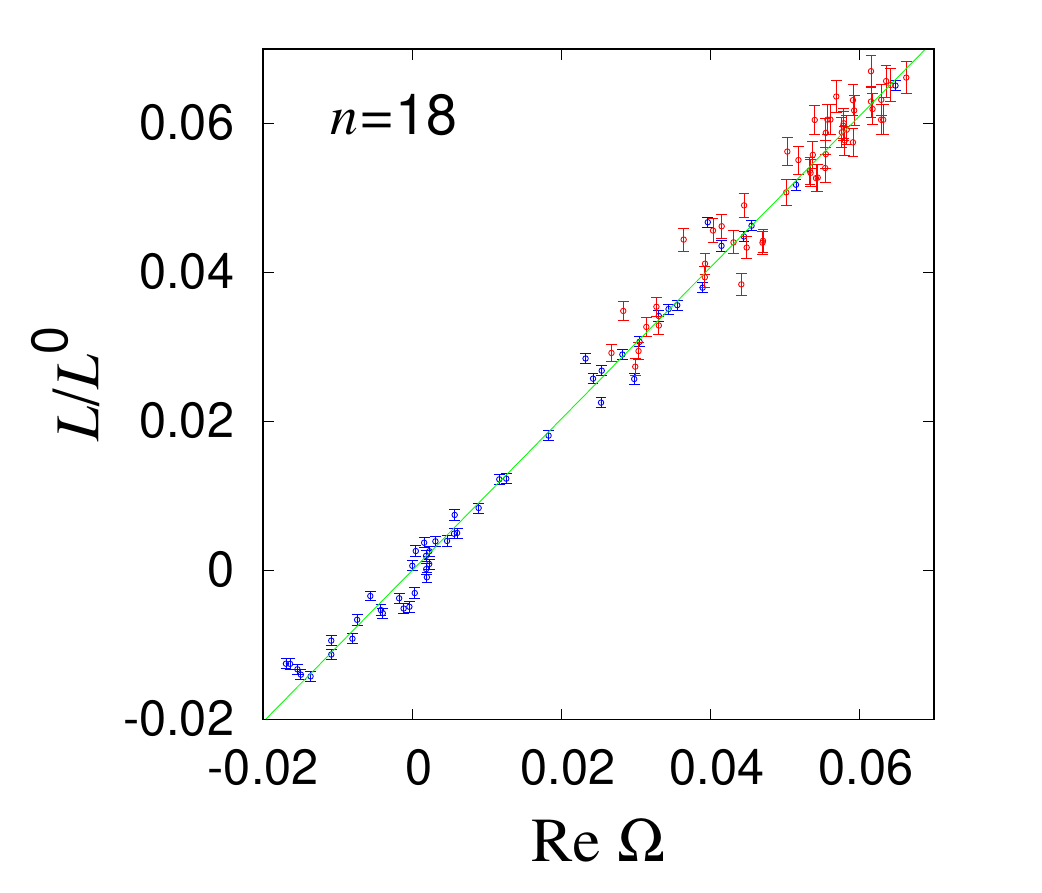}
\hspace*{-7mm}
\includegraphics[width=5.8cm]{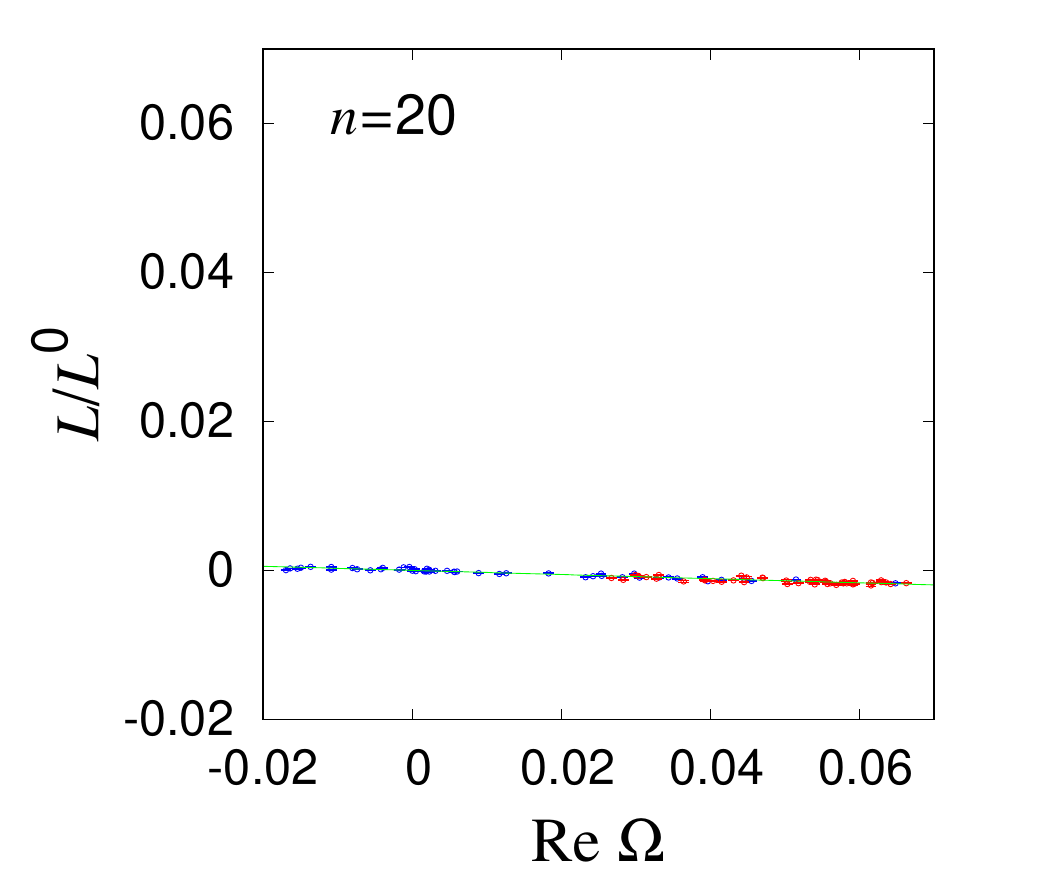}
\vspace{0mm}
\caption{Double distribution of $L(N_t, n)/L^0(N_t,n)$ and ${\rm Re} \hat{\Omega}$ obtained on $32^3 \times 6$ lattice. 
The top left, top middle, $\cdots$, and bottom right panels show the results of $n=6$, 8, $\cdots$, and 20, respectively.
The blue and red symbols are the results obtained at $\beta = 5.8810$ and $5.9000$, respectively.
The green lines are the results of linear fits with Eq.~(\ref{eq:lco}).
As discussed in the text, the cases $n = 18$ and 20 are a little special 
because $L^0 (N_t, n)$ in the denominator changes sign between $n=18$ and 20. 
}
\label{fig:corrln}
\end{center}
\end{figure}

\begin{figure}[tb]
\begin{center}
\vspace{0mm}
\includegraphics[width=5.8cm]{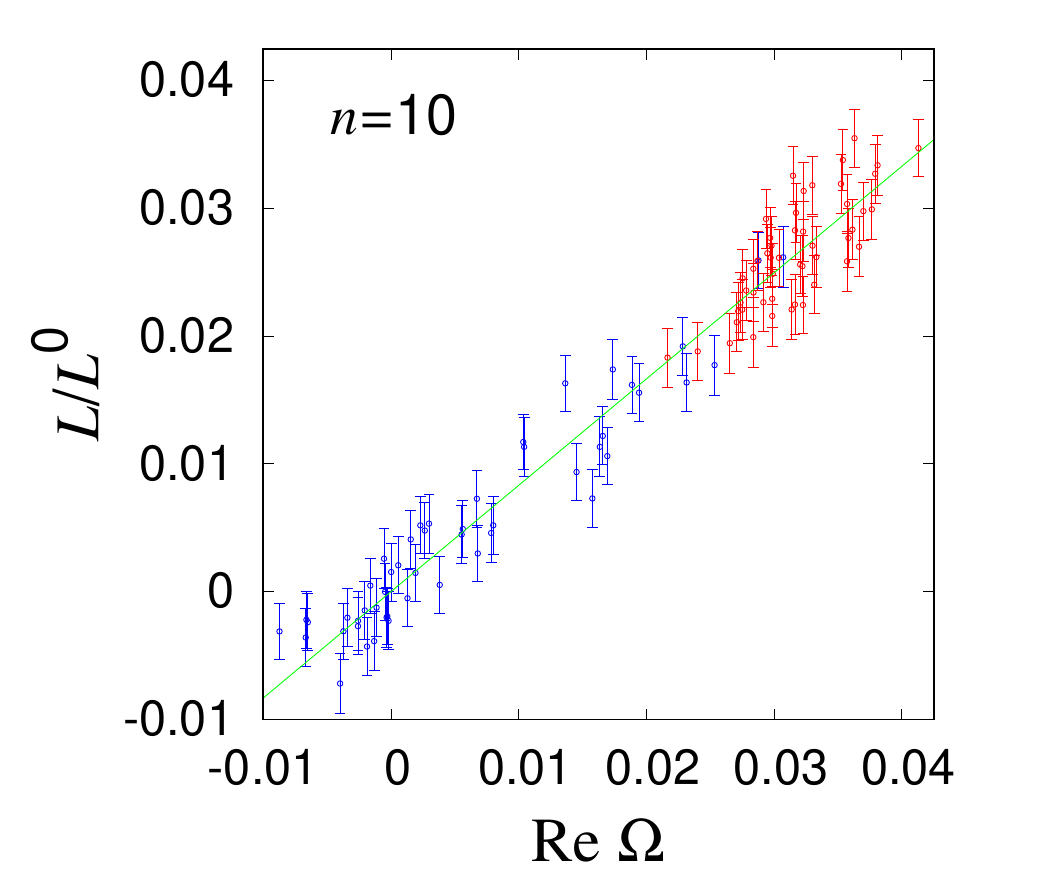}
\hspace*{-7mm}
\includegraphics[width=5.8cm]{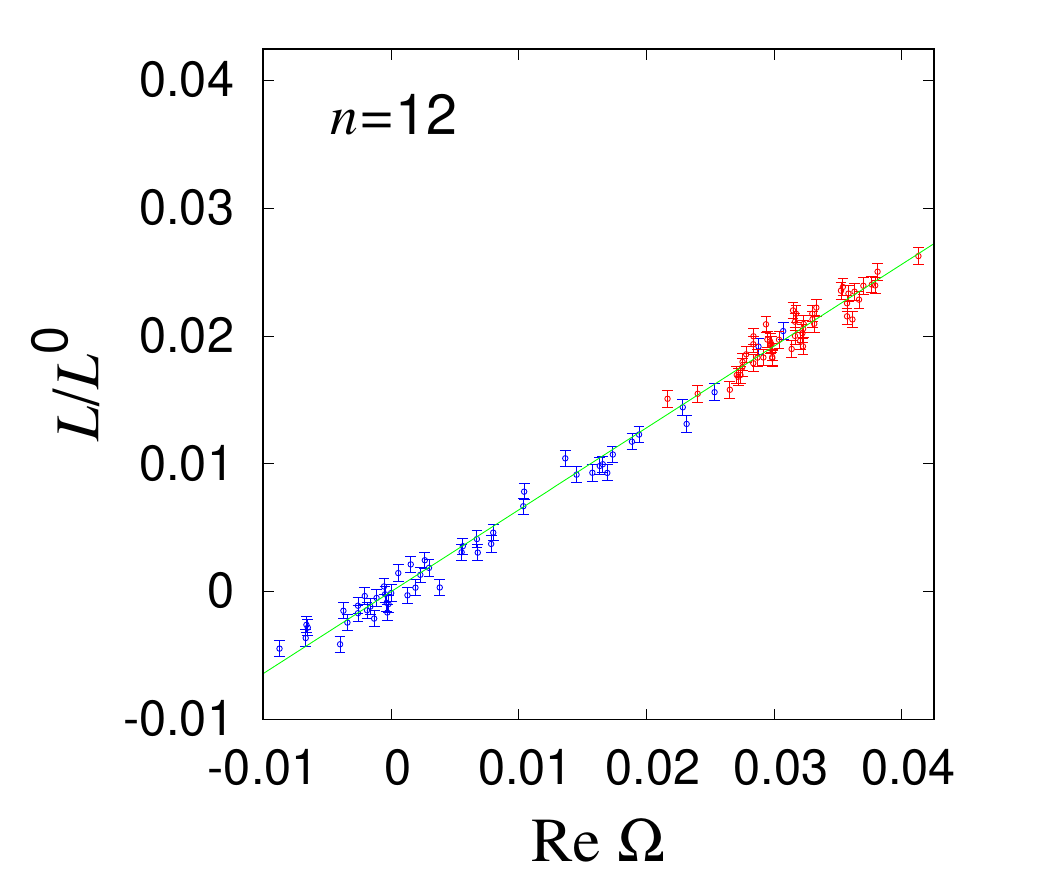}
\hspace*{-7mm}
\includegraphics[width=5.8cm]{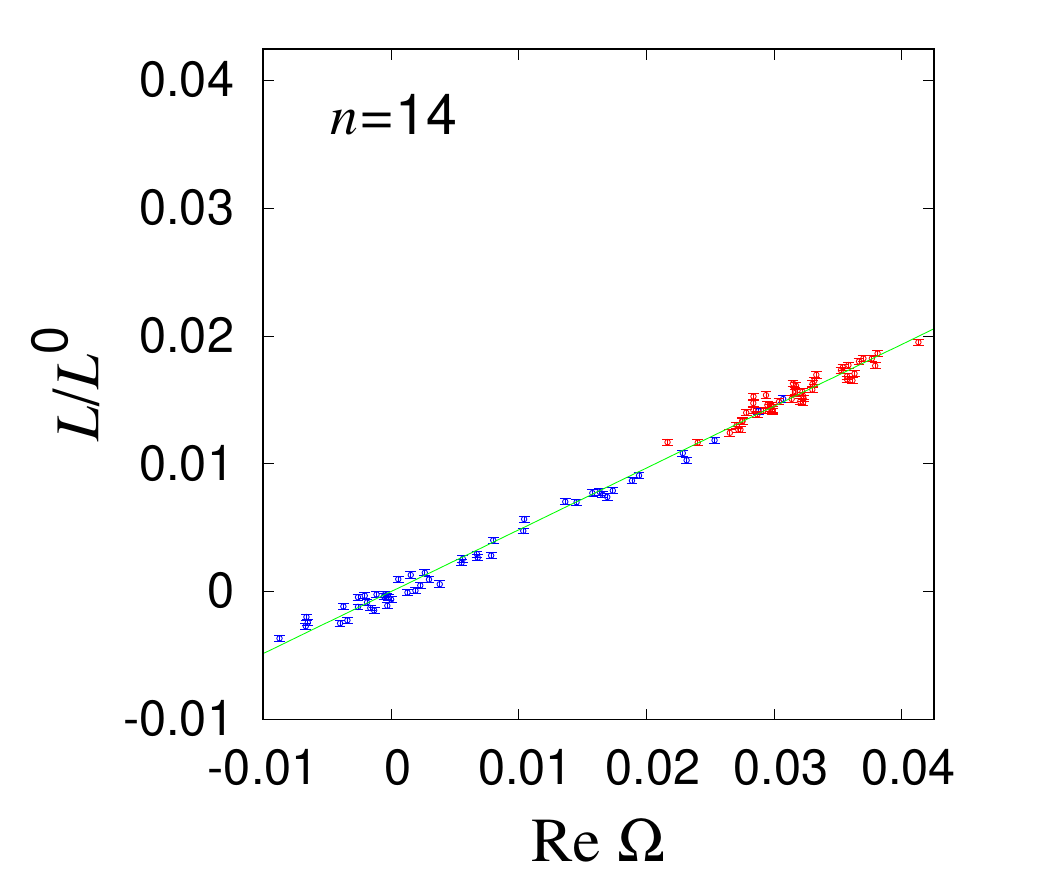} \\
\includegraphics[width=5.8cm]{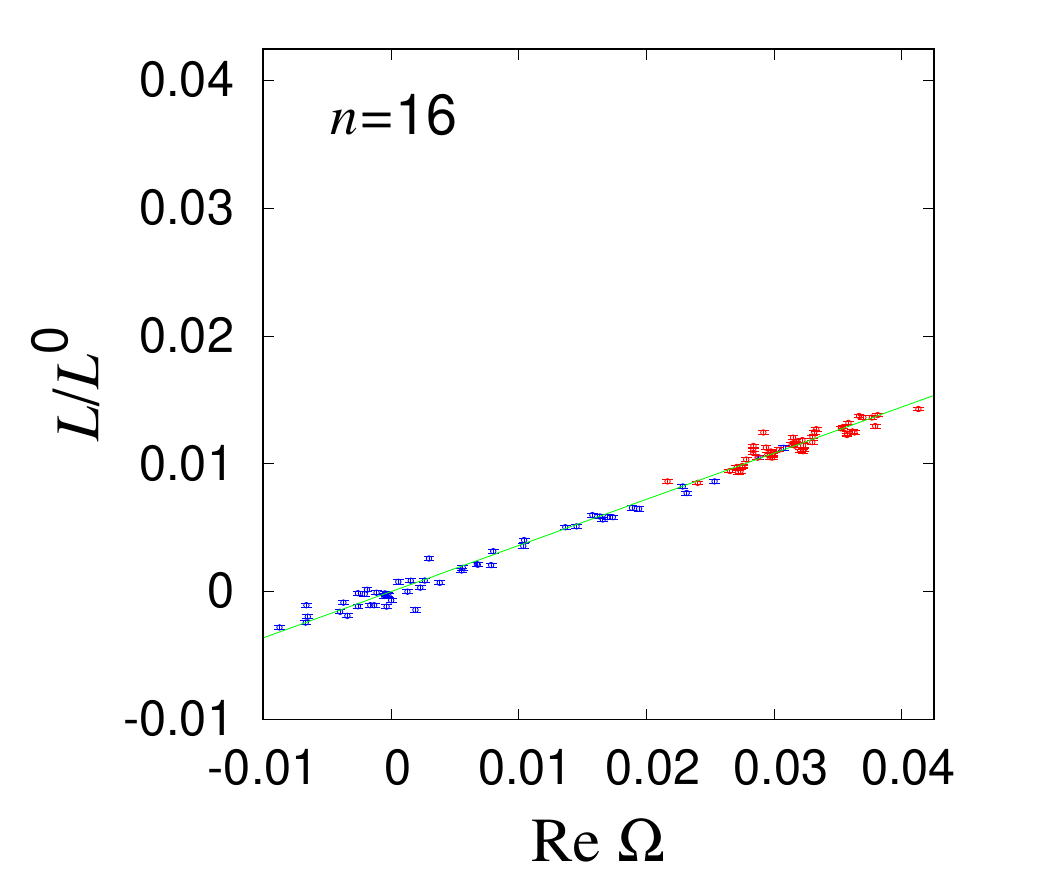}
\hspace*{-7mm}
\includegraphics[width=5.8cm]{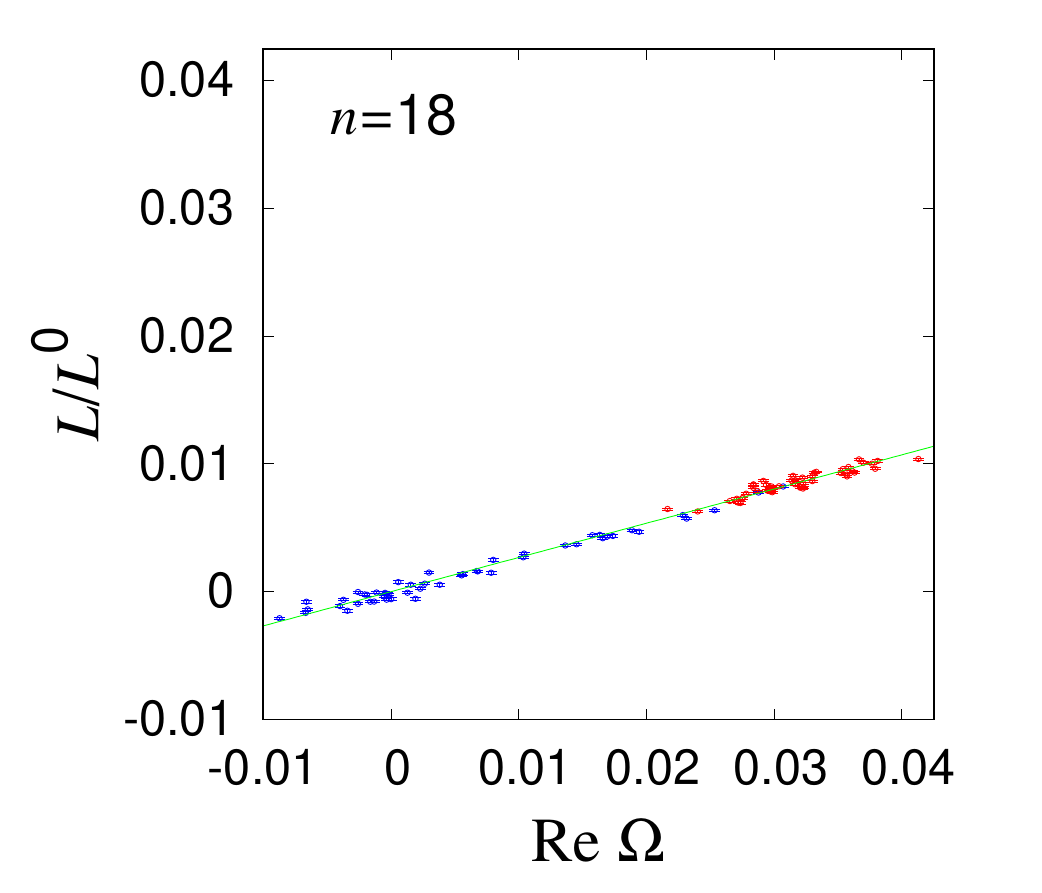}
\hspace*{-7mm}
\includegraphics[width=5.8cm]{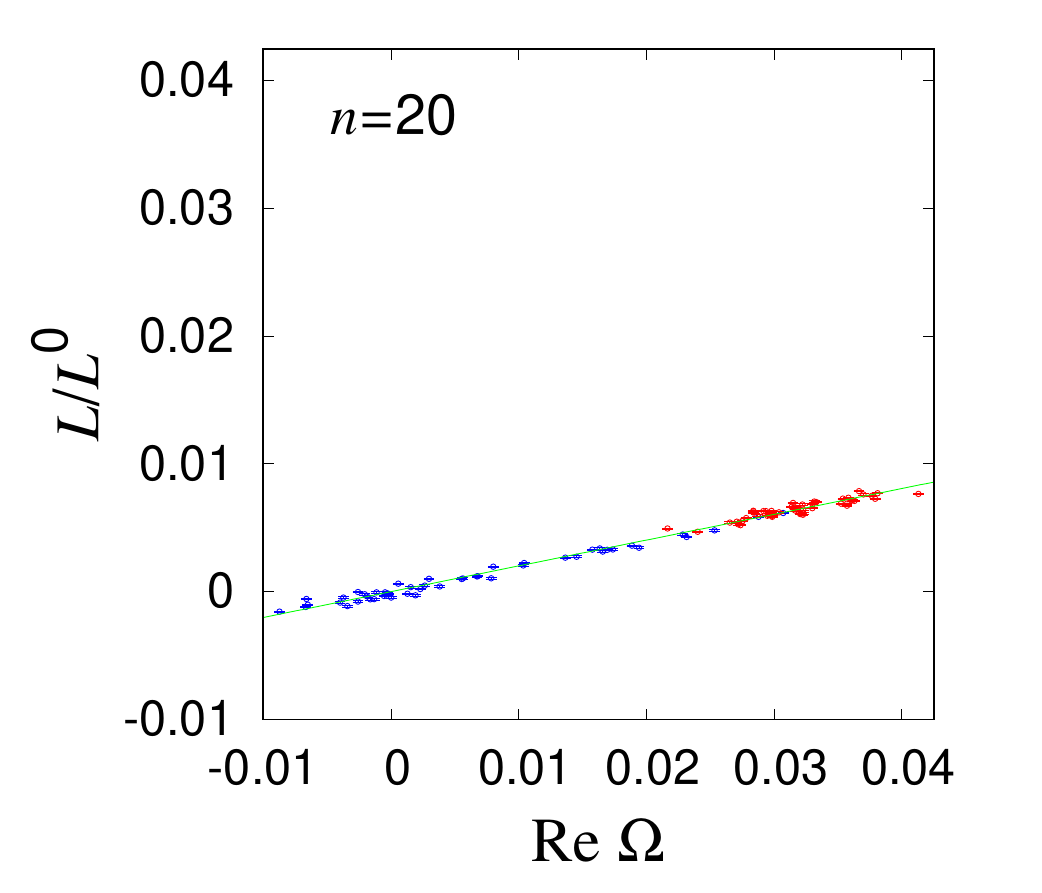}
\vspace{0mm}
\caption{The same as Fig.~\ref{fig:corrln} but on $32^3 \times 8$ lattice 
at $\beta =  6.0320$ (blue) and $6.0660$ (red). 
The top left, top middle, $\cdots$, and bottom right panels show the results of $n=10$, 12, $\cdots$, and 20, respectively.}
\label{fig:corrln8}
\end{center}
\end{figure}

\begin{table}[t]
\begin{center}
\caption{Coefficients $c_n$ of Eq.~(\ref{eq:lco}) for $n=N_t$--$20$ on $N_t=6$ and 8 lattices.}
\label{tab:cn}
\begin{tabular}{ccc}
\hline
\hspace{12mm} & \hspace{10mm} $N_t=6$ \hspace{10mm} & \hspace{10mm} $N_t=8$ \hspace{10mm} \\
\hline
$c_6$    & 1        & \\
$c_8$    & 0.8112(20)(7)   & 1  \\
$c_{10}$ & 0.6280(15)(3)   & 0.8327(114)(95) \\
$c_{12}$ & 0.4736(29)(15)  & 0.6408(36)(27) \\
$c_{14}$ & 0.3609(26)(11)  & 0.4841(22)(10) \\
$c_{16}$ & 0.3106(25)(10)  & 0.3616(21)(6) \\
$c_{18}$ & 1.0159(90)(33)  & 0.2679(16)(3) \\
$c_{20}$ & $-0.02771(57)(13)$ & 0.2020(13)(2) \\
\hline
\end{tabular}
\end{center}
\end{table}

\subsection{Correlation among Polyakov-type loop terms}

In Fig.~\ref{fig:corrln}, we plot the double distribution of $L (N_t, n)/ L^0 (N_t, n)$ (verticl axis) and 
${\rm Re} \hat{\Omega} = L (N_t, N_t)/ L^0 (N_t, N_t)$ (horizontal axis) obtained on each configuration of the $N_t=6$ lattice.
The blue and red symbols are for $\beta = 5.8810$ and $5.9000$, respectively.
The error bar represents the error caused by the finite $N_{\rm noise}$ on each configuration, 
which decreases as $n$ increases since $L (N_t, n)/ L^0 (N_t, n)$ is a weighted average of Polyakov-type loops and the number of Polyakov-type loops to be averaged increases as $n$ increases.
Although the results of these figures include contributions from $L_1(N_t,n)$, $L_2(N_t,n)$ and $L_3(N_t,n)$, the contributions from $L_2(N_t,n)$ and $L_3(N_t,n)$ are negligibly small compared to $L_1(N_t,n)$.  
The top left panel of~Fig.~\ref{fig:corrln} shows the result of $n=6$. 
Because $L(6,6)=L^0(6,6)\, {\rm Re} \hat{\Omega}$, both axes are ${\rm Re} \hat{\Omega}$, and the figure confirms the accuracy of the noise method.
The other panels are the results of $n=8$ (top middle), 10 (top right), 12 (middle left), 14 (middle), 16 (middle right), 18 (bottom left), and 20 (bottom right). 

Figure~\ref{fig:corrln8} shows the double distribution of $L (N_t, n)/ L^0 (N_t, n)$ and ${\rm Re} \hat{\Omega}$ obtained on the $N_t=8$ lattice.
Each panel shows the result for $n=10$ (top left), 12 (top middle), 14 (top right), 16 (bottom left), 18 (bottom middle) and 20 (bottom right), respectively, obtained at $\beta = 6.0320$ (blue) and $6.0660$ (red).

These figures show that $L (N_t, n)$ has a strong linear correlation with ${\rm Re} \hat{\Omega}$, 
i.e., Eq.~(\ref{eq:lco}) is well satisfied.
We thus calculate the coefficient $c_n$ by fitting the $L (N_t, n)$ data by Eq.~(\ref{eq:lco}), for each $n$ and $N_t$.
Minimizing $\chi^2=\sum_i [(L(i) - L^0(N_t,n) \, c_n {\rm Re} \hat{\Omega}(i))/ \Delta L(i)]^2$,
with $L(i)$ the result of $L(N_t, n)$ on the $i$th configuration and $\Delta L(i)$ the error of $L(i)$ due to the noise method, 
the best value of $c_n$ is given by
\begin{eqnarray}
c_n = \frac{1}{L^0(N_t, n)} \frac{\langle L(N_t, n) {\rm Re} \hat{\Omega}/ \Delta L^2 \rangle}{\langle ({\rm Re} \hat{\Omega})^2 / \Delta L^2 \rangle} .
\end{eqnarray}
The error propagation from $\Delta L$ to $c_n$ is given by 
\begin{eqnarray}
\Delta c_n = \frac{1}{\sqrt{N_{\rm conf}} \, |L^0(N_t, n)|} \left\langle \frac{({\rm Re} \hat{\Omega})^2}{\Delta L^2} \right\rangle^{-1/2}, \nonumber
\end{eqnarray}
where $N_{\rm conf}$ is the number of configurations.

The results of the fits are shown by the green lines in~Figs.~\ref{fig:corrln} and \ref{fig:corrln8}.
We summarize the results of $c_n$ in Table~\ref{tab:cn},
in which the numbers in the first parenthesis are the statistical errors and those in the second parenthesis are the errors propagated from the error of the noise method.
We note that $c_n$ decreases by a factor of about $0.8$ as the order of $\kappa^2$ increases by one.
Thus, our results of $c_n$ are well approximated by $c_{n} \approx (0.8)^{(n-N_t)/2}$, except for $c_{18}$ and $c_{20}$ for $N_t=6$
--- the cases $n = 18$ and 20 for $N_t=6$ are a little special 
because the sign of $L^0 (N_t, n)$ changes between 18 and 20 and thus the positive sign terms and the negative sign terms cancel with each other in $L^0 (N_t, n)$ there. 
For $N_t = 8$, the sign of $L^0(N_t, n)$ changes first at $n = 28$.

\begin{figure}[tb]
\begin{center}
\vspace{0mm}
\includegraphics[width=7.8cm]{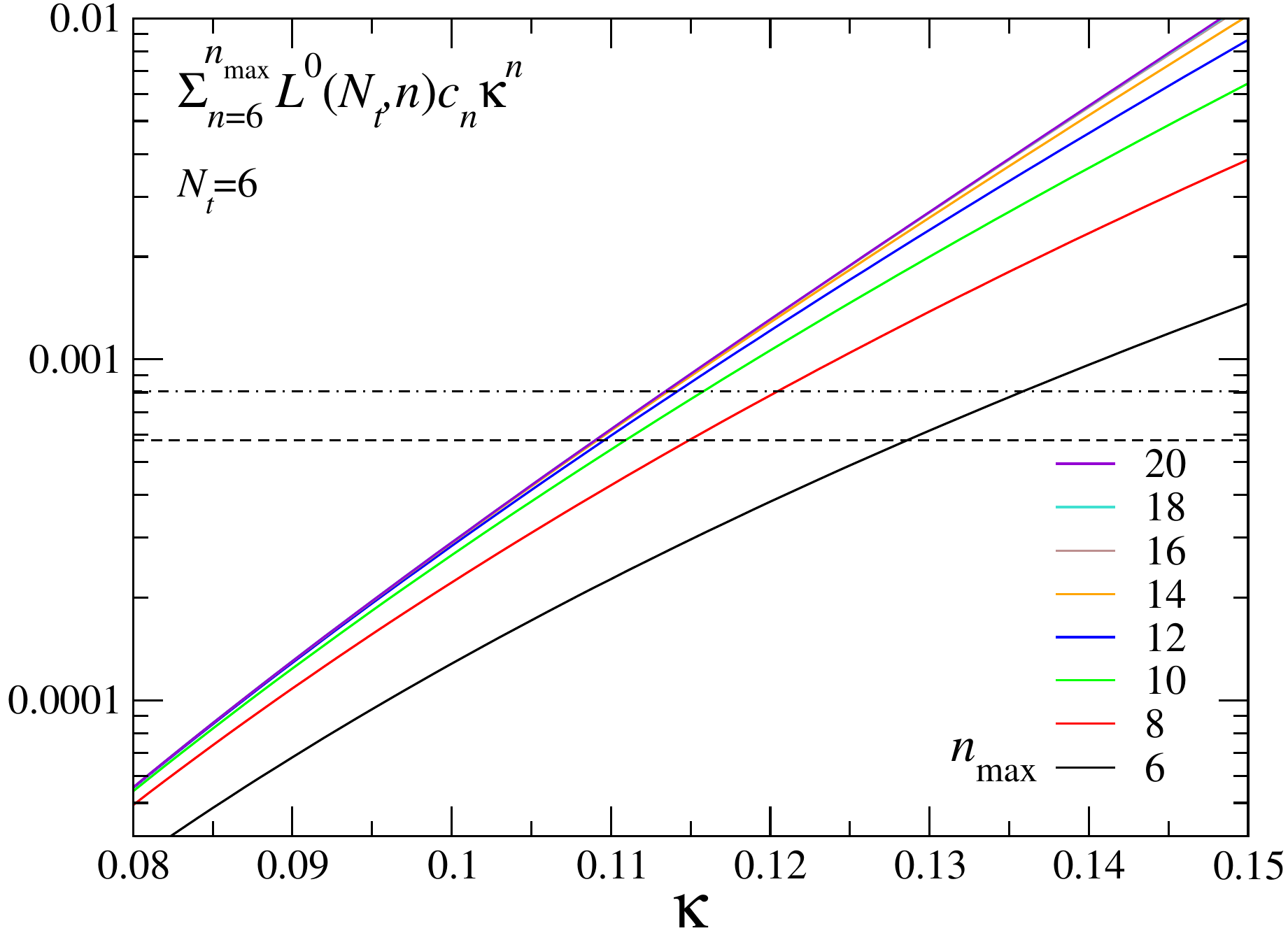}
\hspace*{2mm}
\includegraphics[width=7.8cm]{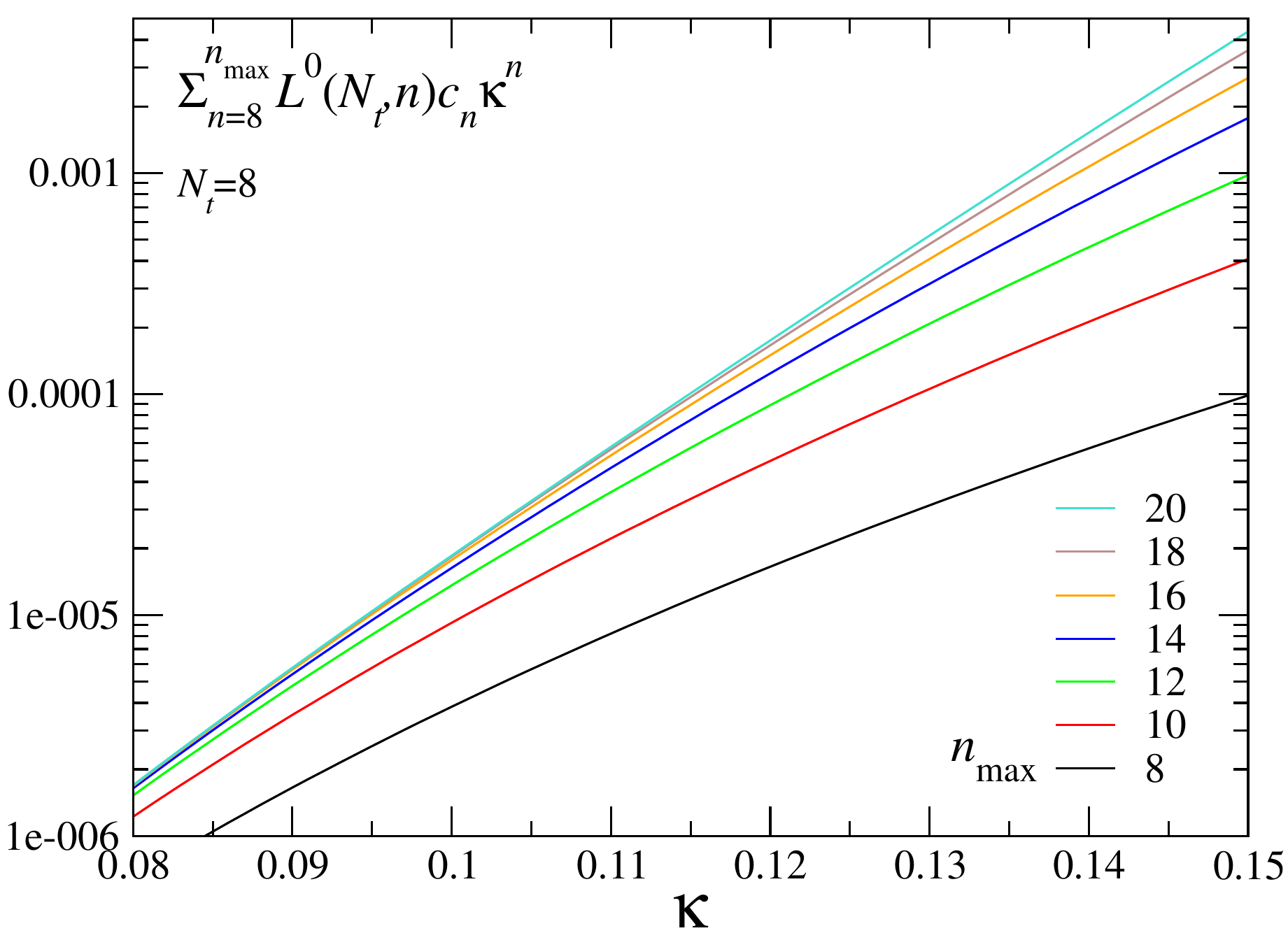}
\vspace{0mm}
\caption{$\sum_{n=N_t}^{n_{\rm max}} L^0(N_t, n) \, c_n \kappa^n$ for $N_t=6$ (left) and 8 (right) as functions of $n_{\rm max}$. 
The horizontal dot-dashed and dashed lines in the left panel are $L^0 (N_t, N_t) \, \kappa_{\rm c, LO}^{N_t}$ obtained on $24^3 \times 6$ and $32^3 \times 6$ lattices, respectively.}
\label{fig:hexcn8}
\end{center}
\end{figure}

\begin{table}[t]
\begin{center}
\caption{Effective critical point $\kappa_{\rm c, eff}$ in degenerate $N_{\rm f}$-flavor QCD on $N_t=6$ lattices. }
\label{tab:keff}
\begin{tabular}{cccc}
\hline
$N_s^3 \times N_t$ & $N_{\rm f}=1$ &  $N_{\rm f}=2$ &  $N_{\rm f}=3$ \\
\hline
$24^3 \times 6$ & 0.1228(18) & 0.1134(18) & 0.1080(17) \\
$32^3 \times 6$ & 0.1183(25) & 0.1090(25) & 0.1037(24) \\
\hline
\end{tabular}
\end{center}
\end{table}

\begin{figure}[tb]
\begin{center}
\vspace{0mm}
\includegraphics[width=7.8cm]{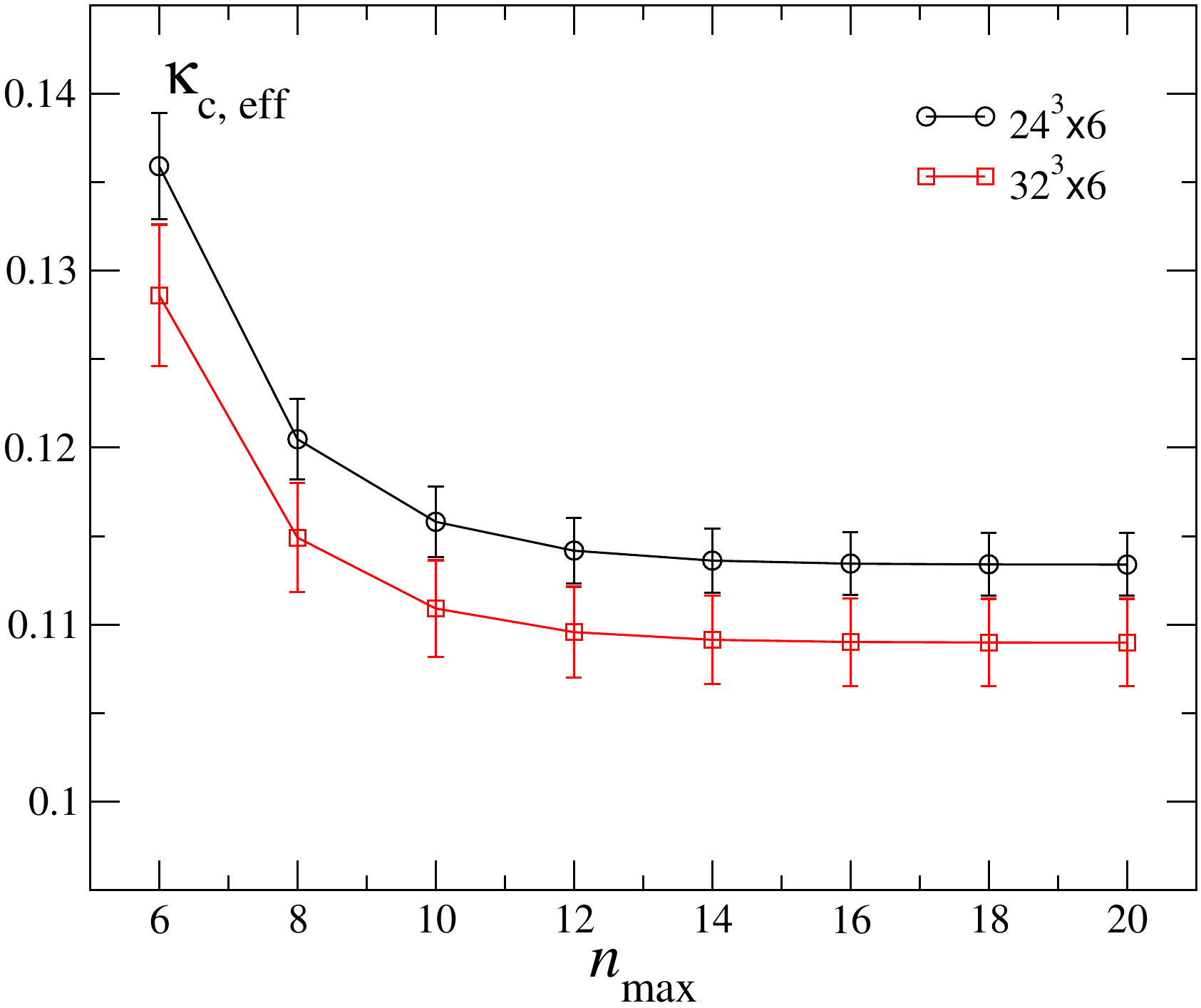}
\vspace{0mm}
\caption{Effective critical point $\kappa_{\rm c, eff}$ in two flavor QCD for $N_t=6$ as function of $n_{\rm max}$.
Black circle and red square symbols are for the results using $\kappa_{\rm c, LO}$ obtained on $24^3 \times 6$ and $32^3 \times 6$ lattices, respectively.
}
\label{fig:keff6}
\end{center}
\end{figure}

\subsection{Location of critical point incorporating high order effects}

We now determine the effective critical point $\kappa_{\rm c, eff}$ 
by substituting $c_n$ of Table~\ref{tab:cn} into Eq.~(\ref{eq:keff}), i.e.,
\begin{eqnarray}
\sum_{n=N_t}^{n_{\rm max}} L^0 (N_t, n) \, c_n \kappa_{\rm c, eff}^n = L^0 (N_t, N_t)\, \kappa_{\rm c, LO}^{N_t} .
\nonumber
\end{eqnarray}
In Fig.~\ref{fig:hexcn8},  
we show $\sum_{n=N_t}^{n_{\rm max}} L^0 (N_t, n) \, c_n \kappa^n$ for $N_t=6$ (left panel) and 8 (right panel) using $c_{n}$ of Table~\ref{tab:cn}. 
For $\kappa_{\rm c, LO}$ in the right hand side of Eq.~(\ref{eq:keff}),
we adopt the results of leading-order calculation in two flavor QCD, 
$\kappa_{\rm c, LO} = 0.1359(30)$ obtained on a $24^3 \times 6$ lattice, and also
$\kappa_{\rm c, LO} = 0.1286(40)$ obtained on a $32^3 \times 6$ lattice~\cite{Ejiri:2019csa}.
In the left panel, the horizontal dot-dashed and dashed lines are the $L^0 (N_t, N_t) \, \kappa_{\rm c, LO}^{N_t}$ obtained on $24^3 \times 6$ and $32^3 \times 6$ lattices, respectively.
The crossing points of the horizontal line with colored solid curves give the values of the effective critical point $\kappa_{\rm c, eff}$ for various $n_{\rm max}$.

Figure~\ref{fig:keff6} shows the results of $\kappa_{\rm c, eff}$ for $N_t=6$ as function of $n_{\rm max}$.
Black and red data are obtained using $\kappa_{\rm c, LO}$ on $24^3 \times 6$ and $32^3 \times 6$ lattices, respectively.
Our result $\kappa_{\rm c, eff} = 0.1205(23)$ for $n_{\rm max}=N_t+2$ on the $24^3 \times 6$ lattice is consistent with the result of the next-to-leading-order calculation given in Ref.~\cite{Ejiri:2019csa}.
As $n_{\rm max}$ increases, $\kappa_{\rm c, eff}$ converges to 
$\kappa_{\rm c, eff} = 0.1134(18)$ for the $24^3 \times 6$ lattice and 
$0.1090(25)$ for $32^3 \times 6$. 
The effective critical point $\kappa_{\rm c, eff}$ of the degenerate $N_{\rm f}$-flavor QCD on $N_t = 6$ lattices in the $n_{\rm max}$ infinity limit are shown in Table.~\ref{tab:keff}.
We find that the hopping parameter expansion works well for $N_t=6$ when high order effects are incorporated up to a sufficiently high order. 
At the same time, we note sizable finite volume effect in $\kappa_{\rm c, eff}$.
Large finite volume effect was reported also in the leading-order determination of the critical point by the shape of the histogram of the Polyakov loop~\cite{Kiyohara:2021smr}.
It is important to perform a systematic study with large spatial volumes to remove the finite volume effect.

\begin{figure}[tb]
\begin{center}
\vspace{0mm}
\includegraphics[width=7.2cm]{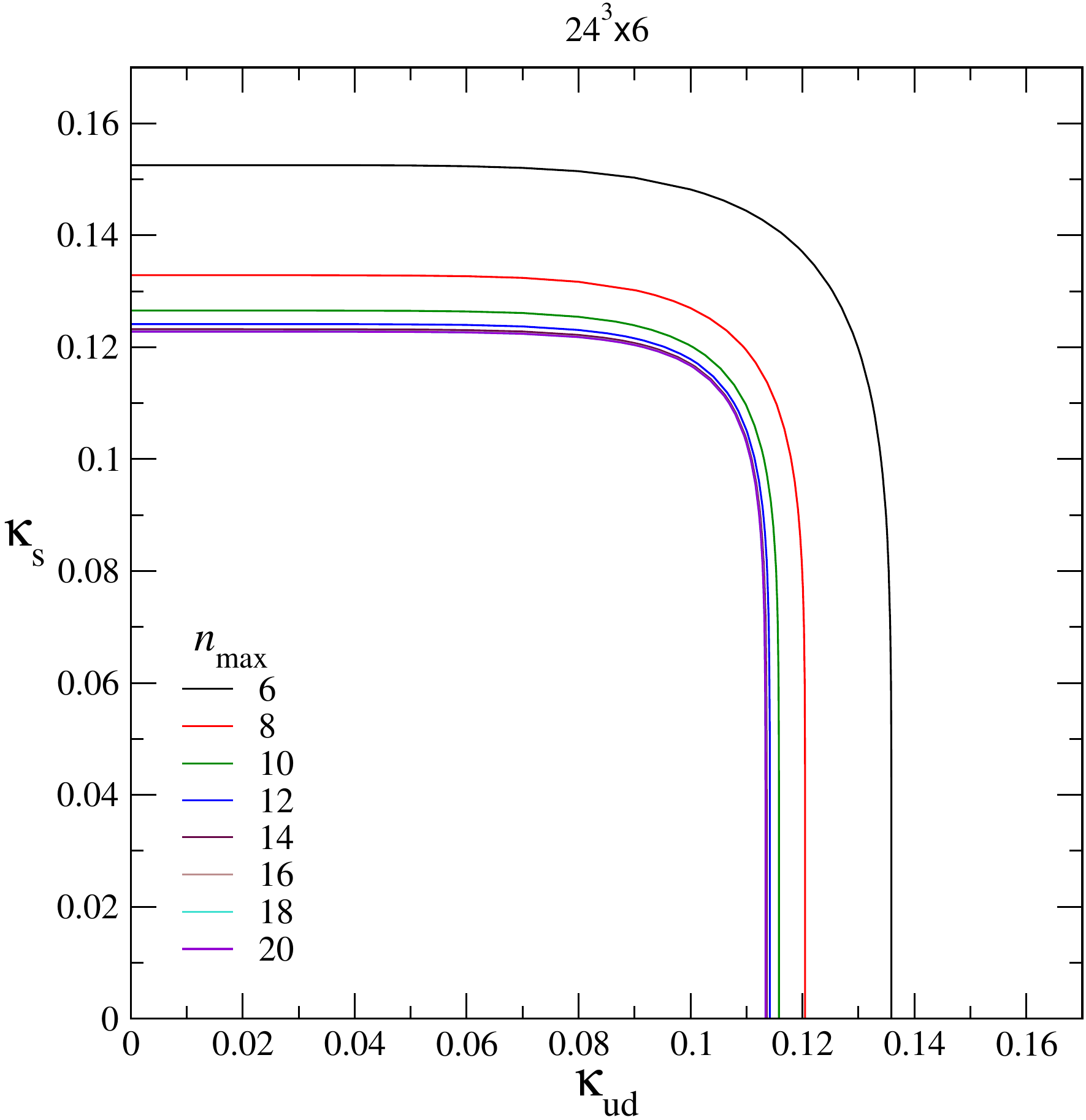}
\hspace*{5mm}
\includegraphics[width=7.2cm]{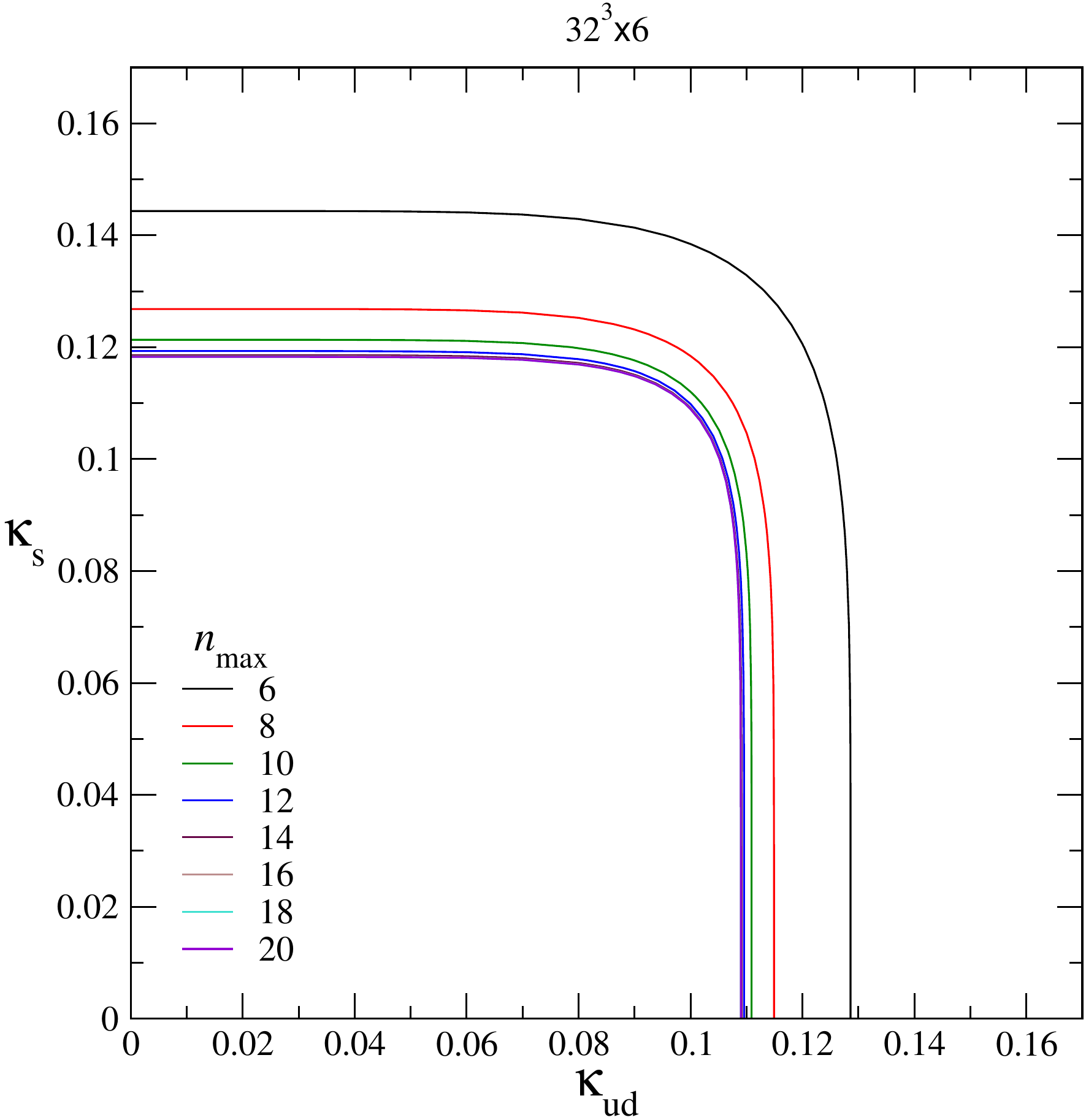}
\vspace{0mm}
\caption{Critical line in 2+1-flavor QCD calculated with various $n_{\rm max}$ for $24^3 \times 6$ (left) and $32^3 \times 6$ (right) lattices.}
\label{fig:21fkeff6}
\end{center}
\end{figure}

It is easy to generalize the argument to the case of non-degenerate quarks.
For 2+1-flavor QCD, denoting the hopping parameter for the up and down quarks as $\kappa_{ud}$ and that for the strange quark as $\kappa_s$, the critical line in the $(\kappa_{ud}, \kappa_s)$ plane is obtained by finding $(\kappa_{c, ud}, \kappa_{c, s})$ that satisfies the following equation 
\begin{eqnarray}
2 \sum_{n=N_t}^{n_{\rm max}} L^0 (N_t, n) \, c_n \kappa_{c, ud}^n 
+ \sum_{n=N_t}^{n_{\rm max}} L^0 (N_t, n) \, c_n \kappa_{c, s}^n 
= 2 L^0 (N_t, N_t) \, \kappa_{\rm c, LO}^{N_t} ,
\label{eq:keff3f}
\end{eqnarray}
where $\kappa_{\rm c, LO}$ is the leading-order critical point in two-flavor QCD.
The critical lines calculated with $n_{\rm max} = 6$--20 are shown in Fig.~\ref{fig:21fkeff6} using $\kappa_{\rm c, LO}$ obtained on the $24^3 \times 6$ (left) and $32^3 \times 6$ (right) lattices.
The critical line converges well when $n_{\rm max} \simge 10$.

\subsection{Correlation among Wilson loops}
\label{wilsonloopc}

\begin{figure}[tb]
\begin{center}
\vspace{0mm}
\includegraphics[width=5.8cm]{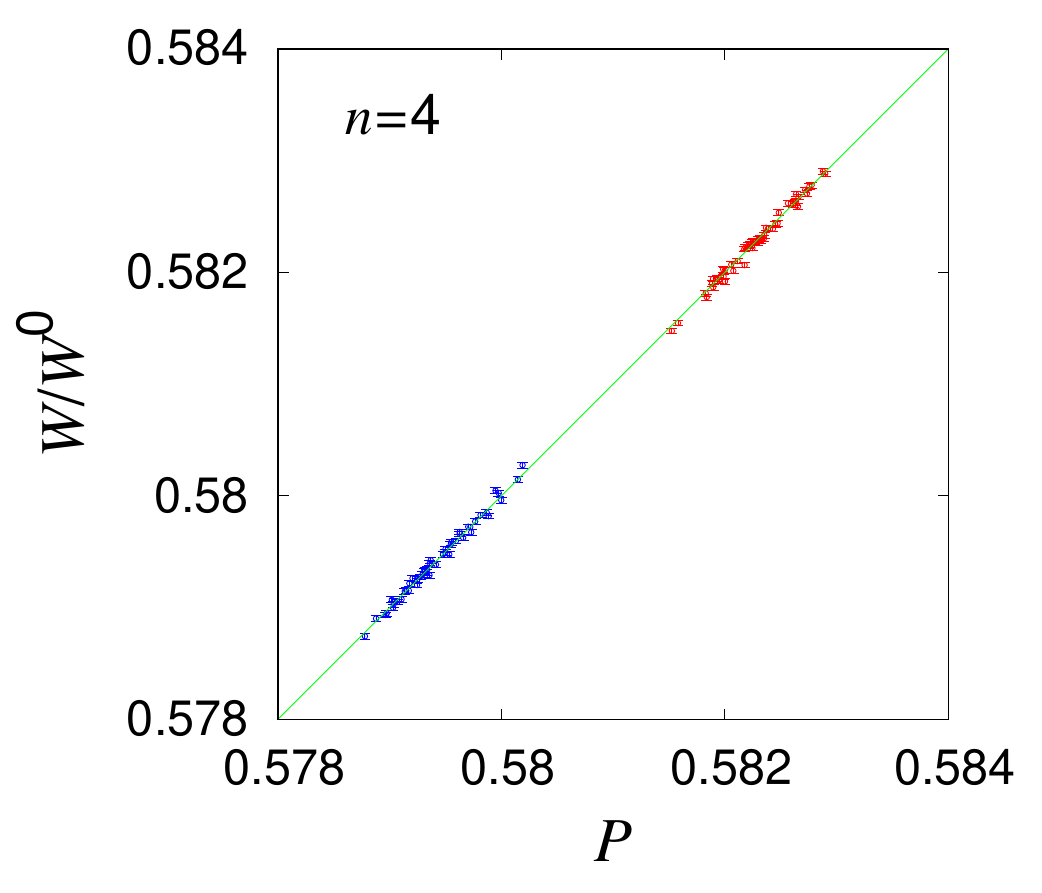}
\hspace*{-7mm}
\includegraphics[width=5.8cm]{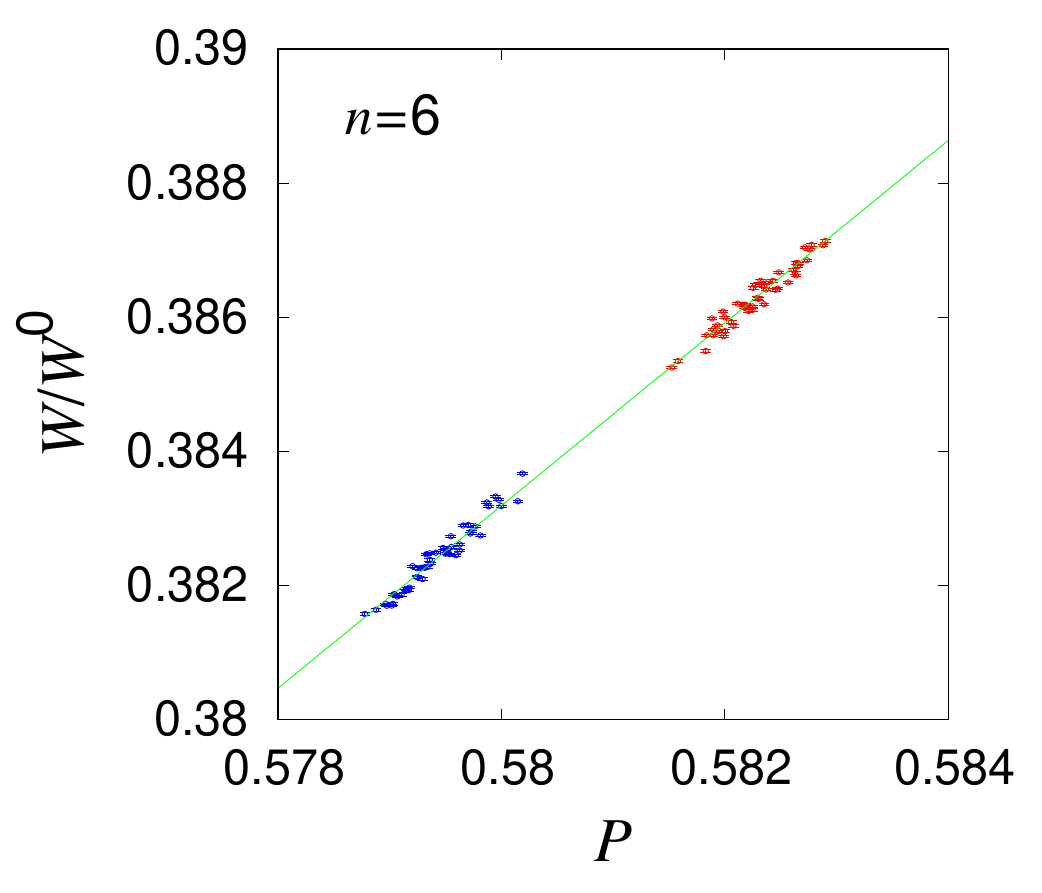}
\hspace*{-7mm}
\includegraphics[width=5.8cm]{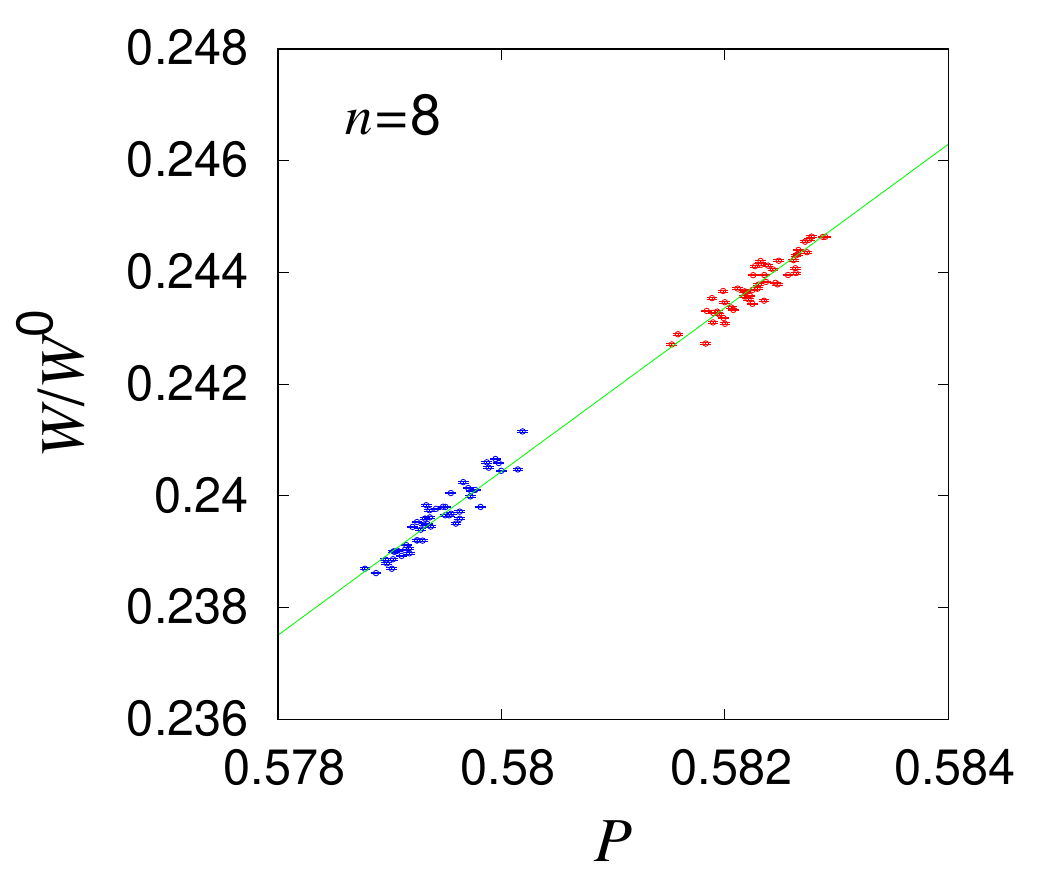} \\
\includegraphics[width=5.8cm]{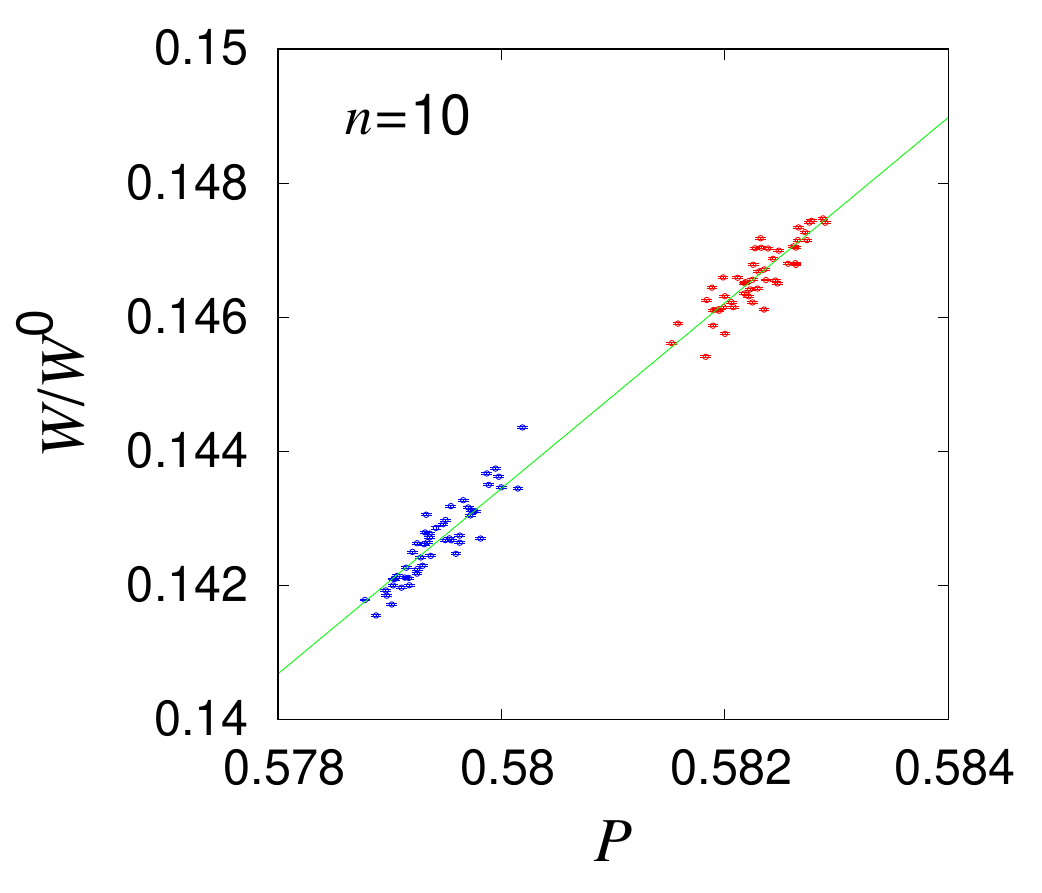}
\hspace*{-7mm}
\includegraphics[width=5.8cm]{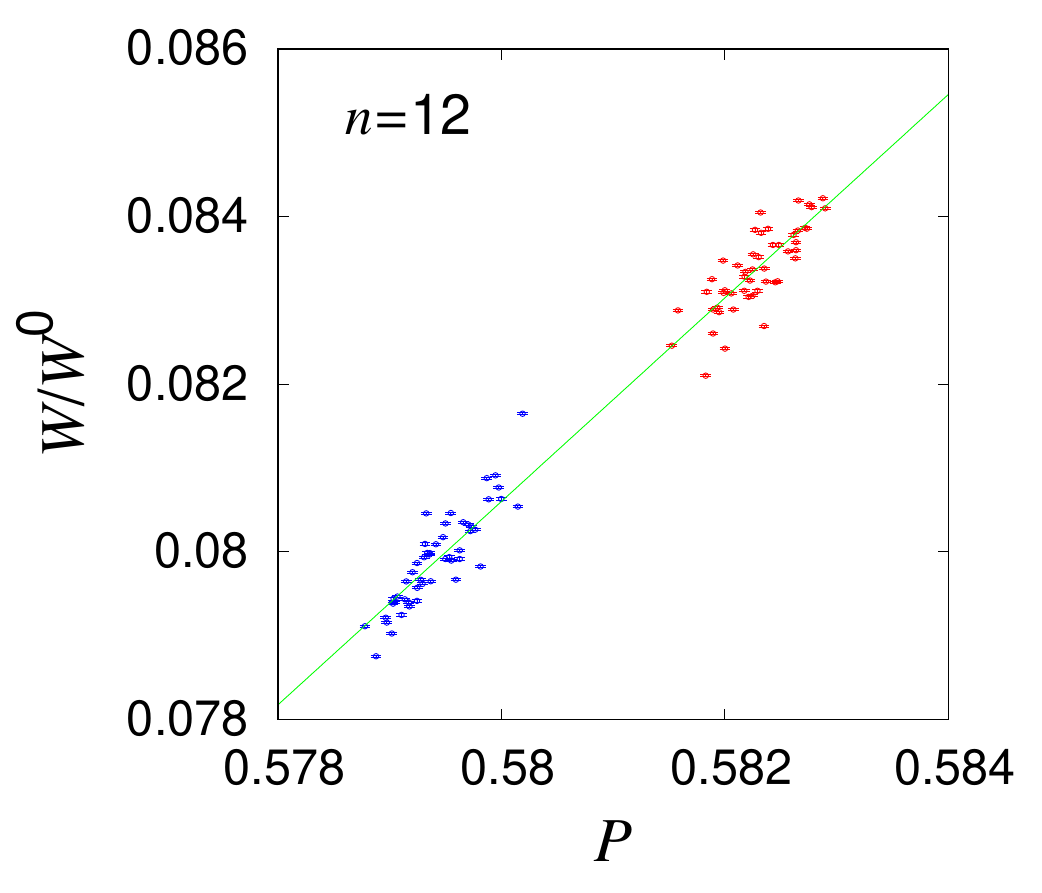}
\hspace*{-7mm}
\includegraphics[width=5.8cm]{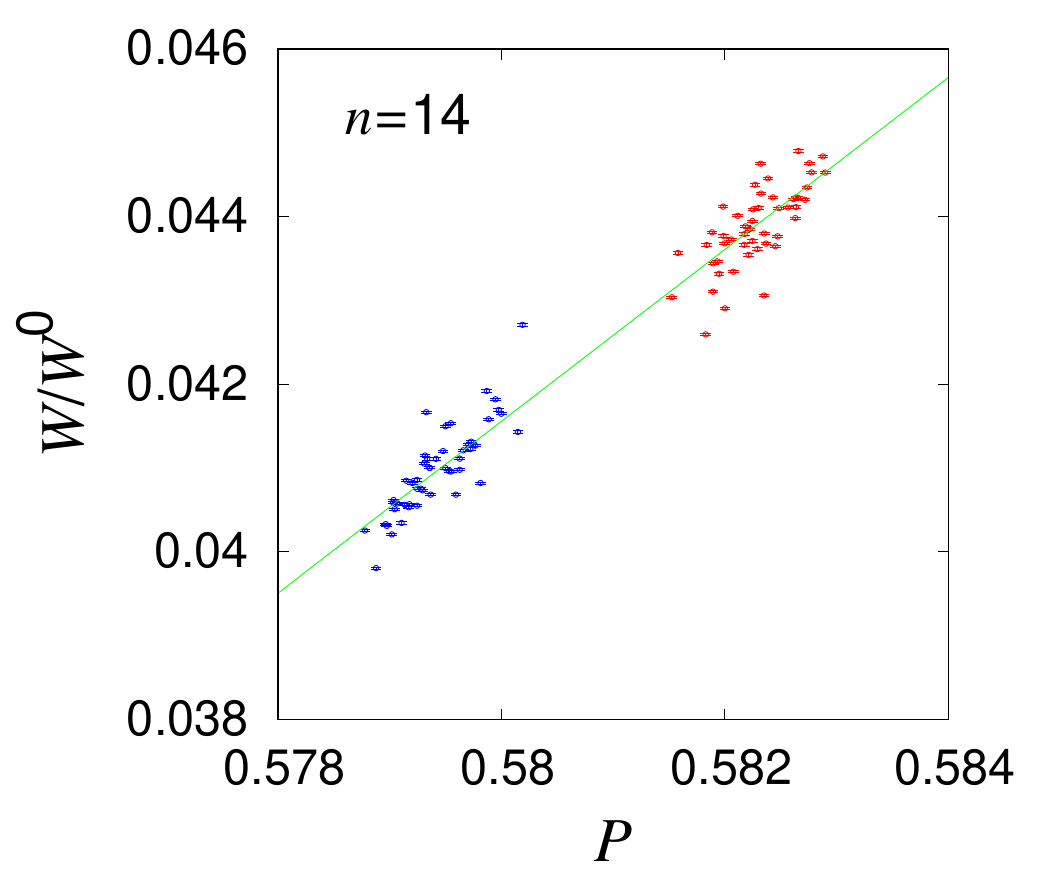} \\
\includegraphics[width=5.8cm]{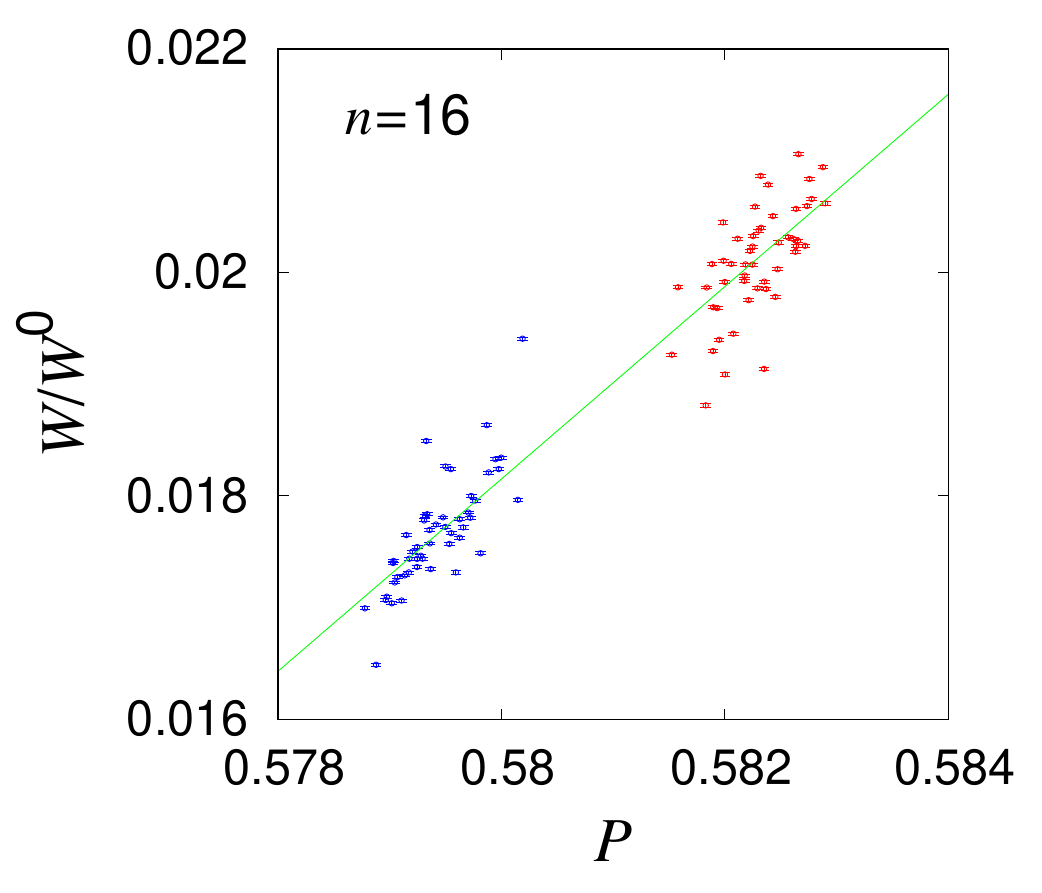}
\hspace*{-7mm}
\includegraphics[width=5.8cm]{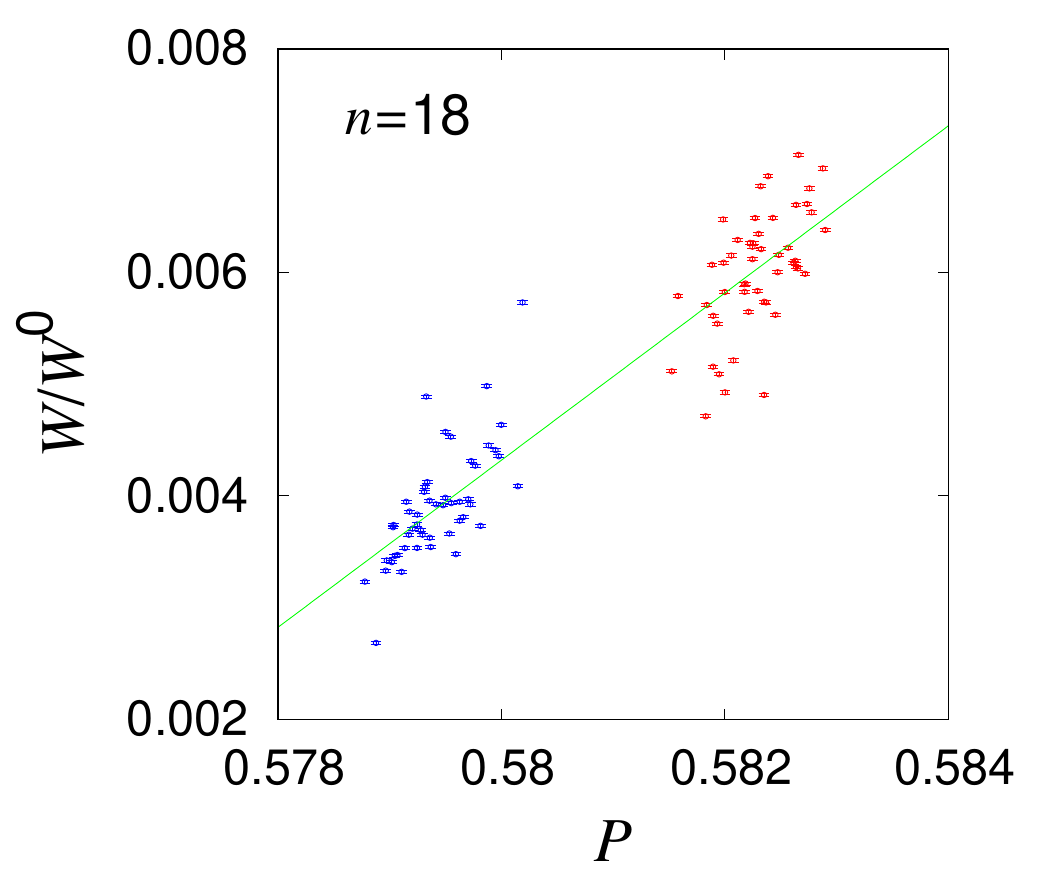}
\hspace*{-7mm}
\includegraphics[width=5.8cm]{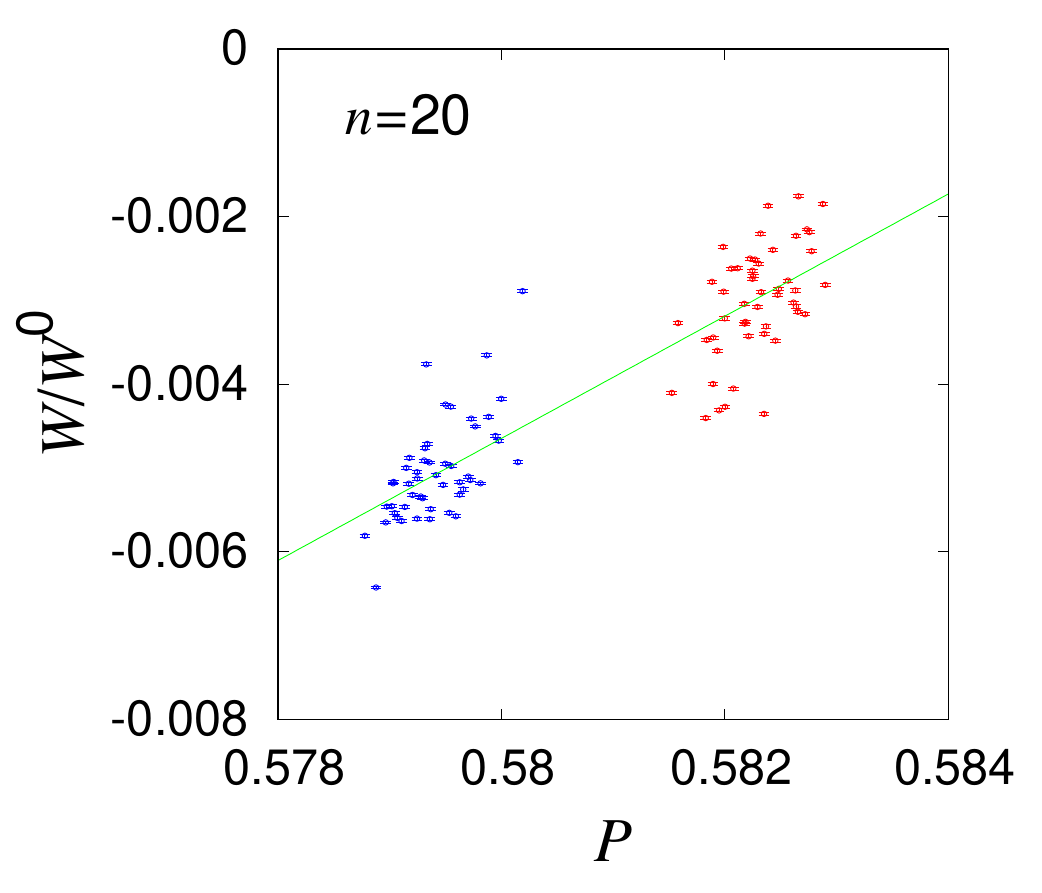}
\vspace{0mm}
\caption{Double distribution of the Wilson loop term $W(n)/W^0(n)$ and the plaquette $\hat{P}$ in quenched QCD on a $32^3\times6$ lattice.
The top left, top middle, $\cdots$, and bottom right panels show the results of $n=4$, 6, $\cdots$, and 20, respectively.
Blue symbols are obtained at $\beta=5.8810$, and red symbols at $\beta=5.9000$, which are slightly below and above the phase transition point.
}
\label{fig:corrwn}
\end{center}
\end{figure}

\begin{table}[t]
\begin{center}
\caption{Coefficients $d_n$ and $f_n$ of Eq.~(\ref{eq:wlinear}) obtained on $N_t=6$ and 8 lattices. 
The numbers in the first parenthesis are the statistical errors by the jackknife method and those in the second parenthesis are the errors of the noise method.}
\label{tab:dn}
\begin{tabular}{ccccc}
\hline
$n$ & \hspace{4mm} $d_n (N_t=6)$ \hspace{4mm} & \hspace{4mm} $f_n  (N_t=6)$ \hspace{4mm} & \hspace{4mm} $d_n (N_t=8)$ \hspace{4mm} & \hspace{4mm} $f_n  (N_t=8)$  \hspace{4mm}\\
\hline
$4$  & 1   & 0  & 1  & 0 \\
$6$  & 1.3625(73)(12)  & $-0.4070(42)(7) $ & 1.3366(66)(8)   & $-0.3922(39)(5) $ \\
$8$  & 1.4644(123)(11) & $-0.6089(72)(6) $ & 1.4256(96)(8)   & $-0.5869(57)(5) $ \\
$10$ & 1.3835(156)(10) & $-0.6590(91)(6) $ & 1.3433(117)(8)  & $-0.6367(70)(5) $ \\
$12$ & 1.2140(178)(9)  & $-0.6235(103)(5)$ & 1.1752(130)(7)  & $-0.6025(78)(4) $ \\
$14$ & 1.0256(196)(9)  & $-0.5533(114)(5)$ & 0.9825(141)(7)  & $-0.5303(85)(4) $ \\
$16$ & 0.8607(219)(9)  & $-0.4811(127)(5)$ & 0.8052(153)(8)  & $-0.4512(92)(5) $ \\
$18$ & 0.7481(258)(10) & $-0.4296(150)(6)$ & 0.6698(173)(9)  & $-0.3870(103)(5)$ \\
$20$ & 0.7290(337)(12) & $-0.4275(196)(7)$ & 0.6071(219)(12) & $-0.3606(131)(7)$ \\
\hline
\end{tabular}
\end{center}
\end{table}

Finally, we study the correlation among Wilson loops.
Figure~\ref{fig:corrwn} shows the double distribution of $W(n)/W^0(n)$ and the plaquette $\hat{P}$ measured at $\beta=5.881$ and 5.900 on the $32^3\times6$ lattice.
In the top left, top middle, $\cdots$, and bottom right panels, the results for $n=4$, 6, 8, $\cdots$, and 20 are shown, respectively,
The top left panel shows that the numerical result of $W(4)/W^0(4)$ is equal to the plaquette $\hat{P}$ within the error of the noise method. 
We find that, though the $W(6)/W^0(6)$ data (the top middle panel) shows a strong linear correlation with $\hat{P}$, the correlation becomes gradually weak as $n$ increases. 
Similar results are obtained in the calculation on the $32^3 \times 8$ lattice.

For small $n$, say $n\simle10$, for which the linear correlation with $\hat{P}$ is strong, we may approximate $W (n)$ as 
\begin{eqnarray}
W(n) \; \approx \; W^0 (n) \, (d_n \hat{P} + f_n),
\label{eq:wlinear}
\end{eqnarray}
where $(d_4, f_4)=(1, 0)$ for the leading term $n=4$.
For $n \ge 6$, the coefficients $d_n$ and $f_n$ are obtained by fitting the data shown in Fig.~\ref{fig:corrwn} by~Eq.~(\ref{eq:wlinear}).
The results of the fits are shown by the green lines in~Figs.~\ref{fig:corrwn} and listed in Table~\ref{tab:dn} together with the results of $N_t=8$.
As expected from the fact that $W^0(n)$ does not depend on $N_t$, 
the $N_t$ dependence of the fit parameters $d_n$ and $f_n$ is quite small. 

Then, in a similar fashion to the effective theory discussed in Sec.~\ref{keffect}, we may incorporate the effect of higher-order Wilson loop terms by a shift of $\beta$ 
\begin{eqnarray}
\beta \; \longrightarrow  \; 
\beta^* = \beta + \frac{1}{6} \, N_{\rm f} \sum_{n=4}^{n_{\rm max}} \, W^0 (n) \, d_n \, \kappa^n 
\label{eq:replaceW}
\end{eqnarray}
in the gauge action.
Here, the first correction term is for $n=4$ and is just the $48 N_{\rm f} \kappa^4$ term in the right hand side of Eq.~(\ref{eq:wltermseffects}).

For $n$ larger than about 10, because the correlation between $W (n)/W^0(n)$ and $\hat{P}$ is not quite strong, 
the effect of $W(n)$ may not be well replaced by the shift of Eq.~(\ref{eq:replaceW}) only.
However, as discussed in Sec.~\ref{wilsonloop}, 
the contribution of $n\simge10$ Wilson loops is small in the effective quark action up to $\kappa \simle 0.125$, 
and their remaining effect will be effectively absorbed by a small shift of improvement parameters of improved gauge action.

\begin{figure}[tb]
\begin{center}
\vspace{0mm}
\includegraphics[width=7.8cm]{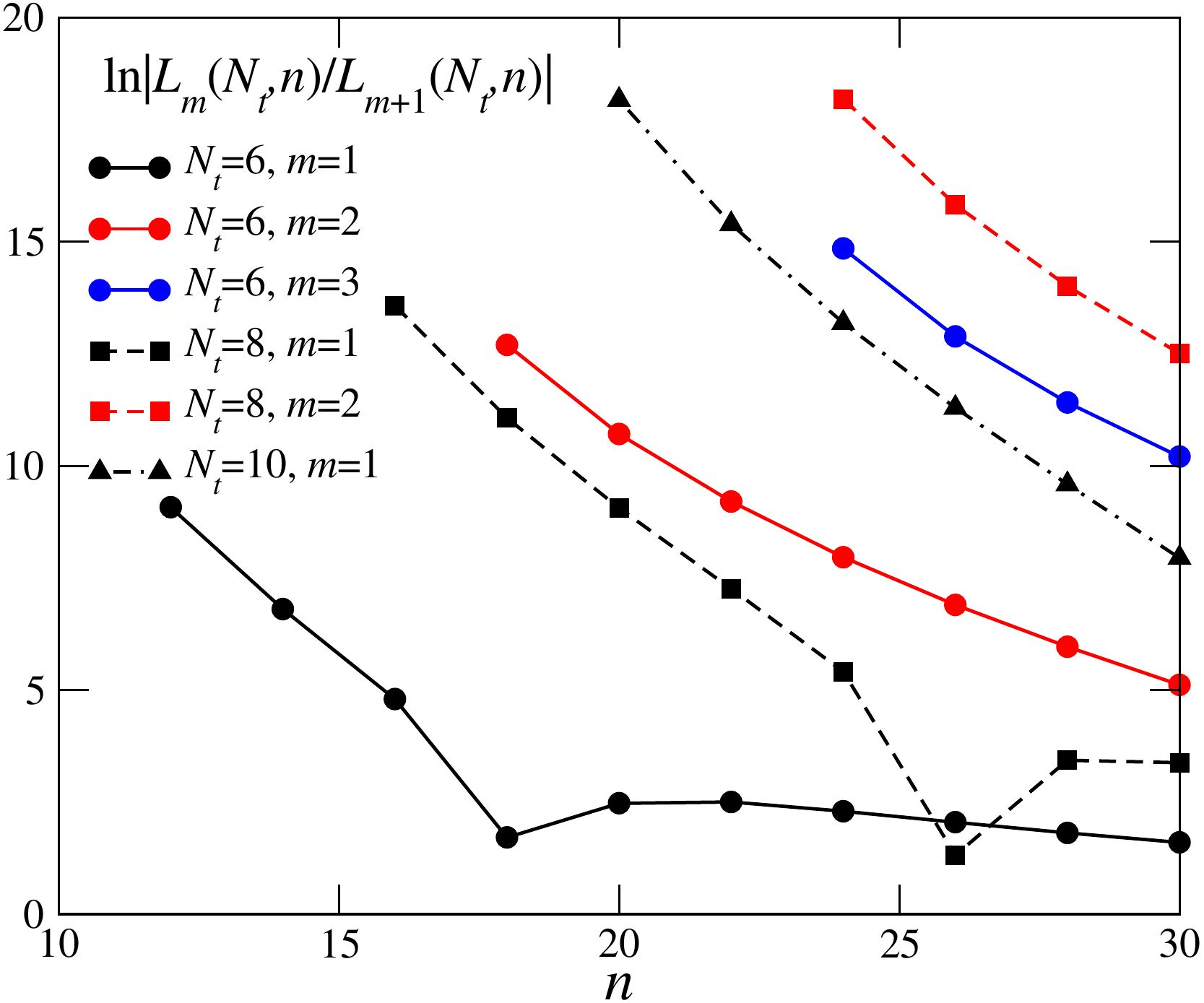}
\vspace{0mm}
\caption{Upper bound of $\mu/T$ that higher-$m$ term is small, given in~Eq.~(\ref{eq:rmut}).}
\label{fig:muconv}
\end{center}
\end{figure}

\section{Critical point at non-zero densities}
\label{sec:nonzeromu}

In Ref.~\cite{Saito:2013vja}, we studied the critical point of heavy quark QCD at non-zero density on an $N_t = 4$ lattice adopting the leading-order approximation of the hopping parameter expansion. 
As discussed in previous sections, the leading-order term is dominant around the critical point for $N_t = 4$ at zero density.
Because the critical point moves toward smaller $\kappa$ as the chemical potential increases, the results of Ref.~\cite{Saito:2013vja} are reliable.
To extend the study to larger values of $N_t$, however, we need to estimate the influence of higher-order terms.

With finite chemical potential $\mu$, the Wilson quark kernel changes to 
\begin{eqnarray}
M_{xy} (\kappa) \;=\; \delta_{xy} 
& - \kappa & \left[ \sum_{\mu=1}^3 
\left\{  (1-\gamma_{\mu})\,U_{x,\mu}\,\delta_{y,x+\hat{\mu}} 
+ (1+\gamma_{\mu})\,U_{y,\mu}^{\dagger}\,\delta_{y,x-\hat{\mu}} \right\} 
\right. \nonumber \\ && \left.
+ \; (1-\gamma_4)\,U_{x,4}\; e^{+\mu a} \,\delta_{y,x+\hat{\mu}} 
+ (1+\gamma_4)\,U_{y,4}^{\dagger}\; e^{-\mu a} \,\delta_{y,x-\hat{\mu}}
\right].
\label{eq:Mxydf}
\end{eqnarray}
Because the link variables change as $U_{x,4} \to U_{x,4} e^{+\mu a}$ and 
$U_{x,4}^{\dagger} \to U_{x,4}^{\dagger} e^{-\mu a}$ in the quark kernel, $L_m^+$ and $L_m^-$ change as
$L_m^+ \to (e^{+\mu a})^{m N_t}  L_m^+ = e^{+m \mu/T} L_m^+ $ and 
$L_m^- \to (e^{-\mu a})^{m N_t}  L_m^{-}  = e^{-m \mu/T} L_m^- $, respectively, 
where, as introduced in Sec.~\ref{hopping}, $L_m^+$ and $L_m^- = (L_m^+)^*$ are the parts of $L_m$ going in the positive and negative directions, respectively.
Therefore, the equation to determine the critical point is changed as follows, 
\begin{eqnarray}
\sum_{n=N_t}^{n_{\rm max}} \sum_{m=1}^{\infty} \left[ L_m^+ (N_t, n) e^{+m \mu/T} 
+ L_m^- (N_t, n) e^{-m \mu/T} \right] \kappa_c^n = L^0 (N_t, N_t) \, \kappa_{\rm c, LO}^n \, {\rm Re} \hat{\Omega} .
\label{eq:keffft}
\end{eqnarray}

Corresponding to the effective theory discussed in Sec.~\ref{keffect} based on the strong correlation among the Polyakov-type loops,
let us assume that
\begin{eqnarray}
L_m^+ (N_t, n) \approx \frac{1}{2} L^0_m (N_t, n) \, c_{n,m} \, {\rm Re} \hat{\Omega} ,
\end{eqnarray}
where $c_{n,m}$ is a constant to be determined by a Monte Carlo simulation. 
Then, Eq.~(\ref{eq:keffft}) becomes 
\begin{eqnarray}
\frac{1}{2} \sum_{n=N_t}^{n_{\rm max}} \sum_{m=1}^{\infty} 
L_m^0 (N_t, n) \left( c_{n, m} e^{+m \mu/T} 
+ c_{n, m}^* e^{-m \mu/T} \right) \kappa_c^n = L^0 (N_t, N_t) \, \kappa_{\rm c,  LO}^n .
\label{eq:kefffta}
\end{eqnarray}

The leading-order calculation of Ref.~\cite{Saito:2013vja} for $N_t=4$ corresponds to the case where the terms $m \geq 2$ are absent.
To judge the magnitude of the effect from higher-$m$ terms, we again consider the case of the worst convergence with $U_{x,\mu}=\mathbf{1}$. 
In this case, because $c_{n, m}=1$, Eq.~(\ref{eq:kefffta}) reads 
\begin{eqnarray}
\sum_{n=N_t}^{n_{\rm max}} \sum_{m=1}^{\infty} 
L_m^0 (N_t, n) \left( \cosh{ \frac{m \mu}{T}} \right) \kappa_c^n 
= L^0 (N_t, N_t) \, \kappa_{\rm c, LO}^n .
\end{eqnarray}
Though $L_m^0 (N_t, n)$ decreases as $m$ increases, $\cosh(m \mu/T)$ may be large when $\mu/T$ is not small.
Approximating $\cosh(m \mu/T) \approx e^{m \mu/T}$ for $m \mu/T >1$, 
we find
\begin{eqnarray}
\frac{\mu}{T} < \ln \left| \frac{L_m^0 (N_t, n)}{L_{m+1}^0 (N_t, n)} \right| 
\label{eq:rmut}
\end{eqnarray}
as a condition that the effect of higher-$m$ term is small. 
In Fig.~\ref{fig:muconv}, we show the right hand side of Eq.~(\ref{eq:rmut}) computed from Table \ref{tab:lmn}. 
The circle, square and triangle symbols are the results for $N_t=6$, 8 and 10, respectively.
The black, red and blue lines mean $m=1$, 2 and 3.
When $\mu/T$ exceeds these values, effects of higher-$m$ terms should be incorporated.

\section{Summary and conclusions}
\label{conclusion}

We studied the convergence and the valid range of the hopping parameter expansion in the determination of the critical point (critical quark mass) of finite-temperature QCD with heavy quarks at which the first-order deconfinement transition in the heavy quark limit turns into crossover at intermediate quark masses. 
Adopting the standard plaquette gauge action and the standard Wilson quark action, 
we expand the effective quark action $\ln \det M$ by the hopping parameter $\kappa$ around the heavy quark limit $\kappa=0$, with $M(\kappa)$ the Wilson quark kernel.
Non-vanishing contributions to the expansion terms are given by closed loops of the hopping term $B=- \partial M/\partial \kappa$.
We classified the closed loops by the winding number $m$ in the temporal direction, and decomposed each expansion term into Wilson loop term ($m=0$) and Polyakov-type loop terms ($m\ne0$).
We developed a general method to calculate Wilson and Polyakov-type loop terms from the expansion terms with various twisted boundary conditions in the temporal direction.

To study the convergence of the hopping parameter expansion, 
we first studied the case of the worst convergence in which all the gauge link variables are unit matrices and thus the Wilson loops and the Polyakov-type loops get their maximum values.
Our explicit calculation of the Wilson and Polyakov-type loop terms up to the 100$^{\rm th}$ order of the hopping parameter expansion shows that the hopping parameter expansion is convergent up to around the chiral limit of free Wilson quarks, $\kappa=0.125$, meaning that the convergence radius of the hopping parameter expansion is not small. 

In practice, however, we need to truncate the expansion at some finite order and have to take into account the systematic error due to the truncation.
We thus studied the issue of the truncation error of the hopping parameter expansion, focusing on the determination of the critical point $\kappa_c$ in heavy quark QCD.
In the case of worst convergence, we found that, the truncation error on $N_t=4$ lattices is well under control up to around $\kappa \sim 0.1$, ensuring the previous next-to-leading order calculations of $\kappa_c$ for $N_t=4$~\cite{Saito:2011fs,Saito:2013vja}. 
We also found that the truncation error increases as $N_t$ increases, such that, already for $N_t=6$, significant effect from higher-order terms exist around $\kappa_c$ determined by a next-to-leading order calculation.

To extend the valid range of the hopping parameter expansion, 
we thus revisit the effective theory of Refs.~\cite{Saito:2013vja,Ejiri:2019csa} 
which incorporates the next-to-leading effect into the leading-order calculation, 
and extend it to higher-orders of the hopping parameter expansion.
We also discussed that the effect of Wilson loop terms can be represented by a shift of coupling parameters in the gauge action.
The effective theory is based on the strong correlation between the leading-order Polyakov loop and next-to-leading bend Polyakov loops. 
By a Monte-Carlo simulation, we showed that the strong correlation holds also for higher-order Polyakov-type loops. 
We thus extended the effective theory to include higher-order terms of the hopping parameter expansion, 
and determined the coefficients needed in the effective theory.
Using the effective theory, we discussed that the truncation error of the hopping parameter expansion is well under control for $\kappa\simle0.125$ when the higher-order effect is incorporated into the effective theory up to sufficiently high orders. 
We evaluated the higher-order correction of the critical point $\kappa_c$ for $N_t=6$. 
We also derived formulae for the critical point with general number of flavors at zero and finite densities.

In this paper, we have discussed the application of the hopping parameter expansion to the reweighting factor Eq.~(\ref{eq:hpeIV}) 
for quenched QCD configurations. 
Besides the truncation error of the hopping parameter expansion,
applicability range of the method is limited also by the overlapping problem of the reweighting method~\cite{Saito:2011fs,Saito:2013vja,Ejiri:2019csa} . 
Here, we note that the effective theory we developed is applicable also to generate configurations 
effectively incorporating dynamical quark effect up to $n_{\rm max}$th order of the hopping parameter expansion:
\begin{eqnarray}
S_{\rm eff} = -6 N_{\rm site} \, \beta^* \hat{P} - N_s^3 \lambda \, {\rm Re} \hat{\Omega}
\quad 
\textrm{with} \hspace{5mm} \lambda = N_{\rm f} N_t \, L^0 (N_t, N_t) \, (\kappa^*)^{N_t},
\label{eq:lamk}
\end{eqnarray}
where $\beta^*$ and $\kappa^*$ are given by Eqs.~(\ref{eq:replaceW}) and~(\ref{eq:replaceL}), respectively.
As performed in~Ref.~\cite{Kiyohara:2021smr}, Monte Carlo simulation with this action can be carried out efficiently.
Because the Polyakov loop $\hat{\Omega}$ is the order parameter of the deconfinement transition of QCD in the heavy quark limit, incorporation of its effect into the configuration can lead to drastic improvements in the lattice study of the QCD phase transition.
In~Ref.~\cite{Kiyohara:2021smr}, it was shown that the configuration generated by Eq.~(\ref{eq:lamk}) with the leading-order $\beta^*$ and $\kappa^*$ removes the overlapping problem in the reweighting to incorporate the next-to-leading order effect.
Effective inclusion of higher-order effect in the configuration will help achieving the high orders of the hopping-parameter expansion required in a study of $\kappa_c$ for large values of $N_t$, thus extending the scope of the hopping parameter expansion.
As noted in Refs.~\cite{Ejiri:2019csa,Kiyohara:2021smr}, $\kappa_{c}$ has visible finite volume effect.
To eliminate the finite volume effect, we need to repeat determination of $\kappa_{\rm c, LO}$ for various spatial volumes.

\section*{Acknowledgments}

The authors thank the members of the WHOT-QCD Collaboration for useful discussions.
This work was in part supported by JSPS KAKENHI Grant Numbers JP21K03550, JP20H01903,
JP19K03819, JP19H05146, and JP19H05598, 
the HPCI System Research project (Project ID: hp200089, hp210039), 
and Joint Usage/Research Center for Interdisciplinary Large-scale Information Infrastructures in Japan (JHPCN) (Project ID: jh200049).

\bibliographystyle{apsrev4-1}
\bibliography{refs.bib}

\end{document}